\documentclass[prl, twocolumn, noshowpacs]{revtex4}
\usepackage{amssymb}
\usepackage{amsmath}
\usepackage{epsfig}
\usepackage{bm}

\begin{document}

\title{ Viscosity of two-dimensional electrons }
\author{ P. S. Alekseev and A. P. Dmitriev }
\affiliation{
 Ioffe  Institute, Politekhnicheskaya 26,
  194021,   St.~Petersburg,   Russia }

 \begin{abstract}

The hydrodynamic regime of electron transport has been recently realized in conductors with ultra-low densities of defects. Although relaxation processes in two-dimensional (2D) fluids have been studied  in many theoretical works, the viscosity of the realistic Fermi gas of 2D electrons having the quadratic energy spectrum and interacting by Coulomb's law has not been reliably determined either in theory or in experiment up to now. Here we construct a theory of viscosity and thermal conductivity in such system. We compare the calculated viscosity of the 2D electron Fermi gas and the previously known viscosity of a 2D Fermi liquid with available experimental data extracted from the hydrodynamic negative magnetoresistance of the best-quality GaAs quantum wells. Based on this comparison, we argue that measurements of the temperature dependence of the viscosity
can allow to trace the transition between an electron Fermi liquid and a Fermi gas.


 \end{abstract}

 \maketitle

{\em 1. Introduction.}   At low density of defects,
conduction electrons    in solids can form a viscous fluid,
which transfers heat and charge along the  sample.
 Such hydrodynamic regime of transport was recently discovered
in novel ultra-pure conductors:  graphene
 \cite{grahene}-\cite{rev},  monovalent layered metal
 PdCoO$ _2 $ \cite{Weyl_sem_1}, bulk Weyl metal WP$_2 $
\cite{Weyl_sem_2},  and high-mobility  GaAs quantum wells
 \cite{exps_neg_1}-\cite{Alekseev_Alekseeva}.

 In dissipative  hydrodynamic  phenomena,
  the key physical process, controlling the magnitude
of viscosity,  is the relaxation  of the shear  stress
of the electron fluid.   For Fermi systems
in low-defect  materials, this process is mainly due to
 inter-particle collisions being controlled by the numbers
of filled and empty quantum  particle states. Therefore
 for the shear stress relaxation time $\tau_{ee,2}$
of 2D electrons it is reasonable to expect that
 $ \hbar / \tau_{ee,2} \sim T^2 / \varepsilon_F$,
 as for the scattering times in 3D Fermi systems
\cite{LP_10,LP_9} (here  $T$   is temperature
and $ \varepsilon_F \gg T $ is the Fermi energy).
 However, the actual relaxation times of 2D fermions
 can deviate significantly from this estimate due to
 kinematic constrictions in 2D electron collisions and
 a sharp angular dependence  of the scattering probability
\cite{Novikov}-\cite{Ledwith_1_2}.

Depending on the strength of the interparticle
 interaction, an electron Fermi system in the hydrodynamic
 regime should be treated as a viscous  gas or
 a viscous liquid. The  possibility of propagation of the
 transverse sound waves can be considered as
 a characteristic property that distinguishes
 these cases.  In a  Fermi liquid,   such modes appear
 at sufficiently  large Landau interaction  parameters
 \cite{LP_9}. For an electron  Fermi liquid,
 this corresponds to relatively low electron densities.
The     transverse sound waves in 2D electron
 and electron-hole liquids were studied in
 Refs.~\cite{Khoo_Villadiego,Svintsov,Link,
 Alekseev_Alekseeva,Alekseev_Semiconductors}.
For a 2D electron  fluid in a magnetic field
 such waves can  manifest itself by
 the viscoelastic resonance  at a double cyclotron frequency
  \cite{Alekseev_Alekseeva,Alekseev_Semiconductors,vis_res},
 that   was   apparently observed in the record-quality
GaAs quantum  wells
\cite{exp_GaAs_ac_1,exp_GaAs_ac_2,Alekseev_Alekseeva}.

The viscosity effect in a strongly non-ideal 2D Fermi liquid
  was theoretically  studied in Ref.~\cite{Novikov}.
It was shown that the main contribution  to the relaxation
of the shear stress    comes from   the collisions of the pairs
 of quasi-particles    with the total momentum being close
 to zero (the ``head-on collisions'').  The probability
 of such collisions in a Fermi liquid is strongly
 affected by  the  Cooperon pairing (that is responsible for
superconductivity in the case of the attractive
interparticle interaction).  The resulting stress
 relaxation rate substantially differs from the $T^2$-dependence:
\begin{equation}
  \label{tau_Ferm_liq}
   \frac{ 1 }{ \tau_{ee,2} }
   \propto \frac{ T^2 }{ \ln^2(\varepsilon_F/T) }
  \:.
\end{equation}

 Relaxation   of  perturbations of various types
  in a 2D  Fermi gas  was theoretically  examined
 in Refs.~\cite{Ledwith_1}-\cite{Ledwith_1_2}.
 It was shown that  the head-on collisions also
 play the main role  for  relaxation   of the shear stress.
 However, the oversimplifications in the inter-particle
scattering probability were made in
  Refs.~\cite{Ledwith_1}-\cite{Ledwith_1_2}
not allowing a proper evaluation of the relaxation rates
for the Fermi gas of 2D electrons interacting
by the screened Coulomb potential.

Viscosity of 2D carriers in graphene was theoretically
studied  in Ref.~\cite{Principi_1}.
 The hydrodynamic equations were derived from
  the current-current response functions.
 In particular, it was obtained that the viscous stress  relaxation  rate
  in the limit of low temperatures
 is proportional to the squared temperature.
 The thermal conductivity of 2D electrons
  with Dirac and quadratic spectrums
  was also theoretically examined in Ref.~\cite{Principi_2}.

In this work we study viscosity and thermal conductivity
 of a Fermi gas of 2D electrons with a quadratic spectrum .
   We show that,  in contrast to
3D  Fermi systems \cite{Abr},\cite{Sykes},  the energy part
 of the electron distribution function  describing
 a viscous flow  has a simple structure corresponding
to the relaxation time approximation for the inter-electron
 collision integral.  Based on this result, we calculate
 the shear stress relaxation time $\tau_{ee,2}$,
determining the viscosity:
 \begin{equation}
 \label{tau_2_res}
   \frac{ \hbar }{ \tau_{ee,2} } =
   \frac{T^2}{  \varepsilon_F }
    \, r_s^2  \, \Lambda
 \:, \qquad
   \Lambda =  \frac{8\pi}{3} \:
    \ln\Big( \,  \frac{ 1}{ \zeta+ r_s } \Big) \:,
\end{equation}
 where  $\zeta = T/ \varepsilon_F$,  $ r_s  = 1/(\sqrt{\pi n} a_B)$ is the interparticle
interaction  parameter which is small for the Fermi gas, $ n $ is the 2D electron density,
 and $a_B$ is the Bohr radius. The rate $1/\tau_{ee,2}$~(\ref{tau_2_res})
  depends on temperature as $\sim T^2$ at
  very low temperatures, $\zeta \ll r_s  $,
  and as
  $\sim T^2 \ln(\varepsilon_F/T)$
  at moderately low temperatures,
 $ r_s \ll \zeta \ll1$.
 Such result reflects the character of interparticle scattering in the 2D electron Fermi gas and
 substantially differs from result (\ref{tau_Ferm_liq}) for the  Fermi-liquid.
    We also
 estimated the time $\tau_{ee,h}$ of relaxation  of the heat flow,
determining the thermal conductivity of 2D electron gas.  Such time $\tau_{ee,h}$
 turns out to be much shorter than  $\tau_{ee,2}$ and close
to  the quantum lifetime $\tau_{ee,q}$ at $\zeta \ll r_s$,
in accordance with the results of Ref.~\cite{Principi_2},
 while at
 $ r_s \ll \zeta \ll1$ the time $\tau_{ee,h}$
  turns out to be much longer than $\tau_{ee,q}$.

We carry out an analysis of  temperature dependencies
 of  the 2D electron viscosity,  which we extract from
experimental  data  on the giant negative magnetoresistance
 in high-mobility GaAs  quantum wells
\cite{exps_neg_1,exps_neg_2,exps_neg_3,exps_neg_4,Gusev_1}.
 Such magnetoresistance, as it was demonstrated
in Ref.~\cite{je_visc}, is proportional  to the viscosity
 of 2D electrons.   We show that
the experimental temperature dependencies of
 the  shear stress relaxation time $\tau_2 $  are well
described by laws  (\ref{tau_Ferm_liq}) or (\ref{tau_2_res}),
 depending on the strength  of the interparticle interaction
determined by the electron density. We conclude that studies
 of temperature dependencies of viscosity can allow
to distinguish between the electron hydrodynamics
in a Fermi gas  and a Fermi liquid (see~Fig.~1).

{\em 2.  Collision integral. }  We study  scattering
  of two 2D  electrons with initial momenta
$ \mathbf{p}_1 $, $ \mathbf{p}_2 $ and final momenta
$ \mathbf{p}_3 $, $ \mathbf{p}_4 $, which are characterized
by the angles $\alpha $, $\varphi$, $\theta$, and $\psi$
as shown in Fig.~2(a). The momentum and the energy
 conservation laws for scattering particles have
the usual forms: $ \mathbf{p}_1  + \mathbf{p}_2 =
  \mathbf{p}_3  + \mathbf{p}_4 $ and  $  \varepsilon _1
  + \varepsilon _2 =   \varepsilon _3  + \varepsilon _4 $.
The energy spectrum of  2D electrons  is supposed to be
quadratic:  $ \varepsilon _i = m v_i^2 / 2$  and $ \mathbf{p}_i =
 m \mathbf{v}_i$, $i = 1,2,3,4$.

We consider small perturbations of an electron Fermi gas,
which are described by the distribution function:
 $ f _{\mathbf{p}}  =  f_F(\varepsilon ) +
 \delta f    _{\mathbf{p}}   $, where  $f_F$ is
the Fermi function and $    \delta f  _{\mathbf{p}}       =
  - f_F' ( \varepsilon  ) \, \Psi(\varepsilon  , \alpha)  $.
Relaxation of perturbations $\delta f   _{\mathbf{p}} $
is described by  the linearized inter-particle collision
 integral~\cite{LP_10,Ledwith_2,Novikov,Abr,Sykes}:
\begin{equation}
  \label{col_int}
\begin{array}{c}
 \displaystyle
\mathrm{St} [\Psi ] ( \varepsilon _1 , \alpha)
 =
 -
 \frac{f_F(\varepsilon _1) }{  T }
  \frac{m ^2   }{ (2 \pi \hbar) ^4  }
     \int  d \varepsilon _2  \, d \varepsilon _3   \:
    f_F(\varepsilon _2)
 \\
 \\
 \displaystyle
    \times  [1-f_F(\varepsilon _3)]
    [1-f_F( \varepsilon _4  )]
    \int  d \varphi  \,  d \theta \:\,
        W
    \left[\,
    \Psi (\varepsilon _1 , \alpha)
    \right.
    \\
    \\
    \displaystyle
    \left.
    + \Psi (\varepsilon _2 , \alpha +\varphi)-
    \Psi (\varepsilon _3 , \alpha +\theta)
    -
        \Psi ( \varepsilon _4  ,
       \alpha +    \psi
        \, ) \,
    \right]
    \:,
    \end{array}
\end{equation}
 where  $\varepsilon _4 = \varepsilon _1  + \varepsilon _2
 - \varepsilon _3 $, the scattering angle $\psi = \psi
 ( \varphi , \theta ; \breve{\varepsilon}   ) $ is calculated
 from the momentum  conservation law (for exact form of $\psi$
 see \cite{SI}), $\breve{\varepsilon}=(\varepsilon_1,\varepsilon_2,
  \varepsilon_3)$, and $   W =   W( \varphi, \theta ;
 \breve{\varepsilon} ) $  is the scattering probability:
\begin{equation}
\label{w}
 W = \frac{2\pi}{\hbar} \, M^2 \,     \delta\Big[
    \varepsilon _1 + \varepsilon _2 - \varepsilon _3
     - \frac{(\mathbf{p} _1 +
      \mathbf{p}_2 - \mathbf{p} _3 )^2}{2m}
        \Big]
        \:.
\end{equation}
  The delta-function in this formula, taking into account
the conservation of energy, leads to the equation
 for  the other scattering angle $\theta = \theta
(\varphi, \breve{\varepsilon })$ (see~\cite{SI}). The squared
matrix element  $M^2 =    M^2 (\varphi, \theta ;
\breve{\varepsilon}) $ in Eq.~(\ref{w}) for spin-unpolarized
 electrons  at sufficiently large scattering angles,
 $\theta \gg \zeta $, $\psi \gg \zeta  $, has
the form \cite{Ledwith_2}:
 \begin{equation}
   \label{matr_el}
   M^2 =
   2   \,
 [ \,  V(q)^2 + V(w)^2 -
   V(q) V (w) \, ]
  \:.
\end{equation}
Here the processes of the forward, $1,2 \to 3,4$,
 and the exchange, $1,2 \to 4,3$,  scattering  with
the transmitted momenta $\mathbf{q}=\mathbf{p}_3-\mathbf{p}_1$
and $ \mathbf{w} = \mathbf{p}_3 - \mathbf{p}_2 $ are
taken into account;   and  $ V(q)$ is the static screened
2D Coulomb potential in the Fourier representation within
the random phase approximation (RPA) \cite{SI}:
\begin{equation}
\label{scre_pot}
V(q) =
    2 \pi e^2 \,[\,
 q /  \hbar  + 2/  a_B \,] ^{-1}
  \:.
\end{equation}

 \begin{figure}[t!]
\centerline{\includegraphics[width=1\linewidth]{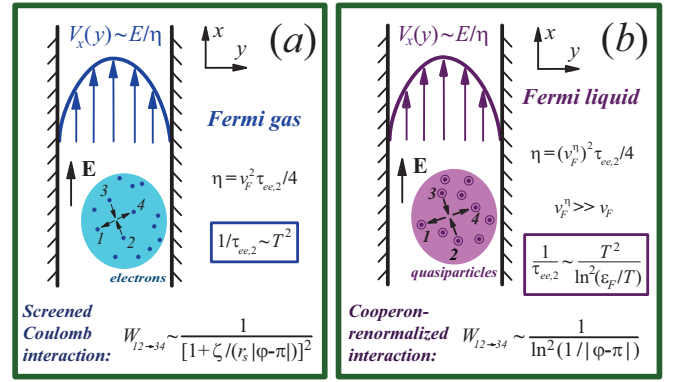}}
\caption{
Poiseuille flow
of an electron  Fermi gas (a) and a Fermi  liquid (b).
 In a Fermi liquid, the viscosity is strongly renormalized
 due to the interaction between quasiparticles,
that is expressed by the parameter $v_F^\eta \gg v_F$.
 }
\end{figure}

Further consideration is based on the smallness
of the two parameters: the  dimensionless temperature,
$   \zeta  \ll 1$, and  the interaction parameter, $r_s \ll 1$.
We will use  the random phase approximation,
being valid at $ r_s \ll 1 $ and any ratio $ \zeta / r_s $.
 The magnitude of  $   \zeta $ is always very small in
realistic  GaAs quantum wells at typical experiment conditions:
 $\zeta \sim 10^{-2} $ at $T=1$~K and $ n=10^{11 }  $~cm$^{-2}$.
while the parameter $r_s$
is smaller than unity only at rather  high electron densities:
 at $ n=10^{11 }  - 10^{12 } $~cm$^{-2}$
the magnitude of  $r_s$ varies  in the interval $1.7 - 0.55$.
So, when discussing experiments,
we will keep in mind that  $r_s$ is not too small and
the case  $\zeta / r_s \ll 1 $  is typically  realized.

The delta-function in Eq.~(\ref{w}) can be written as
the  sum of the  expressions,  corresponding to
 the two solutions $\theta_{\pm}   =  \theta_{\pm}
 (\varphi, \breve{\varepsilon}) $ of  the energy
conservation equation:
\begin{equation}
 \label{singul_factors}
 \sum \limits_{\pm}
  \frac{\delta (\theta -\theta _{\pm} ) }{
2 \, \sqrt{\varepsilon _3}  \, \left|
 \sqrt{\varepsilon _1 }
  \,
 \sin
 (
  \theta_{\pm}
 )
 +
 \sqrt{\varepsilon _2 } \,
 \sin
  (
  \theta_{\pm}
  -
 \varphi
  )
\right|
 }
  \:.
\end{equation}
 Analysis  \cite{SI} demonstrates that, after integration
in the collision integral~(\ref{col_int}) by the
angle $\theta$,   expressions (\ref{singul_factors})
 have the singularities at  $\varphi \to 0$ and
$\varphi \to \pi$, where the values of the $\pm$ terms
in Eq.~(\ref{singul_factors}) are much larger than
at $\varphi \sim 1 $.

The singularity in Eq.~(\ref{singul_factors}) at  $\varphi \to 0 $
refers to the collisions of electrons  with almost collinear
momenta $\mathbf{p}_{1} \approx \mathbf{p}_{2}$. The formulas for
the actual scattering angles  $\theta _{\pm} = \theta _{\pm}
(\varphi ,\breve{\varepsilon}) $  and $ \psi _{\pm}= \psi
[ \varphi  , \theta _{\pm}  (\varphi ,\breve{\varepsilon}),
\breve{\varepsilon}] $  leads  to the estimate:  $ \zeta \lesssim
|\theta _{\pm} |,|\psi _{\pm} |  \ll 1 $  at the small angles
of incident electron,   $ \zeta \lesssim  | \varphi | \ll 1 $
(see Fig.~2(b) and \cite{SI}). So the contribution  from
the $\varphi \to 0 $ singularity corresponds   to
the  almost-collinear small-angle scattering. In this case,
 one should take into account the dependence  of
$ M^2 $  on the particle energies $ \check{\varepsilon} $
which is neglected in Eq.~(\ref{matr_el}) \cite{SI}.

The singularity at $ \varphi \to \pi $  in Eq.~(\ref{singul_factors}) corresponds to scattering
of electrons with  almost oppositely directed momenta,
 $ \mathbf{p}_{1} \approx -\mathbf{p}_{2}$, generally speaking,
 on arbitrary angles, $0 < | \theta _{\pm}  |< \pi  $.
Such ``head-on'' collisions are most important for relaxation of
 the distribution function describing a viscous flow
 \cite{Ledwith_1},\cite{Ledwith_2}. Due to the form
of matrix element~(\ref{matr_el}),    the most significant
are again  the collisions   in which  one of the scattering  angles
is small:    $\zeta \ll | \theta _{\pm} | \ll 1  $ or
 $ \zeta \ll |\psi _{\pm} | \ll 1 $ (see Fig.~2(b) and \cite{SI}).
 For such $ \theta_{\pm} $ and   $\psi_{\pm} $, one can use
 the static matrix element (\ref{matr_el}) with the omitted
dependence on the energies \cite{SI}.

{ \em 3. Relaxation of shear stress  and heat flow. }
  The effect of shear viscosity consists in arising of
a shear stress in a gas or in a liquid due to the presence of
 an inhomogeneity of the hydrodynamic  velocity
$\mathbf{V}(\mathbf{r})$.  This effect is described by
the distribution function $\Psi = \mathbf{p} \cdot
\mathbf{V}  + \Psi _s  $, where $\mathbf{p} \cdot \mathbf{V} $
is the locally equilibrium part  corresponding to
the mean fluid velocity $\mathbf{V} =  \mathbf{V}   (\mathbf{r}) $,
 while $\Psi _s  $ is the nonequilibrium part being proportional
to the second angular harmonic by $\mathbf{p}$ and leading to
a nonzero shear stress tensor   $\sigma_{ik} =
\int  \, f_F' (\varepsilon)    \Psi_s(\varepsilon, \alpha)
p_i  v_k   d ^2 \mathbf{p}/(2\pi^2 \hbar^2) $.  For
 an incompressible flow, where  $\mathrm{div } \,
 \mathbf{V} = 0$,  the kinetic equation  $ \mathbf{v}  \cdot
 \partial f_{\mathbf{p}} / \partial \mathbf{r}
 = \mathrm{St}[f_{\mathbf{p}}] $ leads to
the following form of $\Psi_s$:
 \begin{equation}
  \label{Psi_s}
   \Psi_s(\varepsilon,\alpha)  =   \frac{m v_F^2  }{4}
   \, \Big( \, \frac{   \partial V_i }{ \partial x_k} +
        \frac{ \partial V_ k}{  \partial x_i  } \, \Big)
        \: F_{  ik }
      \:,
    \end{equation}
where $F _{  xx,yy} = \pm F(\varepsilon ) \cos (2\alpha) $,
$ F_{xy}   = F(\varepsilon ) \sin (2\alpha) $. The operator
$\mathrm{St}$ is diagonal by angular harmonics,
for example, $\mathrm{St} [F_{  xx} ] (\varepsilon_1,\alpha)
 \propto \cos (2\alpha) $. Thus the kinetic equation
for the function $F(\varepsilon ) $ takes the form:
\begin{equation}
 \label{int_eq}
       \mathrm{St} [F(\varepsilon )\cos (2\phi) ]
       \Big|_{\alpha=0}
=      - f_F ' (\varepsilon _1)
 \, .
 \end{equation}
The viscosity coefficient $\eta$ is defined as the coefficient
of  proportionality   between the stress tensor $ \sigma_{ik}$
and   the derivatives of $\mathbf{V}(\mathbf{r})$:
 $ \sigma_{ik} =   \eta  \, ( \partial V_i /\partial x_k  +
\partial V_ k /\partial x_i   ) $, $ \eta = m n v_F^2 \tau_{ee,2}
/ 4 $, where $\tau_{ee,2}$ is the shear stress relaxation time.
From this formula for  $\eta$, Eq.~(\ref{Psi_s}), and
the  above  formula  for  $\sigma_{ik}$  we obtain:
 $   \tau_{ee,2}  =  \int d \varepsilon  \,   f_F '
(\varepsilon )   F(\varepsilon ) $.

In the presence
 of a magnetic field, the term  $\omega_c \, \partial
  f_{\mathbf{p}}  / \partial \alpha$ arises in
the kinetic equation. This leads to arising of
 the two viscosity coefficients: the diagonal viscosity $\eta_{xx}$ and the non-diagonal (Hall) viscosity
  $\eta_{xy}$ in the relation between  $ \sigma_{ik}$ and $ \partial V_i /\partial x_k $~\cite{je_visc}.
The time $\tau_{ee,2}$
  determines  the magnitude as well as the magnetic field  dependencies
of the both viscosities: $\eta_{xx} = \eta /(1+4 \omega_c ^2\tau_{ee,2}^2)$
and $\eta_{xy} = 2 \omega_c \tau_{ee,2} \eta  /(1+4 \omega_c ^2\tau_{ee,2}^2)$~\cite{je_visc}.

 \begin{figure}[t!]
\centerline{\includegraphics[width=1\linewidth]{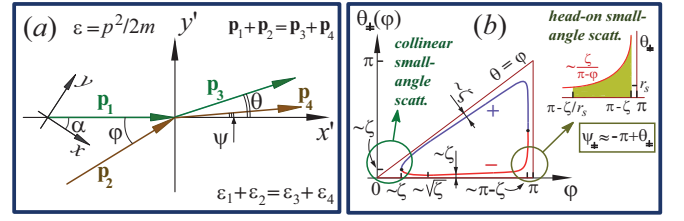}}
\caption{
 Panel~(a):  Initial $\mathbf{p}_1$, $\mathbf{p}_2$
and final $\mathbf{p}_3 $, $\mathbf{p}_4$ momenta
 of two scattering electrons in the reference frame
$x'y'$ related to the direction of the initial momentum
$\mathbf{p}_1$ of the probe electron ``1''.
Panel~(b):
Scattering angles $\theta_{\pm} (\varphi ,\breve{\varepsilon} )$
 of the electron ``1'' as functions of
the angle $\varphi $ of the momentum $\mathbf{p}_2$
of the incident electron ``2''. The case   $ \varepsilon_1 >
 \varepsilon_2 $, $\varepsilon_3 > \varepsilon_4 $,
 and  $\varepsilon_3\varepsilon_4 <\varepsilon_1 \varepsilon_2 $
is shown.  Inset demonstrates the function $\theta_{-}
(\varphi , \breve{\varepsilon} )$  in the vicinity of the point
$\varphi= \pi$, $\theta =0 $, providing the main contribution
to relaxation of the shear stress  at  $\zeta \ll r_s$~\cite{SI}.
  }
\end{figure}

We have shown in \cite{SI} that, due to properties
of scattering of 2D electrons interacting by potential
(\ref{scre_pot}),  the collision integrals $\mathrm{St}$
consists of  the two parts, the first of which acts
 only on the energy variable of $\Psi(\varepsilon, \alpha)$,
 while the second one, being much smaller,  acts
on the angular as well as  energy variable of $\Psi
(\varepsilon, \alpha)$.  Such structure  of $\mathrm{St}$
leads to the fact that   the function $ F(\varepsilon)$
 obtained from the solution of Eq.~(\ref{int_eq}) turns out
to be mainly proportional to a constant.   Therefore
we obtain the formula:
   \begin{equation}
   \frac{1}{
   \tau_{ee,2}} =  - \int d \varepsilon _1 \,
 \{ \,
 \mathrm{St}[ \cos(2 \phi)  ]
\}
   \big|_{\alpha =0 }
   \;,
 \end{equation}
corresponding to applicability of the relaxation time
approximation   for the problem of the viscous transport.
This formula together with Eqs.~(\ref{col_int})-(\ref{singul_factors})
yields   result~(\ref{tau_2_res}) \cite{SI}.
 The main
 contribution to    the  rate   $1/\tau_{ee,2}$~(\ref{tau_2_res})    arises
from the head-on collisions for which $\varphi\approx \pi$ and herewith one of the scattering angles is small: $\theta \ll 1$ or $\psi \ll1$.
[see Fig.~2(b)].
  Such small-angle character
 of  electron-electron collisions leads to arising of the   Coulomb logarithm
  $\Lambda \gg 1 $ in Eq.~(\ref{tau_2_res}).

    The difference of results (\ref{tau_Ferm_liq}) and (\ref{tau_2_res}) for a Fermi gas and a Fermi liquid
    originates from
    different angular dependence of the scattering probabilities for
    weakly ($r_s \ll 1$) and strongly ($r_s \sim 1$)
    interacting electrons (see Fig.~1 and 2).
      The perturbation theory result~(\ref{tau_2_res}) is applicable when  $r_s$ is smaller than
  some value $r_{s0} \ll 1$, which is estimated in Ref.~\cite{SI}.

We also estimated
  the relaxation times $\tau_{ee,m}$ of the distribution functions
 proportional   to the higher angular harmonics,
$\Psi_m (\varepsilon,\phi)= \cos(m\phi)$, $m\geq3$ \cite{SI}.
 In particular, for the relaxation rates
 of the odd harmonics $\Psi_m$ at  $\zeta \ll r_s$ and  $m\ll 1/r_s$
 we obtained \cite{SI}:
% \begin{equation}
  % \frac{
 $  \hbar  / \tau_{ee,m}
    \sim
 (T^2/\varepsilon_F ) \,
   \zeta ^2 \,\ln(r_s/\zeta)\,  m^2
% \end{equation}
$.
 This rate differs
from the result of  Ref.~\cite{Ledwith_2} for
 a 2D Fermi gas with a model inter-particle interaction
 by the logarithm
$\ln(r_s/\zeta)$  due to  taking into account the sharp dependence
  of the Coulomb matrix element $M^2 $~(\ref{matr_el})
on the scattering angles.

 Another effect in 2D electron fluids  being controlled
by the inter-particle scattering is the heat transport.
 The heat flow is described by the term
 \begin{equation}
 \label{Psi_h}
 \Psi_{h}(\alpha, \varepsilon ) =
  \mathbf{v} \cdot    \nabla T \: G(\varepsilon )
 \end{equation}
in the distribution  function, where  $G(\varepsilon ) $ is odd
by  $\varepsilon - \mu $.
% Such $\Psi_h$    describes
% different directions of the velocities of the ``cold''
% ($\varepsilon < \mu$) and the ``hot'' ($\varepsilon > \mu$)
% electrons.
The solution of the kinetic equation
for $\Psi_h$  allows to find the thermal conductivity
 $\kappa$ that relates the heat flux and the temperature
gradient  by the Fourier law:  $\mathbf{q}=-\kappa \nabla T$.
The heat flow   relaxation time is determined via
for the thermal conductivity as $\kappa  = T n v_F^2
 \tau_{ee,h}$.
 The simplest term in $\Psi_{h}$
corresponding a nonzero heat flow $\mathbf{q}$
  is proportional to the function  $ \Psi_{h1}   = \cos ( \alpha )  (\varepsilon - \mu )/T $.
 As a result, the heat flow relaxation rate is estimated
via  $\Psi_{h1} $ as \cite{SI}:
  \begin{equation}
  \label{tau_h__gen_formula}
         \frac{1}{
         \tau_{ee,h}  } \sim
    - \int d \varepsilon _1  \,  \Psi_{h1}
 \mathrm{St}[ \Psi_{h1} ]   \big |_{\alpha =0}
   \:.
 \end{equation}
Calculation   based  on this formula and
Eqs.~(\ref{col_int})-(\ref{singul_factors})
% and (\ref{tau_h__gen_formula})
yields~\cite{SI}:
% \begin{equation}
 % \label{tau_h_ee}
$ \hbar /\tau_{ee,h}
 =
(T^2 / \varepsilon_F  )\,
L_h
$, where $L_h \sim
 \ln( \, r_s /\zeta \,   )  $ at low temperatures, $\zeta \ll r_s $, and
 $ L_h \sim  r_s^2/\zeta^2 $ at intermediate temperatures, $ \zeta \gg r_s
 $.
% \end{equation}
 In the first case,
 such rate originates from the head-on collisions
with the angles $\zeta/r_s \ll \pi - | \varphi| \ll 1 $
 [see Fig.~2(b)] and  coincides  by the order of
 magnitude  with the quantum width of electron
levels due to the electro-electron scattering,
$ \hbar / \tau_{ee,q}$ \cite{SI}.  In the case
 $\zeta \gg r_s$, the rate $ 1 /\tau_{ee,h}$ comes
from intermediate angles $ | \varphi | \sim  1$
 and is much smaller than the levelwidth, which is
estimates as $ (T^2 /\varepsilon_F )
 \ln(\zeta/r_s) \, r_s^2/\zeta^2$.

{\em 4. Comparison with experiment. } When the hydrodynamic
regime of transport in a 2D electron fluid   is realized,
the dependence of resistance on magnetic field $\varrho (B) $
 arises due to the dependence of the diagonal viscosity
 coefficient $\eta_{xx}$  of the fluid on magnetic field
 \cite{je_visc},\cite{SI}:
\begin{equation}
\label{MR}
 \varrho (B) - \varrho (\infty)
 \propto
 \eta_{xx} (B)
 =
  \frac{\eta }{1+(2\omega_c \tau_{2})^2}
 \:.
\end{equation}
The width of this Lorentzian curve is controlled
 by the total shear stress relaxation time $\tau_{2}$.
The Hall viscosity coefficient $\eta+{xy}$ is not substantial for the
 observable  longitudinal resistance $\varrho $, however, possibly,
  can give a contribution
 to the observable   value of the Hall resistance
 \cite{grahene_3},\cite{Gusev_3},\cite{je_visc}.

We compare result (\ref{tau_2_res}) for the interparticle
scattering time $\tau_{ee,2}$ with available experimental
data~\cite{exps_neg_1}-\cite{Gusev_1},\cite{Gusev_3} on magnetotransport
in high-mobility GaAs quantum wells.  In real samples,
the rate $ 1/ \tau_{2} $  contains the two
 contributions \cite{je_visc},\cite{SI}:
\begin{equation}
  \label{tau_2_im_ee}
   \frac{ 1 }{ \tau_{2} (T) }
     = \frac{ 1 }{ \tau_{imp,2} }
  + \frac{ 1 }{ \tau_{ee,2} (T) }
  \:.
\end{equation}
The first term is the ``residual'' relaxation  rate
of the shear  stress at $T \to 0 $ due to scattering
of electrons  on disorder,  while  the second one,
 being the inter-electron relaxation rate,  determines
the temperature dependence of   $1/ \tau_{2}$.

\begin{figure}[t!]
\centerline{\includegraphics[width=1\linewidth]{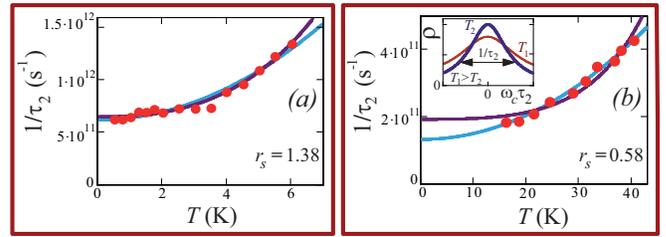}}
\caption{ The shear stress relaxation  rates
as functions of temperature  for the high-mobility
GaAs quantum wells studied  in   Ref.~\cite{exps_neg_1}
(a)   and   in Ref.~\cite{Gusev_3} (b).  Red dots
are the experimental data;  curves are the best fits
by Eq.~(\ref{tau_2_im_ee})    with $\tau_{ee,2}(T) $
corresponding  to a strongly non-ideal Fermi liquid
[purple curves plotted by   Eq.~(\ref{tau_Ferm_liq})]
and to an almost ideal Fermi gas  [light-blue curves
plotted by Eq.~(\ref{tau_2_res})].  Inset in panel (b)
schematically demonstrate
the Lorentzian magnetoresistance curves $\varrho(B)$,   from which
 the data on $\tau_2(T)$ were extracted.
 }
\end{figure}

We fitted the magnetoresistance  measured  in
 Refs.~\cite{exps_neg_1}-\cite{Gusev_1},\cite{Gusev_3}  by the
Lorenzian profile (\ref{MR}). The experimental dependencies
$1/\tau_{2} (T)$ were extracted from the analysis of
  the temperature dependencies of the widths   of
the magnetoresistance curves $ \varrho(B) $ \cite{SI}.
 In Fig.~3 we present the results for the two samples
 with the maximal and minimal electron densities:
$n=1.6\cdot 10^{ 11 }$~cm$^{-2}$ for  one of the samples
studied   in Ref.~\cite{exps_neg_1} (sample A)  and
 $n=9.1\cdot 10^{11}$~cm$^{-2}$ for one of the samples
studied in Ref.~\cite{Gusev_3} (sample B).  For these
two densities, the parameter  $r_s$  is equal to
 $1.38 $ and $ 0.58$, respectively.

 In  Fig.~3 we plotted the experimental data
  for samples A and B together with the  theoretical
curves given by   Eq.~(\ref{tau_2_im_ee})  with
 $ 1 /\tau_{ee,2}^{liq} (T)  =  A^{liq}
 T^2/\ln(\varepsilon_F/T)^2  $ corresponding to
 a strongly non-ideal Fermi liquid
[Eq.~(\ref{tau_Ferm_liq})]
and with   $ 1 / \tau_{ee,2}^ {gas} (T)
  =A^ {gas}T^2  $ corresponding to an almost ideal
   Fermi gas at $\zeta \ll r_s$ [Eq.~(\ref{tau_2_res})].
The values $\tau_{imp,2}\,$, $  A^{gas} $,
 $ A^{liq}$ were the fitting parameters.

We see from Fig.~3 that the  dependence   $1/ \tau_{ee,2}
^{gas} (T)  $ better describes  the experimental data
than $ 1 / \tau_{ee,2}^{liq} (T) $  for the high-density
sample B, for which the interaction parameter $r_s$
is relatively small,  while for the low-density  sample A,
for which $r_s$ is relatively large,
visa versa, the experimental points   better
correspond the curve   $ 1 / \tau_{ee,2}^{liq} (T) $
than to $1/ \tau_{ee,2} ^{gas} (T)  $.

We also compare the amplitude $A^{gas} $  extracted
in this fitting procedure for sample B  with  our theoretical
 result~(\ref{tau_2_res}).  The fitting
 yields $C = 4.0$ for the numeric factor $C $
defined as   $  C = \hbar \varepsilon _ F  A^{gas}
  / [r_s^2 \ln (1/r_s)]  $  [see Eq.~(\ref{tau_2_res})].
 This value  of $ C $
 coincides in the order of magnitude
 with the result  of calculation,  $C = 8\pi/3 =8.4 $,
 corresponding to Eq.~(\ref{tau_2_res}).
Such relation between the theoretical and the experimental $C$
 reasonably corresponds
 to not very small $r_s$ for this sample ($ r_s = 0.58$).

{\em 5. Conclusion and acknowledgements.} We have
evaluated
 the shear stress  relaxation time  and
  the heat flow relaxation time,  those determine
the viscosity and the thermal conductivity of a 2D electron Fermi
gas.
   We have succeeded in distinguishing of the temperature
 dependencies of the first time and the analogous time for a 2D electron Fermi-liquid
  in the experimental data  for high-mobility GaAs quantum wells.

\begin{acknowledgments}
 We thank  M.~I.~Dyakonov for drawing our attention
to the problem of the current study. We also thank
I.~V.~Gornyi  and D.~G.~Polyakov for fruitful discussions.  One of us
(P.~S.~A.) thanks V.~M.~Chistyakov, tragically deceased,
for his lectures on mathematics and for the experience
gained in working together with him, those were reflected,
 to some extent,  in the current study.
This work  was carried out with financial support from
the Russian Science Foundation (Project No.~17-12-01182).
 \end{acknowledgments}

\newpage

 \setcounter{equation}{0}

\onecolumngrid
\begin{center}
{\Large  {\bf Supplemental  material to ``Viscosity of two-dimensional electrons'' }
\linebreak
}

{\large P. S. Alekseev and A. P. Dmitriev
\linebreak \linebreak}
{\small
 Ioffe  Institute, Politekhnicheskaya 26,
  194021,   St.~Petersburg,   Russia
\linebreak
}
\end{center}

{\small
In this Supplemental  material, first,  we  present
 the formulas for the scattering angles
 and the momentum transferred in the inter-particle
 collisions    of  2D electrons in a   Fermi gas;
 the kernel of the electron-electron interaction
with taking into account  the energy transferred
 in collisions;  the asymptotic expressions for
 the inter-particle  collision integral.  Second,
  we develop a method of solution of the  kinetic equation
  for weakly interacting 2D electrons and   perform
 the calculations of the kinetic coefficients
and the relaxation times   of different types
 of perturbations. Third, we discuss  the difficulties
in analysis of the experiments on hydrodynamic magnetotransport
 within the developed theory.
\linebreak
\linebreak}
\twocolumngrid

\section{1.
Expressions for
the characteristics of inter-particle collisions }
\subsection{1.1. The scattering angles $\psi$ and $\theta$}
Below we present the formulas for the relations between
 the scattering angles $\varphi$,  $\theta$, and $\psi$
 at different initial and final electron energies
  $\varepsilon _1$, $\varepsilon _2$ and $\varepsilon _3$.
Herewith we imply that all the angles
 are defined in the range  $(-\pi,\pi)$.

The energy dependence of the kernel of
 the collision integral (M3) is mainly determined by
the Fermi function factor:
\begin{equation}
   \label{Fermi}
     \begin{array}{c}
  \displaystyle
 f_F(\varepsilon_ 1) f_F(\varepsilon_ 2)[ 1-f_F(\varepsilon_ 3) ]
 [ 1-f_F(\varepsilon_ 4)]
 =
 \\
 \\
 \displaystyle
  \frac{1}{2^4\cosh(x_1/2)\cosh(x_2/2)
  \cosh(x_3/2)\cosh(x_4/2)}
 \, ,
   \end{array}
\end{equation}
 where we introduce  the dimensionless electron energies
 counted from the chemical potential $\mu$:
 \begin{equation}
 \label{not}
   x_i = \frac{\varepsilon _i - \mu   }{T }\, ,
    \quad \quad i=1,2,3,4
  \:.
\end{equation}
 Let us remind that $\mu$  differs from the Fermi energy
 $\varepsilon_F$ on the value of the second order
by temperature, $\sim (T/\mu)^2$.  Above and below
 references  (M1), (M2), (M3), and so on denote
the formulas in the main text.

 Due to factor (\ref{Fermi}) the initial and
 final electron energies $\varepsilon _1$,
 $\varepsilon _2$ and $\varepsilon _3$, $\varepsilon _4
 = \varepsilon _1 + \varepsilon _2 -\varepsilon _3$
are  close to the chemical potential
$\mu \approx \varepsilon_F $:
 \begin{equation}
 \label{en}
\frac{\varepsilon _i - \mu  }{\varepsilon _F }
 \lesssim  \zeta
 \, ,  \quad \quad \zeta = \frac{T}{ \varepsilon _F}
 \: .
\end{equation}
These inequalities in the dimensionless energies
$x_i$~(\ref{not}) takes the form:
 \begin{equation}
 \label{en2}
 x_i \lesssim  1   \:.
 \end{equation}

 The momentum conservation  law $\mathbf{p}_1
 +\mathbf{p}_2 =\mathbf{p}_3+\mathbf{p}_4 $
 projected on the $y'$ axis leads to
the following equation for the scattering angle
 $ \psi  =\psi (\varphi,  \theta , \breve{\varepsilon })$
[see~Fig.~2(a) in the main text]:
\begin{equation}
  \label{eq}
 \sqrt{\varepsilon _1 + \varepsilon _2 -\varepsilon _3}
 \, \sin
  \psi
  =
   \sqrt{\varepsilon _2 } \, \sin \varphi
  -
  \sqrt{\varepsilon _3 } \, \sin \theta
  \:.
  \end{equation}
In the arguments of the function
  $ \psi (\varphi,  \theta , \breve{\varepsilon })$
we introduced the shortened designation
 $ \breve{\varepsilon } =   ( \varepsilon _1 , \varepsilon _2
   ,   \varepsilon _3  )$.  The solution  of Eq.~(\ref{eq})
  $\psi = \psi (\varphi,  \theta , \breve{\varepsilon }) $
 has the form    $\psi = \psi_0$, where
 \begin{equation}
 \label{fi_4}
      \psi_0(\varphi,  \theta , \breve{\varepsilon })
 =
\arcsin \Big(
 \frac{
  \sqrt{\varepsilon _2 } \, \sin \varphi
  -
  \sqrt{\varepsilon _3 } \, \sin \theta
 }{\sqrt{\varepsilon _1 + \varepsilon _2 -\varepsilon _3}}
\Big)
\:,
\end{equation}
 $\psi  = \pi - \psi_0 $, or $\psi = -\pi - \psi_0 $,
depending on the positions of the angles $\varphi $ and $ \theta $
 on the unit circle. One should choose one of these three expressions
 for  $ \psi (\varphi,  \theta , \breve{\varepsilon }) $  that
  allow to satisfy the $x'$ component of the momentum conservation law at
  given $\varphi $ and $ \theta $.

 The energy conservation law   $\varepsilon _1 +
 \varepsilon _2  =   \varepsilon _3 + \varepsilon _4 $
 in the variables $\varphi$,   $ \theta $, and $\check{\varepsilon }$
 takes the form:
 \begin{equation}
   \label{ener_cons}
       \varepsilon _3 +  \sqrt {\varepsilon _1 \varepsilon _2 }
     \, \cos \varphi  =
     \sqrt {\varepsilon _1  \varepsilon _3 }
    \, \cos \theta
    +
      \sqrt {\varepsilon _2  \varepsilon _3 }
    \,  \cos (\varphi  - \theta )
    \:.
\end{equation}
 It provides  the relation, $\theta  = \theta  _{\pm}
 ( \varphi ,  \breve{\varepsilon }  )  $, between
 the  scattering angles $\varphi $ and $\theta$
  at given energies $\breve{\varepsilon } $. This solution
 of Eq.~(\ref{ener_cons}) can be presented as:
\begin{equation}
  \label{fi_pm_exact}
\begin{array}{c}
 \displaystyle
\theta_{\pm} ( \varphi ,  \breve{\varepsilon }  )
 =
 \arctan   \Big( \,
 \frac{\sqrt{\varepsilon _2 }  \, \sin \varphi }
 {\sqrt{\varepsilon _1 } +  \sqrt{\varepsilon _2 } \, \cos \varphi  }
  \, \Big) \,
 \pm
 \\
 \\
  \displaystyle
  \pm
  \arccos   \Big( \,
\frac{
 \sqrt{\varepsilon _3 }
    + \sqrt{\varepsilon _1 \varepsilon _2/ \varepsilon _3 }
    \, \cos \varphi
 }
{ \sqrt{
\varepsilon _1 + \varepsilon _2
     + 2 \sqrt{\varepsilon _1 \varepsilon _2} \, \cos \varphi
}}
 \, \Big)
 +N_{\pm}
 \:,
 \end{array}
\end{equation}
where the functions $ N_{\pm}= N_{\pm}(\varphi,
\breve{\varepsilon })$ takes one of the values 0, $\pi$,
or $-\pi$ for which the angles $ \theta_{\pm} (\varphi
 , \breve{\varepsilon}) $  and the corresponding
  other scattering angle  $\psi[\varphi , \theta_{\pm}
(\varphi ,   \breve{\varepsilon}), \breve{\varepsilon}] $  satisfy
  the momentum conservation  law  projected on the $x'$ axis.

By substituting of Eqs.~(\ref{fi_pm_exact})
 into  equation (\ref{fi_4}),  we obtain the angle
  $\psi = \psi_{\pm} ( \varphi ,  \breve{\varepsilon }  ) $
  as a function of only the angle $\varphi$
  and the energies $\breve{\varepsilon }$:
\begin{equation}
 \label{psi_pm}
\psi_{\pm}
 ( \varphi , \breve{\varepsilon }  )
  =
\psi  [\varphi , \theta_{\pm}
  ( \varphi , \breve{\varepsilon }   ) ,
 \breve{\varepsilon }  ]
  \:.
\end{equation}

 Provided Eq.~(\ref{en}) is fulfilled, functions~(\ref{fi_pm_exact})
in the interval of the intermediate angles
 $ |\varphi |\sim 1$ (more precisely,  $ |\varphi| \gg \sqrt{\zeta}  $,
  $|\pi - \varphi|  \gg \zeta $) have the asymptotes:
\begin{equation}
 \label{fi_pm_asimpt}
\begin{array}{l}
\theta_{+}
  = \varphi + \Delta \theta _+
\:,
    \quad
 \theta_{-}
 =  \Delta \theta _-
\:,
 \end{array}
\end{equation}
where $\Delta \theta _{\pm}  = \Delta \theta _{\pm} (\varphi,
\breve{\varepsilon }  ) \sim \zeta $. From Eqs.~(\ref{fi_pm_asimpt})
 and (\ref{fi_4}) we see that   in   the same interval
 of $\varphi$ the similar asymptotes for
 $\psi_{\pm} (\varphi ,\breve{\varepsilon }) $
 take place:
 \begin{equation}
  \psi_{+}
  = \Delta \psi _+
\:,
    \quad
 \psi_{-}
 =  \varphi +  \Delta \psi _-
\:,
  \end{equation}
 where $ \Delta \psi _{\pm} = \Delta \psi _{\pm}
 (\varphi, \breve{\varepsilon }  )  \sim \zeta$.
 So at the intermediate angles $\varphi  $,
 $| \varphi  | \sim 1 $, the scattering angles
 $\theta$ and $\psi $  are close to 0 and $\varphi$,
 or visa versa.

  In other words, the collision processes $(1,2) \to (3,4)$
  at $| \varphi  | \sim 1$ are the  small-angle scattering,
\begin{equation}
   (0 \, ,\:\varphi )
   \; \to \;
  \Big( \:
  \theta = \Delta \theta _{-}
   \; , \;\;
  \psi = \varphi + \Delta \psi _{-}
  \: \Big)
  \:,
\end{equation}
 and  the exchange  of the two scattering electrons
  accompanied by  a small-angle scattering,
\begin{equation}
  ( 0 \, , \: \varphi )
  \; \to \;
   \Big( \:
    \theta = \varphi + \Delta \theta _{+}
   \; , \; \; \psi =
  \Delta \psi_+
  \: \Big)
  \: .
\end{equation}

The functions $\theta_{\pm} ( \varphi ;  \breve{\varepsilon }   )$
  given by  Eqs.~(\ref{fi_pm_exact}) and (\ref{fi_pm_asimpt})
  are drawn in Fig.~2(b) in the main text  for
 the simplest case when $\varepsilon_1 > \varepsilon_2$,
 $\varepsilon_3> \varepsilon_4$, and   $\Delta <0 $
 Here and below we introduce the value:
 \begin{equation}
  \Delta  = \Delta  (\check{\varepsilon}) =
 \frac{\varepsilon_3\varepsilon_4
   - \varepsilon_1 \varepsilon_2 }{T^2} =x_3x_4-x_1x_2
   \:.
 \end{equation}
It is noteworthy that in this case $N_{\pm} \equiv 0 $
 in Eq.~(\ref{fi_pm_exact}) and the parameter
  $\Delta  $  determines the gaps between the curves
 $\theta_{\pm} ( \varphi ,  \breve{\varepsilon }   )$
 and the  vertical lines $\varphi =0$, $\varphi = \pi$
 [see Fig.~2(b) in the main text].
 For other relations between
 $\varepsilon_1$, $ \varepsilon_2 $,
 and $ \varepsilon_3$, the functions
 $\theta_{\pm} ( \varphi ,  \breve{\varepsilon }   )$
 are defined on other intervals  of angles
 $\varphi$ including   the points $\varphi =0$
 and  $\varphi =\pi$ and exhibit other possible types of
  connectivity near the points  $(0,0)$, $(\pi,0)$,
  and $(\pi,\pi)$  on the $(\varphi,\theta)$-plane.

  In the limit  $ \varphi \to 0$, functions (\ref{fi_4})
 and (\ref{fi_pm_exact}) determining the scattering angles
$\psi = \psi ( \varphi ,  \theta , \breve{x} ) $
 and  $\theta  = \theta   _{\pm} ( \varphi ,  \breve{x} )$
 take the forms:
 \begin{equation}
  \label{varphi_4__at_0}
 \psi  ( \varphi ,  \theta , \breve{x} )
 = \varphi - \theta
  \end{equation}
and
\begin{equation}
  \label{varphi'__at_0}
\theta_{\pm} ( \varphi ,  \breve{x} )
 =
 \frac{
 \varphi  \pm  \sqrt{
 \varphi  ^2 + \zeta^2 \Delta  }
  }{ 2}
 \:.
\end{equation}
 From  Eqs. (\ref{varphi_4__at_0}) and (\ref{varphi'__at_0})
for the functions $\psi_{\pm}( \varphi , \breve{x} )$~(\ref{psi_pm})
 we obtain:
\begin{equation}
  \label{varphi_4__at_0_short}
  \begin{array}{c}
  \displaystyle
\psi_{\pm} ( \varphi , \breve{x} )
 =
  \theta_{\mp} ( \varphi  , \breve{x})
 \:.
 \end{array}
\end{equation}
We see that at $|\varphi |\ll 1$ the scattering angles
 $\theta$ and $\psi $ are also small:
  $|\theta|, \, |\psi | \ll 1 $.

 The asymptotic expressions
 (\ref{varphi_4__at_0})-(\ref{varphi_4__at_0_short})
  are valid until $|\varphi | \ll \sqrt{\zeta}$.

From Eqs.~(\ref{fi_4}) and (\ref{fi_pm_exact}) we also derive
the asymptotic formulas for the  scattering angles
 $\psi _{\pm}  =\psi _{\pm} (\varphi , \breve{x}) $
  and $\theta _{\pm} = \theta _{\pm} (\varphi, \breve{x})$
  determining the collision integral in the limit
  $|\varphi| \to \pi $ in the main order by   $\zeta \ll 1$.
  For the interval  of positive $\varphi$ the formulas
takes the form:
 \begin{equation}
  \label{varphi_4_at_pi}
 \psi (\varphi,\theta,\breve{x})
  = \pi +  \theta  _{\pm}
   \quad \mathrm{or }\quad
    \psi (\varphi,\theta,\breve{x}) = - \pi +  \theta
    \:,
\end{equation}
corresponding to negative and positive
 values of  $ \theta   = \theta  _{\pm} (\varphi,\breve{x})$,
 respectively [see Fig.~2(a) in the main text], and
\begin{equation}
 \label{varphi'__at_pi}
 \begin{array}{c}
 \displaystyle
 \theta _{\pm}(\varphi,\breve{x})
  =
  \arctan \Big( \, \frac{ s }{a}   \, \Big)
  \pm
     \\
   \\
   \displaystyle \pm
  \arccos \Big( \, \frac{b}{ \sqrt{a^2 +  s^2 \,  } }  \, \Big)
   + N_{\pm}
  \,.
  \end{array}
  \end{equation}
Here we introduced the notations: $s= (\pi -\varphi)/\zeta$,
$ a =  ( x_1    -x_2  ) /2    $ and $  b=   ( x _3  -
x  _4   ) /2  $.

Formulas (\ref{varphi_4_at_pi}) and (\ref{varphi'__at_pi})
are valid up   to the angles $\varphi  \sim 1$ (that is,
at $s \lesssim 1/\zeta$).

\subsection{1.2. The transferred momenta $q$ and $w$ }
The absolute value  $q=q(\theta, \breve{\varepsilon})$ of the
transferred momentum $\mathbf{q}  = \mathbf{p} _3 - \mathbf{p}_1$
is expressed via the particle   energies $\varepsilon_1$,
$\varepsilon_3$  and the scattering angle $\theta $
 as:
\begin{equation}
  \label{q_abs}
\frac{ q } { \sqrt{2m}  } =
  \sqrt {\varepsilon _1 + \varepsilon_3
 - 2 \sqrt {\varepsilon _1  \varepsilon_3}
  \, \cos \theta }
  \:.
\end{equation}
Analogously, for the absolute value $w=w(\varphi,\theta,
\breve{\varepsilon})$  of the indirect transferred momenta,
 $ \mathbf{w} = \mathbf{p}_3 - \mathbf{p}_2 $, entering
 the exchange terms in Eq.~(M5),
  we have:
\begin{equation}
  \label{obmen_analog__q}
\frac{ w}{ \sqrt{ 2 m}  } =
  \sqrt {\varepsilon _2 + \varepsilon_3
 - 2 \sqrt {\varepsilon _2  \varepsilon_3}
    \, \cos (\varphi - \theta )}
    \:.
\end{equation}

In order to derive the asymptotic expressions
 for the collision integral near
  the angles $\varphi = 0 $ and $\varphi = \pi $,
one needs to evaluate
the  values $q  $    and   $w  $
  in the zero, the first, and the second  orders
    by the small parameter $\zeta$
   in the limits  $\varphi \to 0  $ and $\varphi \to  \pi $.

   In the first limit,   $|\varphi| \ll 1  $,
   for the functions $q=q(\theta , \breve{x})$
    and $w=w( \varphi , \theta  , \breve{x})$
     we get  from
   Eqs.~(\ref{fi_4}), (\ref{fi_pm_exact}),   (\ref{q_abs}),
   (\ref{obmen_analog__q}):
   \begin{equation}
\label{q_w_appr_0}
\begin{array}{c}
\displaystyle
q= p_F \,\sqrt{ \theta^2 + \zeta^2 (x_1 - x_3 )^2/ 4}
\:,
\\
\\
\displaystyle
w=p_F\,\sqrt{\psi^2+ \zeta^2 (x_2 - x_3 )^2/ 4}
\:,
\end{array}
\end{equation}
where $p_F = \sqrt{2 m \varepsilon_F}$;
$\psi = \psi (\varphi , \theta  , \breve{x}) $
is given by Eq.~(\ref{fi_4}); and
$\breve{x} =(x_1,x_2,x_3)$
 are the dimensionless electron energies (\ref{not}).

It is seen from Eqs.~(\ref{fi_4})-(\ref{fi_pm_asimpt}),
 (\ref{q_w_appr_0}) that at the actual scattering angles
 $\theta = \theta _{\pm}$ and $\psi = \psi _{\pm}$
 we have the estimate
  \begin{equation}
  q, \, w\sim p_F \zeta
  \end{equation}
   in  the range  of very small $\varphi $, $| \varphi | \sim \zeta$,
  and the estimates
  \begin{equation}
  q\sim p_F \zeta \:, \quad  w \sim p_F |\varphi|
    \end{equation}
  or, visa versa,
   \begin{equation}
    q \sim  p_F |\varphi |  \: , \quad w \sim p_F \zeta
      \end{equation}
  in  the range  of moderately small   $\varphi $,
  $ \zeta \ll | \varphi| \ll 1$.

 In the limit, corresponding to the head-on collisions,
  $\pi -|\varphi | \ll 1$,
 the functions $q = q(\theta , \breve{x}) $
  and $w =  w (\varphi, \theta , \breve{x})$
  from (\ref{q_abs})  and (\ref{obmen_analog__q})  take
   the forms:
\begin{equation}
\label{q_w_appr_pi}
\begin{array}{c}
\displaystyle
q= p_F \,\sqrt{2\, (1 - \cos\theta ) + \zeta^2 (x_1 - x_3 )^2/ 4}
\:,
\\
\\
\displaystyle
w=p_F\,\sqrt{2\, ( 1 + \cos\theta ) + \zeta^2 (x_2 - x_3 )^2/ 4}
\:.
\end{array}
\end{equation}
Depending on the sign $\pm$ in Eqs. (\ref{fi_pm_exact}), (\ref{psi_pm})
and the value of $\varphi $, the first terms
under the square roots in Eq.~(\ref{q_w_appr_pi}),
 $1 \mp \cos \theta $, at actual scattering angles
 $\theta = \theta _{\pm} (\varphi , \breve{\varepsilon })$
 and given values $\breve{\varepsilon }$
 can be much greater or comparable
with the second terms, $\zeta^2 (x_{1,2} - x_3) ^2 /4$.

\section{2.  The kernel of the inter-particle interaction}
\subsection{2.1. The kernel taking
 into account the transferred energies }
  The use of the static screened potential $ V( q) $  (M6)
 implies the neglect of the dependence
 of the kernel $ V( q,\epsilon ) $  of the inter-particle interaction
 on the energy transferred  in a collision,
\begin{equation}
\label{eps}
  \epsilon = \varepsilon_3 - \varepsilon_1
   \:.
 \end{equation}

 The approach that provides a general description
 of the inter-particle scattering in
 a 2D electron Fermi gas is
 the random phase approximation   (RPA).
In particular, this approach  allows to account for
  the energy transfer (\ref{eps}) in
    the scattering probability  $W$
  by the formula analogous to Eq.~(M4).
 The main element in calculation of   $ V( q,\epsilon ) $
 within RPA is the
 polarization operator $\Pi( q,\epsilon) $ of a 2D Fermi gas.
Accounting of $\Pi( q,\epsilon) $  leads to the change of
 the static screened potential $V(q)$~(M6)
 in the squared scattering matrix element $M^2$
  on the kernel $V(q,\epsilon)$
  \cite{Narozhni_et_al_,Narozhni_et_al_2_}:
 \begin{equation}
   \label{scre_pot___omega}
    \displaystyle
V(q,\epsilon)
 =
 \frac{ \pi e^2 a_B}{
 \displaystyle  \
 1 +  \frac{ a_B q}{ 2\hbar } - i \,
 \frac{  \epsilon }{ \sqrt{  v_F^2 q^2 - \epsilon^2 }}}
 \:.
\end{equation}
Here and below references [1], [2], [3], and so on denote
the items in the reference list at the end of this Supple-
mental material.

 At the absence of spin polarization of 2D electrons,
 the squared matrix element
  $M^2 = M^2 (\varphi , \theta  , \breve{\varepsilon}) $
  corresponding
to the kernel (\ref{scre_pot___omega}) takes the form
\cite{Narozhni_et_al_,Narozhni_et_al_2_,Ledwith_2_}:
   \begin{equation}
\label{matr_el_ome}
\begin{array}{c}
\displaystyle
   M^2
 =
   2   \,
 \{ \:  |V(q,\epsilon)|^2 + |V(w,\epsilon')|^2
 -
 \\
 \\
 \displaystyle
 -
 \mathrm{Re} \, [\, V(q,\epsilon) \, V (w,\epsilon')^*\,] \, \}
 \:,
 \end{array}
\end{equation}
where $\epsilon' = \varepsilon_3 - \varepsilon_2$
 is the indirect transferred energy.
In Eq.~(\ref{matr_el_ome}) the first term
is the direct diagonal term,
while the second and the third
ones are the exchange diagonal
 and  the exchange  interference terms.
The scattering probability $W$ is again
expressed via $M^2$ by the Golden rule-like
formula~(M4).

 Below in this section we will
 analyze at which scattering angles
 it is possible and in which it is impossible
   to neglect the dependence of the matrix element $M^2$
   on the transferred energies $\epsilon$ and $\epsilon ' $.
    We present this consideration in details
 only for the first direct term $2|V(q,\epsilon)|^2$
 in $M^2$~(\ref{matr_el_ome}).
  A similar consideration
 for the second and  the  third terms in $M^2$,
 containing the kernel $V(w,\epsilon')$,
  is performed just by the change of the values
  corresponding to the ``3'' and ``4'' electrons.

  It follows from Eqs.~(\ref{q_abs}) and (M3)
  that the magnitudes of $q$ and $\epsilon $
    independently vary
  in the ranges  from 0 to $ \approx 2p_F$
  and from 0 to $ \sim T$, respectively.
 For the  small
 and the large scattering angles:
\begin{equation}
 \label{diap_thet_0}
  |\theta| \sim  \zeta
  \: ,\qquad
  \zeta \ll |\theta| \ll  1
  \: ,\qquad
   |\theta  | \sim 1
   \:,
   \end{equation}
 we have the estimates:
\begin{equation}
 q \sim  p_F \zeta
  \: ,\qquad
 q \sim    p _F \theta
  \: ,\qquad
 q \sim p_F
 \:,
\end{equation}
  respectively [see Fig.~2(a) in the main text].
 Thus the absolute value of
 the term  $- i \epsilon/ \sqrt{  v_F^2 q^2 - \epsilon^2 }$
  in the denominator of Eq.~(\ref{scre_pot___omega})
  is estimated as:
\begin{equation}
 \label{Pi_term}
\frac{  \epsilon }{  \sqrt{  v_F^2 q^2 - \epsilon^2 } }
    \;\;\; \sim \;\;\;
   1
     \,, \;\;\;
\frac{  \zeta }{ |\theta |}
       \,, \;\;\;
       \zeta
\end{equation}
 in the interval of $\theta$ (\ref{diap_thet_0}).
  Therefore  for the not too small angles $\theta$,
$ |\theta | \gg \zeta $,  term  (\ref{Pi_term}) can be
 neglected in the denominator  of
  Eq.~(\ref{scre_pot___omega})  compared with unity,
 while at the small angles, $|\theta |\sim  \zeta$,
 it  is about unity and cannot be neglected.
 [Moreover, at $|\varphi | \sim \sqrt{\zeta} $ we obtain
  from Eq.~(\ref{varphi'__at_0}) that $\theta_- \ll \zeta$
 therefore term  (\ref{Pi_term}) turns out to be
 even much greater than unity].

 When all the energies   $\varepsilon_{1,2,3}$
 are close to the chemical potential,
   $ | \varepsilon_{1,2,3} - \mu  | \lesssim T $,
it follows from Eqs.~(\ref{fi_4}), (\ref{fi_pm_exact}),
 and  (\ref{fi_pm_asimpt}) that
at the small and the intermediate angles $\varphi$,
 \begin{equation}
 \label{diap_small_med_fi}
  \quad  \pi- |\varphi| \sim 1
 \end{equation}
(in particular, $|\varphi | \sim \zeta $),
 the scattering angles
 $ \theta = \theta_{\pm} (\varphi, \breve{\varepsilon} ) $ and
 $ \psi = \psi _{\pm}( \varphi, \breve{\varepsilon}  ) $
 satisfy the conditions
\begin{equation}
\label{diap_sc_angl_horiz}
 |\theta  | \lesssim \zeta
   \;\,\quad
   | \psi - \varphi | \lesssim \zeta
\end{equation}
 [the ``horizontal'' part of the red curve in Fig.~2(b)
 in the main text] or the conditions
\begin{equation}
 \label{diap_sc_angl_diag}
   | \theta - \varphi | \lesssim \zeta
       \;\;,\quad
    |\psi  | \lesssim \zeta
\end{equation}
 [the ``diagonal'' part of the blue curve in Fig.~2(b)
 in the main text].For the ``head-on'' collisions,
 when the angle  of the incident electron $\varphi$
is close to $ \pm \pi $,
  \begin{equation}
  \label{diap_app_pi_fi}
  \pi- | \varphi | \ll 1 \:,
  \end{equation}
 [the ``vertical'' parts of the blue   and the red curves
  in Fig.~2(b) in the main text],  both  the scattering angles
   $ \theta = \theta_{\pm} (\varphi, \breve{\varepsilon} ) $ and
 $ \psi = \psi _{\pm}( \varphi, \breve{\varepsilon}  ) $
  become large as compared with  the small parameter $\zeta$:
  \begin{equation}
  \label{diap_sc_angl_vert}
|\theta |,
     |\psi| \gg \zeta
     \:.
\end{equation}

So equations  (\ref{diap_sc_angl_horiz}),
(\ref{diap_sc_angl_diag}), and (\ref{diap_sc_angl_vert})
   evidence the following.
   First, the head-on collisions [$\pi - |\varphi| \ll 1$]
 correspond to scattering on the large angles $\theta  $
 and $\psi$ [Eq.~(\ref{diap_sc_angl_vert})].
 In view of Eq.~(\ref{Pi_term}), for the scattering probability
 on them one can use the static potential $V(q)$~(M6).
  Second, the collisions with the incident electrons having a small
   or an intermediate angle $ \varphi $  [$\pi - |\varphi| \sim 1$]
  are small-angle: $|\theta |  \lesssim \zeta $
 or $ |\psi | \lesssim \zeta $. Thus, owing to
 Eq.~(\ref{Pi_term}), it is necessary to use
 the exact inter-particle interaction kernel
 $V(q, \epsilon)$~(\ref{scre_pot___omega})
in order to properly account them.

\subsection{2.2. The factor $S$ and
 importance of accounting for the
  transferred energy at different $\varphi$  }
  In this subsection, we estimate
  the expression $S=S(\alpha;\varphi , \breve{\varepsilon} ) $
  in the last square brackets in the collision integral~(M3):
\begin{equation}
 \label{sq_br}
  \begin{array}{c}
  S =  \Psi(\varepsilon_1,\alpha) +\Psi(\varepsilon_2,\alpha+\varphi) -
  \\\\
  - \Psi(\varepsilon_3,\alpha+\theta) - \Psi(\varepsilon_4,\alpha+\psi)
  \:,
 \end{array}
\end{equation}
 where $\varepsilon_4 = \varepsilon_1 + \varepsilon_2 - \varepsilon_3$,
 $\theta = \theta_{\pm}(\varphi,\breve{\varepsilon})$, and
 $\psi = \psi_{\pm}(\varphi,\breve{\varepsilon})$.  For
  this purpose, it is convenient  to put  $\alpha=0$
 and to consider that the function $ \Psi(\varepsilon,\phi)$
 is even by $\phi$ and is normalized such as
$\Psi(\varepsilon,\phi) \sim 1$  at general $\phi \sim 1 $
 and $x = (\varepsilon - \mu )/T \sim \zeta $.
 Thus  $S$ becomes the function of the variables
  $\varphi$ and $\breve{\varepsilon}$,
  $S=S_\pm(\varphi,\breve{\varepsilon})$,
and should be compared with unity.

  Owing to  Eqs.~(\ref{fi_pm_asimpt}) and
   (\ref{diap_small_med_fi})-(\ref{diap_sc_angl_diag}),
 we can decompose such
   $S(\varphi , \breve{\varepsilon} )  =S(\varphi , \breve{x} )  $
 at each small or medium $\varphi$, $\pi - |\varphi| \sim 1$,
  in the Taylor series by the small values
 $\theta$ and $(\psi-\varphi)$    for the ``$-$''-solutions
 [or by the small values $\psi$ and
  $(\theta-\varphi)$ for the ``$+$''-solutions].
 It follows from Eqs.~(\ref{fi_4}) and (\ref{fi_pm_exact})
 that the angles $\theta_{\pm}(\varphi,  \breve{x}=0 )$
  and $ \psi_{\pm}(\varphi, \breve{x}= 0 ) $
 are equal to $0$ and $\varphi$ or visa versa.
  This means that we decompose $ S $ with respect to the deviations
    of the angle functions $\theta_\pm$ and $\psi_\pm$
 from their value at zero energy variable $\breve{x}=0$.
  In Fig.~2(b) in the main text
   the points $[  \varphi,\theta_{\pm}(\varphi,  0 )]$
 on the $(\varphi, \theta)$-plane
 are drawn    by the brown  thin lines;
 while the functions  $\theta_{\pm}(\varphi,  \breve{x}  )$
 at some nonzero $\breve{x}$ are  drawn
  by thick blue and red curves.

 We express the value $S$ in the form
 being convenient for the Taylor decomposition
  by the small deviation of the scatting angles $\theta $
and $\psi$   from their values at $\varepsilon_i = \varepsilon _F$.
For case of the sign ``$-$'' in the solutions $\theta=\theta_\pm$
and  $\psi = \psi_\pm$ we write:
\begin{equation}
 \label{sq_br_fi_sim_1}
  \begin{array}{c}
  S_- =  \Psi(x_1,0) +\Psi(x_2,\varphi)
    - \Psi(x_3,0) - \Psi(x,\varphi)
  \\\\ + \,
[\,  \Psi(x_3,0) -\Psi(x_3,\theta)\,] \,+
\\\\+   \,
[\,
 \Psi(x_4,\varphi) -  \Psi(x_4,\psi)
\,]
\:.
  \end{array}
  \end{equation}
 Such form  allows to make the decomposition
 of $S$ by  $\theta$ and $(\psi-\varphi)$.
 In the case of the sign ``$+$''    in the solutions
  $\theta=\theta_\pm$ and   $\psi = \psi_\pm$
it is convenient to  write:
        \begin{equation}
  \label{sq_br_fi_sim_2}
  \begin{array}{c}
  S_+ =  \Psi(x_1,0) +\Psi(x_2,\varphi)
    - \Psi(x_3,\varphi) - \Psi(x_4,0)
  \\\\ + \,
[\,  \Psi(x_3,\varphi) -\Psi(x_3,\theta)\,] \,+
\\\\+   \,
[\,
 \Psi(x_4,0) -  \Psi(x_4,\psi)
\,]
  \:,
  \end{array}
\end{equation}
instead of Eq.~(\ref{sq_br_fi_sim_1})
 in order to make   the Taylor decomposition
by the small values  $(\theta-\varphi)$
 and  $\psi$.

 If $ \Psi(\varepsilon,\alpha)$  substantially depends
  on the angle variable  $\alpha$ and
  weakly depends on $\varepsilon$
 [$ \Psi(\varepsilon,\alpha) \approx \Psi(\alpha) $],
  the terms in the first lines of Eqs.~(\ref{sq_br_fi_sim_1})
  and  (\ref{sq_br_fi_sim_2})  are close to zero and
the decomposition of the other terms
 in $S$~(\ref{sq_br_fi_sim_1}), (\ref{sq_br_fi_sim_2})
in the Taylor series leads to the following estimate.
  At small and intermediate $\varphi $,
  $\pi -|\varphi | \sim  1 $ [where the scattering angles
 $\theta$ and $\psi$ lie in the intervals
  (\ref{diap_sc_angl_horiz}) and (\ref{diap_sc_angl_diag})],
  $S$ is a linear combination of $x_i$  in the first order
 by $ \zeta $  and is a quadratic function of $x_i$
in the second order by $ \zeta $:
 \begin{equation}
  \label{S3}
    S (\varphi  ,
   \breve{x }) \approx \zeta  \, A_1( \varphi , \breve{x })
   + \zeta^2   A_2 (\varphi , \breve{x })\:,
 \end{equation}
 where
\begin{equation}
 \label{S3_}
   A_1( \varphi , \breve{x }) =
   \sum \limits _{i=1}^3 a_i \, x_i
   \:, \quad
    A_2 ( \varphi , \breve{x })=
    \sum \limits_{i,j=1}^3 a_{ij}\, x_i\,x_j \:,
\end{equation}
$a_i =  a_i ( \varphi ) \sim 1$,  and
 $     a_{ij}   = a_{ij}  ( \varphi )
 \sim 1 $.

 For  the head-on collisions corresponding
  to $  \varphi $ near $ \pm \pi $
  [Eq.~(\ref{diap_app_pi_fi})],
 the scattering  angles   are large  [Eq.~(\ref{diap_sc_angl_vert})].
  Provided $\Psi(0)$ is not close to $ - \Psi(\pi)$,
 the combination $S$~(\ref{sq_br}) cannot be decomposed
  in the Taylor series by $\theta$, $(\theta-\varphi)$,
 $\psi$, and $(\psi-\varphi)$
and is generally estimated as:
     \begin{equation}
     \label{S_large}
  S   (\varphi   ,    \breve{x }) \sim  1
  \:.
  \end{equation}
Visa versa, if  $\Psi(0) = -\Psi(\pi)$, we can
 make the decomposition of $S$~(\ref{sq_br_fi_sim_1})
by $\varphi - \pi$ and therefore the inequality
 \begin{equation}
   \label{S_small_uns}
       S   (\varphi   ,    \breve{x }) \ll 1
   \end{equation}
 is valid instead of Eq.~(\ref{S_large}).

From comparison of the magnitudes of $S$
 in Eqs.~(\ref{S3}) and (\ref{S_large})
we conclude that, for the functions
 $\Psi (\varepsilon,\alpha) = \cos(m\alpha) $
with even nonzero     $m $, the contributions
to the collision integral~(M3)
 from the head-on collisions [$ \pi - | \varphi | \ll 1 $]
 can dominate on the contribution from
  the small and the intermediate  angles
   $\varphi$ [$ \pi - |\varphi | \sim  1 $], provided
 the sufficiently large  magnitude of the other factors
 in the operator $\mathrm{St}$~(M3)
 in the region $ \pi - |\varphi | \ll 1 $.
 Such  other factors are the delta-function factors
  (M7) and the matrix element $M^2$.
 Factors (M7) diverge as $1/\varphi$ at $\varphi \to 0$
 and as $1/(\pi-\varphi)$ at  $\varphi \to \pi$,
   while matrix element (M5)  [and (\ref{matr_el_ome})],
 generally speaking, decreases at  $|\varphi |\to \pi$.
 Therefore, the question  of the importance of
 the head-on collisions at $\varphi \to \pi$
 for various $\Psi$ is non-trivial.
 If they actually dominate, one should  use
 the static potential $V(q)$~(M6)
 in the corresponding main contribution in $\mathrm{St}$,
according to the obtained in Section~2.1
 relations between     $ V(q) $~(M6) and the exact kernel
   $ V(q,\epsilon) $~(\ref{scre_pot___omega}) .

 Visa versa,  from Eqs.~(\ref{S3})  and (\ref{S_small_uns})
it is seen that   that   for distribution functions
  $\Psi (\varepsilon,\alpha) = \cos(m\alpha) $   with odd $m$
  [leading to $\Psi(0) = -\Psi(\pi)$)] the factor
  $S(\varphi,\breve{\varepsilon})$ acts in some other way at
  determining the range of the important angles $\varphi$.
   For such $\Psi$  not only the head-on collisions, but,
   generally, the collisions with the angles
   $|\varphi|\ll1$ and $|\varphi|\sim1$ can be also  important.
 In accordance with the consideration of the last subsection,
 in this case, one needs to use the exact kernel
  $ V(q,\epsilon) $~(\ref{scre_pot___omega}).

  If the function   $ \Psi(\varepsilon,\alpha)$
 substantially depends on the energy variable,
  for example:
\begin{equation}
   \label{X}
     \Psi_m^F (\varepsilon , \alpha)=
     F(\varepsilon ) \cos(m  \alpha )
     \:,
 \end{equation}
the factor $S$~(\ref{sq_br_fi_sim_1}) at intermediate angles
 $\varphi$, $| \varphi| \sim 1$ as well as for the head-on
 collisions,  $\pi - | \varphi|  \ll1$,
 is related mainly with the terms
 in the first line in Eq.~(\ref{sq_br_fi_sim_1})
 and is estimated as unity [similarly to Eq.~(\ref{S_large})].
  At small angles, $| \varphi| \ll 1$,
 for such $ \Psi = \Psi_m^F$   we have:
 \begin{equation}
   \label{en_cons_law}
    S (\varphi   ,    \breve{x }) \sim
     F(x_1) +F(x_2) -F(x_3) -F(x_4)
   \:,
 \end{equation}
   which is also a value of the order of unity  except
  the cases  $F(x)=1$ and  $F(x)=x$.  We conclude that
   the main part of   $S(\varphi , \breve{\varepsilon} )$
 is estimated as unity   at any  $\varphi$  and $F$
  except $\varphi \to 0 $ and those $F$ corresponding
 to perturbations of the particle density and energy.

  The terms in Eqs.~(\ref{sq_br_fi_sim_1}), (\ref{sq_br_fi_sim_2})
 in the second and the third lines,   being much  smaller
   than the terms in the first line,   are
  in the case of the energy-dependent distribution
    function $ \Psi_m^F $~(\ref{X})  determine  the dependence
of the factor $S$ and the collision operator $\mathrm{St}[\Psi_m^F]$
 on the angular harmonic number $m$, that is the angular-dependent
  part of $\mathrm{St}[\Psi_m^F]$.  This fact is important
 for the analysis of the general properties of
collision operator (see the next Sections 3 and 4).

\section{ 3. Contributions to the collision integral
   from   different intervals of the angles $\varphi $}
\subsection{3.1. General structure
    of the collision integral }
  The regions   $ 0 \leq  \varphi \leq \pi $ and
    $ - \pi \leq \varphi \leq 0  $ provide identical contributions
  to the collision integral $\mathrm{\mathrm{St}}$, given by Eq.~(M3).
 Thus in the operator $\mathrm{\mathrm{St}}$
 one can calculate the integral by the variable $\varphi$
  over only one of these two intervals,
 for example, over $ 0 \leq  \varphi \leq \pi $,
  and to take into account the contribution
  from  the second interval, $ - \pi \leq \varphi \leq 0 $,
 just by   multiplication of the first contribution
 on  the factor two.

One can distinguish the contributions
 to the collision integral from the collinear collisions
  (corresponding to  $|\varphi |\ll1$
and, at the same time, $|\theta|,|\psi |\ll 1$);
 from head-on collisions (for which $|\varphi |\approx \pi$);
  and from the collisions with electrons
  having the intermediate angles
($|\varphi |\sim 1$):
\begin{equation}\label{sum}
\mathrm{St} = \mathrm{St}_0+\mathrm{St}_\pi+\mathrm{St}_1\:.
\end{equation}
As in was discussed above in Section 2.2,
the estimates of the factor   $S$~(\ref{sq_br})
demonstrates that depending on the type
 of the function $\Psi_m^F$~(\ref{X}), some of the terms,
 $\mathrm{St}_0$, $\mathrm{St}_\pi$, or $\mathrm{St}_1$,
 in  Eq.~(\ref{sum}) can be more or less substantial.

   Further, it is seen from the general expressions~(M3)
    for the collision integral that
  the operator $\mathrm{St}$ is diagonal in angular harmonics:
\begin{equation}
\mathrm{St}[\Psi_m^F (\varepsilon,\phi) \,]
(\varepsilon_1 ,\alpha)
 = \cos(m\alpha) \,\mathrm{St}^{(m)}[ F(\varepsilon)\,]
  (\varepsilon_1 )
\:,\end{equation}
and, correspondingly:
\begin{equation}
\mathrm{St}_\varsigma[\Psi_m^F(\varepsilon,\phi)\,]
(\varepsilon_1 ,\alpha)
 = \cos(m\alpha) \,\mathrm{St}^{(m)}_\varsigma[ F(\varepsilon)\,]
 (\varepsilon_1 )
\:,
\end{equation}
 where $\varsigma=0,\pi,1$.  One can show from Eqs.~(\ref{sq_br_fi_sim_1})
 and (\ref{sq_br_fi_sim_2}) [see detains below in this section]
  that for distribution functions of the  general form,
  $\Psi_m^F$~(\ref{X}),      it is possible to distinguish
 in each of the terms  $\mathrm{St}_\varsigma$
 the angular-independent part and
the angular-dependent parts:
 \begin{equation}
    \label{def_angl_dep_indep}
\mathrm{St}_\varsigma^{(m)}[ F] = \mathrm{St}^{in} _\varsigma[ F]
+ \mathrm{St}_\varsigma^{dep,(m)}[ F]
\:,\end{equation}
 Here it is implied that the angular-dependent part
vanishes at zero harmonics:
 \begin{equation}
  \label{def_angl_dep_indep2}
    \mathrm{St}_\varsigma^{dep,(m=0)} \equiv 0
    \:,
 \end{equation}
 so that  the collision integral on the functions
 being independent  on $\alpha$ is given  by
the the angular-independent part:
 \begin{equation}
  \label{def_an_in}\mathrm{St}_\varsigma
      [ \, c_0 (\phi) F(\varepsilon) \, ]
       \equiv \mathrm{St}^{in}_\varsigma[ \, F(\varepsilon) \, ]
  \:,
\end{equation}
where $c_0 (\phi) = 1 $.

  We will show in next subsections that for energy-dependent
 functions (\ref{X}) with $F(x)\neq \mathrm{const}$,
some of the angle-independent contributions
 $\mathrm{St}^{in} _\varsigma$ are much greater than
 all the angular-dependent parts $\mathrm{St}^{dep} _\varsigma$.
At the same time, for the functions $\Psi_m^F$~(\ref{X})
with $F(x)= \mathrm{const}$ and $m\neq 0$,
the operators $\mathrm{St}^{in} _\varsigma$
 become exact zero [see Eq.~(M3)]  and only the contributions
 $\mathrm{St}^{dep,(m)} _\varsigma$ remain.

 \subsection{3.2.     The collision integral
    at $\varphi \to 0 $ }
According to Eq.~(M7), integration by
the scattering angle  $\theta$  of the expressions
in the collision integral~(M3) containing
the energy delta-function yields:
\begin{equation}
  \label{energ_delta__at_0}
  \begin{array}{c}
  \displaystyle
\int _0 ^{\pi} d \theta \:
\delta \Big[\, \varepsilon_1 + \varepsilon _2
 - \varepsilon_3 -
\frac{( \mathbf{p}_1 + \mathbf{p}_2
- \mathbf{p}_3  )^2}{2m}
 \, \Big] \times
  \\
 \\
 \displaystyle
\times H ( \varphi, \theta ; \breve{x} )
 =
 \sum \limits _{\pm}
 \frac{ H \, [ \, \varphi,\theta_{\pm}(\varphi) ; \breve{x} \, ] }
 { \displaystyle
 2 \, \varepsilon_F  \sqrt { \varphi^2
 + \zeta ^2
 \Delta
 }
 }
 \:,
 \end{array}
\end{equation}
where $ H (\varphi,\theta , \breve{x} ) $ is any regular function
 of the variables $\varphi, \theta $, and  $\breve{x}$.

Substitution of Eqs.~(\ref{varphi_4__at_0})-(\ref{varphi_4__at_0_short})
 into the exact matrix element (\ref{matr_el_ome}),
  taking into account the energy transfer in collisions,
  leads to the following result for
  the dimensionless squared matrix element
  $ \tilde{M}^2_{\pm} (\varphi , \breve{x})  = M^2 [
  \varphi , \theta_{\pm} (\varphi , \breve{x}), \breve{x}
  ] /(\pi e^2 a_B )^2$:
\begin{equation}
  \label{sq_matr_el__at_0}
  \begin{array}{c}
  \displaystyle
 \tilde{M} ^ 2  _{\pm}=
 2
\, \Big\{ \,
\frac{1}{
 \displaystyle
 |\chi_q +   q_0|^2 }
 +
\frac{1}{
 \displaystyle
 |\chi _w +   w_0|^2 }
 -
 \\
 \\
 \displaystyle
 - \mathrm{Re} \, \Big[\,
\frac{1}{
 \displaystyle
( \chi_q +   q_0 ) ( \chi_w +   w_0 )^*
 }\:
 \Big]\: \Big\} \: ,
 \end{array}
 \end{equation}
  where
\begin{equation}
 \chi_q = 1 - \frac{ i\,\epsilon}
 {\sqrt{v_F^2 q^2 - \epsilon^2}}
 =1-\frac{i\, \zeta \, (x_3-x_1)}{
  |\theta_{\pm}| }
 \:,
 \end{equation}
\begin{equation}
 \chi_w = 1 - \frac{ i\,\epsilon ' }
 {\sqrt{v_F^2 w^2 - (\epsilon' )^2}}
 =1-\frac{i\, \zeta \, (x_3-x_2)}{
  |\psi_{\pm}| }
 \:,
 \end{equation}
[here we used Eq.~(\ref{q_w_appr_0})] and
\begin{equation}
 q_0  = \frac{1}{ \sqrt{2} \, r_s}  \, \sqrt {
 \theta_{\pm} ^2
 +
  \frac{1}{4}\,
 \zeta^2  \, ( x_1  - x_3 )^2
 }
 \:,
 \end{equation}
\begin{equation}
 \label{w0}
 w_0 =  \, \frac{1}{ \sqrt{2} \,  r_s}  \sqrt {
 \psi_{\pm} ^2
 +
 \frac{1}{4}
  \,
 \zeta^2\,  ( x_2  - x_3 )^2
 }
 \:.
 \end{equation}

Due  to the singular factor $1/\sqrt{\varphi^2 + \zeta^2 \Delta  } $
in Eq.~(\ref{energ_delta__at_0}), the contribution $\mathrm{St}_0$
 to the collision integral
$\mathrm{St}$~(M3)  from the angles $\varphi$ in the vicinity
 $ \varphi_{\mathrm{min}}  < \varphi <  \varphi_{\mathrm{max}} $
 of the angle  $\varphi = 0$ can be much greater
 that the contribution from the angles $\varphi \sim 1 $.
  Owing to Eq.~(\ref{varphi'__at_0}),
  the minimal angle $ \varphi_{\mathrm{min}}   $
   is equal to zero at $\Delta >0$ and
   to $\zeta \sqrt{-\Delta }$ at $\Delta <0$.
 The maximum angle $ \varphi_{\mathrm{max}} $
 depends on the distribution function $\Psi$
    and the relation between $\zeta$ and $r_s$.
Herewith, in order to use the above formulas
(\ref{varphi_4__at_0})-(\ref{varphi_4__at_0_short})
and
(\ref{sq_matr_el__at_0})-(\ref{w0})
 one needs
 that $ \varphi_{\mathrm{max}} \lesssim \sqrt{\zeta} $.
If it turns out that $ \varphi_{\mathrm{max}} >\sqrt{\zeta} $,
 one needs to use the general expressions (M3),
  (\ref{fi_4}), and (\ref{fi_pm_exact}).

 The analysis shows that the collinear collisions
can provide the substantial contribution $\mathrm{St}^{in}_0$
  in  the angular-independent  $\mathrm{St}^{in}$
 part of the operator  $\mathrm{St}$~(M3).
  It is written as:
\begin{equation}
 \label{St__at_00}
\begin{array}{c}
 \displaystyle
\mathrm{St}^{in}_0 [\Psi]
   (x _1 , \alpha) = -  \frac{T^2}{ \varepsilon _F }
 \frac{1}{ 8 \pi \hbar T }
    \int
    \frac{
    d x _2  \, d x _3
    }{
    \displaystyle
    f_1  f_2 f_3 f_4 }    \times
        \\
 \\
  \displaystyle
                \int \limits _{\varphi  _{\mathrm{min }}
    }
    ^  {\varphi  _{\mathrm{max }}
   }
     \frac{ d \varphi  \,    }
     { \sqrt {\varphi^2 + \zeta^2 \Delta } }
     \, \sum_ {\pm} \tilde{M} _{\pm} ^2
               \left[
    \Psi (x_1 , \alpha)
    +
    \Psi (x_2 , \alpha )
        \right.
            \\
     \\
     \displaystyle \left.
                  -
    \Psi ( x _3 , \alpha )
    -
 \Psi ( x_4  ,
        \alpha    \, ) \,
    \right]
    ,
     \end{array}
\end{equation}
where $f_i = 2\cosh(x_i/2)$, $i=1,2,3,4$; and, as usual,
 $x_4 = x_1+x_2  - x_3 $.  Due to the singular character
  of the integrand in Eq.~(\ref{St__at_00}),
 such  $\mathrm{St}_0^{in}$  must be independent on
  the upper limit $\varphi  _{\mathrm{max }} $
  in the main order by the small parameters
  $\zeta \ll 1 $ and $ r_s \ll1 $.

According to the definition of the angular-independent part
(\ref{def_angl_dep_indep}), in formula (\ref{St__at_00})
 integration over angular variable $\varphi$
 does not affect the unknown distribution function $\Psi$.
 Therefore this integration
  can be performed explicitly.

At moderately   low temperatures,
  $ \zeta  \ll r_s \ll \sqrt{\zeta} $,
   for the  squared matrix element at the angles
   $ \varphi_{\mathrm{min}}  < \varphi \ll r_s$
    we have from (\ref{sq_matr_el__at_0}):
  \begin{equation}
  \label{sq_matr_el__at_0_app}
  \begin{array}{c}
  \displaystyle
 \tilde{M}^2  _{\pm}=
 2
\, \Big[ \,
\frac{1}{
 \displaystyle
 |\chi_q |^2 }
 +
\frac{1}{
 \displaystyle
 |\chi _w |^2 }
 - \mathrm{Re} \, \Big(\,
\frac{1}{
 \displaystyle
\chi_q \chi_w ^*
 }\:
 \Big)\: \Big] \: ,
 \end{array}
 \end{equation}
At $\varphi \sim \zeta$ this expression  as a function
 of $\varphi$ substantivally depends
 on the energy variables $\breve{x}$,
 herewith we have:
   \begin{equation}
   \label{M_sim_1}
    \tilde{M}_{\pm}^2  \sim 1 \:.
    \end{equation}
At the angles
$ \zeta \ll \varphi \ll r_s $ one of
 the terms $2/|\chi_q|^{2}$ or $2/|\chi_w|^{2}$
  (depending on the sign $\pm$)
 dominates in Eq.~(\ref{sq_matr_el__at_0_app}) and
  the squared matrix element  becomes equal to
 $2$ in the main order by $r_s$ and $\zeta$:
\begin{equation}
  \label{M_eq_2}
 \tilde{M}_{+}^2 = 2 \, , \quad \tilde{M}_{-}^2 = 2  \:.
 \end{equation}
 At the larger angles, $ r_s \ll \varphi \ll \sqrt{\zeta}   $
 the matrix element $\tilde{M}^2_{\pm}$ is
   determined by general formula (\ref{sq_matr_el__at_0}).
    A calculation shows that it
  decreases as   $ \propto 1/\varphi^2$:
 \begin{equation}
 \label{M_fi_gr_rs}
   \tilde{M}^2_{\pm}
    \sim  r_s^2 /\varphi^2 \:.
 \end{equation}

 Using these estimates of $\tilde{M}^2_{\pm}$,  one
 can calculate the integral by $d\varphi$
  in the operator $\mathrm{St}_0^{in}$:
 \begin{equation}
 \label{int}
 I_0(\breve{x})= \int \limits
 _ { \varphi  _{ \mathrm{ min } } }
 ^{\varphi  _{ \mathrm{max }  }}
\frac{d \varphi \: \tilde{M}_{+}^2 }
{ \sqrt{\varphi^2+ \zeta^2 \Delta  }}
\approx
\int \limits _ { \varphi  _{ \mathrm{ min } } }
 ^{\varphi  _{ \mathrm{max }  }}
\frac{d \varphi \: \tilde{M}_{-}^2 }
 { \sqrt{\varphi^2+ \zeta^2 \Delta  }}
\:.
 \end{equation}
 Putting  $\varphi  _{ \mathrm{ min } }  = \zeta$
  and $\varphi  _{ \mathrm{max }  } = r_s$ in view
  of the fast decrease of $\tilde{M}^2_{\pm} (\varphi ) $
   at $ \varphi \gg  r_s $ [Eq.~(\ref{M_fi_gr_rs})] and
   and keeping in mind the estimate
  $ \Delta  \sim 1 $ at  $x_{i}\sim 1  $,
  we have:
\begin{equation}
 \label{int1}
I_0  \approx  2 \:
 \int \limits _ {  \zeta }  ^{ r_s }
\frac{d \varphi  }{ \varphi}
   =2 \, \ln(  r_s/\zeta )
  \:.
\end{equation}

In the limit  of very   low temperatures,
 $  \sqrt{\zeta}   \ll r_s $,  result (\ref{M_eq_2})
 for the matrix element  $\tilde{M}^2_{\pm}$ remains valid
 in the whole range, $0<\varphi \ll \sqrt{\zeta}$, for
which equations (\ref{varphi_4__at_0})-(\ref{varphi_4__at_0_short})
and (\ref{sq_matr_el__at_0})-(\ref{w0})  have been
 derived. At larger angles, $\varphi \gtrsim \sqrt{\zeta} $,
   the matrix element  is calculated by the general formulas
    (\ref{fi_4}), (\ref{fi_pm_exact}),
   (\ref{q_abs}), (\ref{obmen_analog__q}), and (\ref{matr_el_ome}).
We performed such calculation and obtained
equation~(\ref{M_sim_1}) for  $\tilde{M}^2_{\pm}$
 at $ \sqrt{\zeta} \ll \varphi \ll r_s $
 and equation~(\ref{M_fi_gr_rs})
at $ \varphi \gtrsim r_s$. This estimates are similar to
 the ones for the matrix element $\tilde{M}^2_{\pm}$
  in the case  $\zeta  \ll r_s \ll \sqrt{\zeta} $
  up to the change of Eq.~(\ref{M_eq_2}) on Eq.~(\ref{M_sim_1}),
therefore we again arrive to  result (\ref{int1})
for the angle integral $I_0$~(\ref{int})
with the sign ``$\sim$'' instead of the sign
  ``$\approx $''.

In the case not too small temperatures (the moderately
 weak interaction), $\zeta ^{3/2} \ll r_s \ll \zeta $,
equation~(\ref{sq_matr_el__at_0}) leads to
 the following estimates of the matrix element:
  \begin{equation}
   \label{M_small_rs_0}
   \tilde{M}_{\pm}^2 \sim  r_s^2/\zeta^2 \:.
 \end{equation}
at the angles $ \zeta \lesssim \varphi \ll \zeta^2/r_s$ and
   \begin{equation}
   \label{M_small_rs_1}
   \tilde{M}_{\pm}^2 \sim   \zeta^2/\varphi^2 \:.
 \end{equation}
at the angles
$\zeta^2/r_s\lesssim \varphi \ll \sqrt{\zeta} $.
 In this case, we introduced the lower bound
  on the value $r_s $, $r_{s} \gg \zeta^{3/2}$,
  in order to  fulfill the inequality
  $\zeta^2/r_s \ll \sqrt{\zeta} $  allowing to use
equations~(\ref{varphi_4__at_0})-(\ref{varphi_4__at_0_short})
and (\ref{sq_matr_el__at_0})-(\ref{w0})
 for $\tilde{M}^2_{\pm}$.

 Formulas~(\ref{M_small_rs_0}) and (\ref{M_small_rs_1})
  lead to the estimate   $\varphi_{\mathrm{max}} \sim \zeta^2 /r_s$
  and    therefore yield the following result
  for  the integral $I_0$:
\begin{equation}
 \label{int2}
I_0
 \sim
\frac{r_s^2}{\zeta^2}   \int \limits _ { \zeta }  ^{ \zeta^2/r_s }
\frac{d \varphi  }{ \varphi}
   \approx  \frac{r_s^2}{\zeta^2}  \, \ln(  \zeta /r_s )
  \:.
\end{equation}

Integration by one of the energy variables
$x_2$ and $x_3$ in Eq.~(\ref{St__at_00})
 leads to the final form of the angular-independent
  contribution  to $\mathrm{St}$ from the collinear
   scattering:
\begin{equation}
 \label{St__at_0}
\begin{array}{c}
 \displaystyle
\mathrm{St}_0^{in} [\Psi]
   (x _1 , \alpha)
  = -  \frac{T^2}{ \varepsilon _F }
  \frac{  1 }{ 4 \pi \hbar T  }
  \int     d x'  \: I_0  \,  K_0 (x_1,x')
 \\
 \\
     \displaystyle
     \times
        [ \,
    \Psi ( x_1 , \alpha)
    +
    \Psi ( - x', \alpha )
    -2 \Psi (  x', \alpha )
   \, ]
      \displaystyle
    \:,
     \end{array}
\end{equation}
where $I_0$ is given by Eqs.~(\ref{int1}) or  (\ref{int2}),
the kernel  $K_0(x_1,x')$  is   an exponentially
decreasing  function  at $|x_1|,|x'| \gtrsim 1$:
\begin{equation}
 \label{K0}
\begin{array}{c}
\displaystyle
K_0 (x_1 , x') = \frac{  1  }{ 4 \cosh (x_1/2) \cosh (x'/2)  }
\times
    \\
       \\
      \displaystyle
\times
\frac{ (x_1 - x')/2 }{ \sinh [(x_1 - x')/2] }\:.
 \end{array}
\end{equation}

According to the definition (\ref{def_angl_dep_indep}),
(\ref{def_an_in}) of the angular independent part,
 we have for functions $\Psi^F_m $ (\ref{X})
  with $F(x) \neq \mathrm{const}$
   \begin{equation}
   \label{St_0_app_F_neq_const}
   \mathrm{St}_0 ^{in}[\Psi ^F_m ] (x_1 , \alpha ) =
    \cos(m  \alpha ) \,
     \mathrm{St}_0  ^{in}[F(x) ] (x_1 , 0)
   \:.
 \end{equation}
For  the functions
 $ \Psi _m  (x , \alpha) = \Psi   (\alpha)
 $,   being independent on the energy variable $x $,
 operator~$  \mathrm{St}_0 ^{in}$  is zero
 [see Eqs.~(\ref{St__at_00}) and  (\ref{St__at_0})].

   \subsection{3.3. The collision integral
   at $\varphi \to \pi $}
 Integration  of the expression in the collision integral
 $\mathrm{St}$~(M3) with the energy delta-function
 by the angle $\theta $ takes the form:
\begin{equation}
  \label{energ_delta__at_pi}
  \begin{array}{c}
  \displaystyle
\int _0 ^{\pi} d \theta \:
\delta \Big[\, \varepsilon_1 + \varepsilon _2
 - \varepsilon_3 -
\frac{( \mathbf{p}_1 + \mathbf{p}_2
- \mathbf{p}_3  )^2}{2m}
 \, \Big] \times
  \\
 \\
 \displaystyle
\times L ( \varphi ,   \theta  , \breve{x} )
 =
 \sum \limits _{\pm}
 \frac{ L \,[ \, \pi - \zeta s
 \, , \, \theta_{\pm}(s , \breve{x})  \,,\, \breve{x} \, ] }
 { \displaystyle
 2 \, \varepsilon_F \, \zeta   \, \sqrt { s^2
 +
 \Delta
 }
 }
 \:.
 \end{array}
\end{equation}
Here $L(\varphi ,   \theta  , \breve{x} )$  is any regular
function of the variables $\varphi$, $ \theta $,
 and $\breve{x}$.  The value $\Delta =\Delta (\breve{x} ) $
   is expressed  via the variables
  $a$ and $b$ as:  $ \Delta =  a^2 - b^2 $.

 As we discussed in Section 2,
  for the contribution $\mathrm{St}_{\pi}$
    to the collision integral $\mathrm{St}$
 from the angles $\pi -|\varphi | \ll 1 $
 the dependence of scattering probability
  on the transferred energy $\epsilon$
  is not substantial and
one can use the matrix element $   M^2$~(M5)
 with the static  inter-particle potential $V(q)$ (M6).

The value  $\cos \theta $,  entering
   Eqs.~(\ref{q_w_appr_pi}) for
   $q $    and   $w $ in the matrix element  $M^2$,
 at the actual scattering angles
  $\theta = \theta _{\pm} $ and $s \sim 1 $  is estimated as unity.
  In  the range $ 1 \ll s \ll 1/\zeta $
  it takes the form:
  \begin{equation}
 \label{cos_varpi'__at_pi}
 \begin{array}{c}
 \displaystyle
 \cos[ \theta _{\pm} (s , \breve{x})]  \approx \mp \Big[ \,  1
 -
 \frac{ (a \pm b)^2  }{2 \,s ^2 } \, \Big]
  \:.
  \end{array}
\end{equation}
 Thus for the the values $ [ 1 \pm \cos (\theta_{\pm}) ]$,
 occurring  in expressions  (\ref{q_w_appr_pi})
 for $q$ and $w $,  we obtain the relations:
 \begin{equation}
 \begin{array}{c}
    \displaystyle
 \zeta^2 \ll  1 + \cos (\theta_+)  \ll 1
  \: ,
  \\\\
     \displaystyle
  \zeta^2 \ll  1 - \cos (\theta_-)  \ll 1
  \: ,
\\\\
    \displaystyle
  1 - \cos (\theta_+) \approx
  1 + \cos (\theta_-)  \approx
  2
  \end{array}
  \end{equation}
 at  $ 1 \ll s \ll 1/\zeta $.
 Therefore the term  $\zeta^2(x_1-x_3)^2/4$
in Eqs.~(\ref{q_w_appr_pi})
 can be omitted, and the momenta
  $q$ and $w$ are expressed as: $q = ( 2 \hbar  /a_B ) \, q_{\pm} $
  and  $w = (2 \hbar  /a_B) \, w_{\pm} $, where
 \begin{equation}
 \label{q_w_pm}
 \begin{array}{c}
    \displaystyle
 q_{\pm} (s , \breve{x})=  \frac{\sqrt{1-\cos (\theta _{\pm})} }
 {  r_s }
 \:
 \approx
 \frac{ \sqrt{2} }{  r_s } \, , \;\;  \frac{ |a-b| }
 { \sqrt{2} \,  r_s  s }
 \, ,
   \\
  \\
    \displaystyle
 w_{\pm} (s , \breve{x})=
 \frac{\sqrt{1+\cos (\theta _{\pm})} }
  { \sqrt{2} \,   r_s }
 \:
 \approx
   \frac{ |a+b| }{ \sqrt{2} \, r_s  s }
   \, , \;\;  \frac{ \sqrt{2}}{  r_s }
  \:.
  \end{array}
  \end{equation}
Let us remind here
 that $r_s = \sqrt{2} \hbar / (p_F a_B) \ll 1$.

 From Eqs.~(\ref{q_w_pm}) we get  the following inequalities
  between the factors in the direct  and
 the exchange contributions  in the matrix element (M5):
\begin{equation}
 \label{kernel_ineq__at_pi}
 \begin{array}{c}
 \displaystyle
\frac{1}
{
 \displaystyle
 1+  q_+  }
 \ll
\frac{1}
{
 \displaystyle
 \,  1+   q_-   }
\:,
\quad
\frac{1}
{
 \displaystyle
 1+  w_-  }
 \ll
\frac{1}
{
 \displaystyle
 \,  1+   w_+   }
 \:,
\\
\\
  \displaystyle
\frac{1}{
 \displaystyle
 ( \,  1+  q_-  )
( \, 1+  w_-  ) }
 \ll
  \frac{1}{
  \displaystyle
 ( \, 1+   q_- \, ) ^2 }
  \:,
\\
\\
  \displaystyle
\frac{1}{
 \displaystyle
( \,  1+  q_+   )
( \, 1+  w_+ )  }
 \ll
  \frac{1}{
  \displaystyle
 ( \, 1+   w_+  \,  ) ^2 }
 \:.
  \end{array}
\end{equation}
  Based on inequalities~(\ref{kernel_ineq__at_pi}),
 we obtain for the main terms
 in the dimensionless squared matrix
 element  $ \tilde{M}^2  _{\pm} (\varphi,\breve{x})
 = M^2[\varphi ,\theta_{\pm}(\varphi,\breve{x}),\breve{x}]
 / (\pi e^2 a_B)^2$:
\begin{equation}
 \label{matr_el__at_pi}
 \begin{array}{c}
   \displaystyle
\tilde{M}^2_- (s, \breve{x} )
=
\frac{2 }{
 \displaystyle
 (
 \,
 1 +  q_- )^2
}
=
 \frac{2 }{
 \displaystyle
 \Big(
 \,
 1+  \frac{|a-b|}{ \sqrt{2} \,  r_s s}
 \, \Big)^2 }
 \:,
\\
\\
 \displaystyle
\tilde{M}^2_+ ( s; \breve{x} )
 =
\frac{2 }{
 \displaystyle
 (
 \,
 1 +  w_+ )^2 }
=
 \frac{2 }{
 \displaystyle
 \Big(
 \,
 1+  \frac{|a+b|}{ \sqrt{2} \,  r_s s}
 \, \Big)^2 }
 \:.
 \end{array}
\end{equation}

The analysis shows that the collisions with the electrons
having the angles $\pi - \varphi   \ll 1$
 can dominate  both the angular-dependent and
 angular-independent parts of $\mathrm{St}$.
  For the sum   $ \mathrm{St}_\pi  = \mathrm{St}_\pi ^{in}
  + \mathrm{St}_\pi ^{dep}$ of such the two  contributions,
 $ \mathrm{St}_\pi ^{in} $ and $  \mathrm{St}_\pi ^{dep}$,
in the main order by $\zeta$ we obtained:
\begin{equation}
 \label{St__at_pi}
\begin{array}{c}
 \displaystyle
\mathrm{St}_{\pi} [\Psi] (x_1 , \alpha)
=
 -  \frac{T^2}{ \varepsilon _F }
  \frac{1}{ 8 \pi \hbar T }
    \int
    \frac{
    d x _2  \, d x_3
    }{
    \displaystyle
           f_1 f_2 f_3 f_4}
     \int \limits _ {
   s _{\mathrm{min }}
    } ^  { s _{\mathrm{max }} }ds
  \\
 \\
 \displaystyle
     \times  \sum \limits _{ \pm }
    \frac{
    \tilde{ M}^2 _{\pm} ( s; \breve{x} )
    }{ \sqrt { s^2 + \Delta } }
     \left\{ \,
    \Psi ( x _1 , \alpha)
  +
    \Psi (x_2 , \alpha + \pi ) -
            \right.
        \\
      \\
    \displaystyle
    \left.
        -  \Psi [\,  x _3 , \alpha + \theta_{\pm}(s) \, ]
    -     \Psi [ \, x_4 ,
        \alpha    + \pi + \theta _ {\pm}(s) \, ]     \,
     \right\}
    \:,
        \end{array}
\end{equation}
 where $f_i = 2 \cosh(x_i/2) $, $i=1,2,3,4$;
 $x_4 = x_1+x_2 - x_3$;  and the lower limit
 $s  _{\mathrm{min }} = s  _{\mathrm{min }} (\breve{x}) $
is  zero at $\Delta >0 $ and $\sqrt{-\Delta}  $
at $\Delta <0 $.  In view of the singular factor
 $1/\sqrt{s^2 + \Delta}$ in $\mathrm{St}_\pi$~(\ref{St__at_pi}),
 for the upper limit  we choose just  the maximum possible
  value when the expressions (\ref{varphi_4_at_pi}),
  (\ref{varphi'__at_pi}),   (\ref{matr_el__at_pi}),
  and (\ref{St__at_pi})   remain asymptotically
   valid:  $s_{\mathrm{max}} = 1/\zeta$.

An analysis based on Eqs.~(\ref{matr_el__at_pi}) and
(\ref{St__at_pi}) shows that at low temperatures,
 $ \zeta \ll r_s $,  the main contribution
 in the operator  $\mathrm{St}_{\pi} $  for
 the functions $\Psi _m^F = F(x) \cos(m\alpha)$~(\ref{X})
with even $m$  originates  from the angles $ \varphi $
 in the interval   $\zeta \ll \pi - |\varphi| \ll 1$,
    that corresponds to $1 \ll s \ll 1/\zeta$.

In this way, one should use in Eq.~(\ref{St__at_pi})
 for the functions  $\theta_{\pm} (s)$
 their asymptotes at $ s \gg 1 $, being equivalent
  to asymptotes~(\ref{cos_varpi'__at_pi}).
For example,    in the case $b>a>0$  we have:
\begin{equation}
\label{assy_theta_pl_mi}
\begin{array}{c}
 \displaystyle
   \theta_+(s,\breve{x}) =  \pi - \frac{a+b }{s}
  \:, \quad\quad
   \theta_-(s,\breve{x}) =  \frac{b -a }{s}
  \:.
  \end{array}
  \end{equation}
It follows from these functions that  at such $s$
 the head-on collisions are  simultaneously small-angle:
 $|\theta_- |\ll 1$ or $|\psi_+ |= |\pi - \theta_+| \ll 1$
   [see also Fig.~2(b) in the main text].

The change of variables $x_3 \leftrightarrow x_4 $
 in the terms ``$+$''  in Eq.~(\ref{St__at_pi})
 leads to the following simplification of $\mathrm{St}_{\pi}$
 on the functions
   $\Psi _m^F = F(x) \cos(m\alpha)$~(\ref{X})
  with even $m$:
  \begin{equation}
 \label{St__at_pi_00}
\begin{array}{c}
 \displaystyle
\mathrm{St}_{\pi} [\,\Psi _m^F \, ] (x_1 , \alpha ) =
 -  \cos (m\alpha  ) \, \frac{T^2}{ \varepsilon _F }
 \,  \frac{1}{ 8 \pi \hbar T }
\\
\\
\displaystyle
  \times
    \int
    \frac{
    d x_2 d x_3
    }{
    f_1f_2f_3f_4       }
         \int \limits _ {1  } ^{ 1/\zeta }
     \frac{ds}{ s}
     \frac{4}{ \Big(\displaystyle  1 +
     \frac{|a-b|}{\sqrt{2} \,  r_ss} \Big )^2}
 \\
 \\
 \displaystyle
 \times    \big \{ \,
    [  F( x _1 ) + F( x _2 ) - 2 F( x _3 ) ]
  + \\
 \\
 \displaystyle
 +
  2  \,  F( x _3 ) \,[    1-\cos (m\theta_-
  )\, ]     \,
    \big \}
    \:.
        \end{array}
\end{equation}
Here, in particular, we changed the interval of integration
from $ s_{\mathrm{min}} <s< 1/\zeta $ on
$ 1 <s< 1/\zeta $, that is valid in the main order
by $\zeta $ due to the presence of  the singular factor
 $1/ \sqrt { s^2 + \Delta } \approx 1/s$
in the source expression~(\ref{St__at_pi}).

We see that the operator $\mathrm{St}_{\pi}$ (\ref{St__at_pi_00})
  is indeed divided on the two parts,
  $\mathrm{St}_{\pi}=\mathrm{St}_{\pi}^{in} + \mathrm{St}_{\pi}^{dep}$,
  the first of which, $\mathrm{St}_{\pi}^{in}$,
 is the integral operator only by the energy variable $x$,
 while the second one, $\mathrm{St}_{\pi}^{dep}$,
 acts on both the angular and the energy variables $s$ and $x$
  and is equal to zero at $m=0$, according
  to definition (\ref{def_angl_dep_indep}), (\ref{def_angl_dep_indep2}).

  The integration  by $d\varphi$ in the both parts
  can be explicitly calculated.
    The integral by $s$ in the first angular-independent part
     of Eq.~(\ref{St__at_pi_00}), $\mathrm{St}_{\pi}^{in}$, is
 \begin{equation}
 \label{asy_angle_int_en}
  I_{\pi} (\breve{x})=    \int \limits _ { 1} ^ {
   1/\zeta}
     \frac{ds}{ s }
          \frac{   r_s^2 s^2   }{
  \displaystyle
 (  \,
  r_s \, s + |a-b|/\sqrt{2}
 \, )^2
 }
 \:,
\end{equation}
At $\zeta \ll r_s $ it turns out into:
  \begin{equation}
 \label{asy_angle_int_en_res}
  I_{\pi} \approx    \int _ { 1/r_s} ^ {
   1/\zeta}
     \frac{ds}{ s } = \ln(r_s / \zeta)
  \:.
\end{equation}

The integral by $s$ in the second angular-dependent
  part of Eq.~(\ref{St__at_pi_00}), $\mathrm{St}_{\pi}^{dep}$,
   takes the form:
\begin{equation}
 \label{asy_angle_int}
  I_s(\breve{x})=    \int \limits _ { 1} ^ {
   1/\zeta}
     \frac{ds}{ s }
          \frac{   r_s^2 (a-b)^2   }{
  \displaystyle
 (  \,
  r_s \, s + |a-b|/\sqrt{2}
 \, )^2
 }
 \:,
\end{equation}
 In all the  three cases $\zeta \ll r_s$, $\zeta \gg r_s$,
 and $\zeta \sim r_s$  the result of its calculation can be presented in the form:
\begin{equation}
 \label{res_angle_int0}
I_s = 2\, r_s^2\,
  \ln\Big( \,
  \frac{1}{\zeta + r_s}
  \, \Big)
\:.
\end{equation}
It is seen from Eq.~(\ref{asy_angle_int}) that at $\zeta \ll r_s$
 the main contribution to $I_s$ comes the angles
  $\zeta \ll \pi - \varphi \ll \zeta/r_s$ [see also inset in Fig.~2(b) in the main text], while at $\zeta \gg r_s$
   all the head-on collisions, $\zeta \ll \pi - \varphi \ll 1$,
   determine the magnitude  of  $I_s$.

After the substitution of  using Eqs.~(\ref{assy_theta_pl_mi}),
 (\ref{asy_angle_int_en}), and (\ref{res_angle_int0})
 into Eq.~(\ref{St__at_pi_00})
and the integration over one of the variables $x_2$ and $x_3$,
  the operator~$\mathrm{St}_{\pi}$
 is written as:
\begin{equation}
 \label{St__at_pi_00_3}
\begin{array}{c}
 \displaystyle
\mathrm{St}_{\pi} [\, \Psi_m^F \, ] (x_1 , \alpha  ) =
 -  \cos ( m \alpha   ) \,\frac{T^2}{ \varepsilon _F }
  \frac{1}{ 2 \pi \hbar T}
          \\
 \\
 \displaystyle
\times
 \int
        d x' \, K_0(x_1,x')
 \,  \big\{
\, I_\pi \,
   [ \,  F( x _1 ) + F( - x ' ) -
      \\
 \\
 \displaystyle
   -2 F(  x ' ) \, ]
        + I_s  \,  m^2
  F( x ' )
\,
     \big\}
    \:.
        \end{array}
\end{equation}
 where the kernel $K_0(x_1,x')$~(\ref{K0}) is the same as
  for the contribution $\mathrm{St}_0$~(\ref{St__at_0}).

It is important that the angular-dependent part
$\mathrm{St}_{\pi}^{dep}$ of the obtained operator
 $\mathrm{St}_\pi$~(\ref{St__at_pi_00_3}),
  being proportional to $I_s \ll 1 $,
 is much small than its angular-independent
  $\mathrm{St}_{\pi}^{in}$,  containing
  the factor $I_\pi \gg 1$.

At the very weak interparticle interaction, corresponding
 to moderately low temperatures, $ r_s \ll \zeta $,
  an analysis of  $\mathrm{St}_\pi$~(\ref{St__at_pi})
applied to the functions $\Psi_m^F$   again allows to
 divide in it the angle-independent and the angle-dependent parts.
 The second one again  originates
  from the head-on collisions corresponding to
 $1 \ll s \ll 1/ \zeta$ and is expressed
 by the last term in the curly brackets in Eq.~(\ref{St__at_pi_00_3})
  with the same $I_s$~(\ref{res_angle_int0}).
 However,
 the  angle-independent part of $\mathrm{St}_\pi$
  now mainly comes not from the interval  $\pi-\varphi \ll 1 $,
  but from the intermediate angles, $\varphi \sim 1$
  [see Eq.~(\ref{asy_angle_int_en})].
  That is, the asymptotic form of the angular-independent terms
  in $\mathrm{St}_\pi$  in the limit $\varphi  \to \pi$
  becomes  unapplicable for exact calculations.
   However, Eq.~(\ref{St__at_pi_00_3}) now provides the estimation
  for the angular-independent part $\mathrm{St}_1^{in}$
  of the operator $\mathrm{St}$  from the angles $\varphi \sim 1$.
   So the integral factor $I_\pi$~(\ref{asy_angle_int_en}),
  becoming in fact the factor $I_1$
  in the contribution $\mathrm{St}_1^{in}$,
   is evaluated  as:
   \begin{equation}
   \label{int_est}
   I_\pi \sim r_s^2 / \zeta^2\:.
   \end{equation}
  This value just corresponds to the estimate
 $M^2_{\pm} \sim  r_s^2 / \zeta^2 $    of the matrix element,
 being valid at the minimal scattering angles
     $\theta \sim \zeta $ and $ \psi \sim \zeta $,
 corresponding to $\varphi \sim 1 $
 [see Eq.~(\ref{fi_pm_asimpt})].

  We see that the contribution $\mathrm{St}_{\pi}^{in}$
  with such $I_\pi \sim I_1$~(\ref{int_est})
  is much smaller as compared
 with the angular-independent part
  $\mathrm{St}_{0}^{in}$~(\ref{St__at_0})
  from the collinear collisions, $\varphi \ll 1$.

From Eq.~(\ref{St__at_pi_00_3}) we see the two properties
 of the operator $\mathrm{St}_{\pi} $ acting on the functions
  $\Psi_m^F$~(\ref{X}) with even $m$. First,
as  the angular-independent part of $\mathrm{St}_{\pi}$
 is much larger than the angular-dependent one,
 for the functions $\Psi_m^F = F(x)\cos(m\alpha)$
 with $F(x)  \neq \mathrm{const}$  the equality
 takes place:
\begin{equation}
   \mathrm{St}_\pi [\Psi ^F_m ] (x_1 , \alpha ) \approx
    \cos(m  \alpha ) \,   \mathrm{St}_\pi [F(x) ] (x_1 , 0)
   \:.
\end{equation}
 Second, the first term   in the curly brackets in
  $\mathrm{St}_\pi $~(\ref{St__at_pi_00_3}),
 being the angular-independent part $\mathrm{St}_\pi ^{in}$,
turns   into zero for the  function
 $F(x) $ equal to unity, $F(x)  \equiv\mathrm{ const}$.

 \subsection{3.4. The collision integral at $\varphi \sim 1$
  and relative roles
  \\ of the contributions  $\mathrm{St}_0$,
     $\mathrm{St}_\pi$, and~$\mathrm{St}_1$}
To estimate the contribution $\mathrm{St}_1$
 to the collision   integral $\mathrm{St}$
  from the intermediate angles, $\varphi \sim 1 $,
 we note the following.

Due to the asymptotic behavior of the scattering angles
 at $\varphi \sim 1 $ [Eq.~(\ref{fi_pm_asimpt})], the exact
 matrix element~(\ref{matr_el_ome}) in this range of $\varphi$
  is estimated as:
 \begin{equation}
   \tilde{M}^2_{\pm} \sim 1
 \end{equation}
 at low temperatures, $\zeta \ll r_s$,
 and as
 \begin{equation}
   \tilde{M}^2_{\pm} \sim r_s^2 /\zeta^2
 \end{equation}
 in the opposite case $\zeta \gg r_s$.
 This is just the typical value of $\tilde{M}^2_{\pm}$
 or the ``typically-small values'' of the scattering
 angle $\theta  $: $\theta \sim \zeta$
  in the range  $\varphi \sim 1 $
 [see Fig.~2(b) in the main text].

The factor (M7) from the energy delta-function
 is simply estimated as
 \begin{equation}
   \int d\theta \:\delta[\varepsilon_1 + \varepsilon _2 -
    \varepsilon _ 3- \varepsilon_4
    (\varphi,\theta,\breve{\varepsilon})\,  ]
    \sim \frac{1}{\varepsilon_F}
 \end{equation}
  compare with Eqs.~(\ref{energ_delta__at_0}) and
 (\ref{energ_delta__at_pi})]. As it was discussed in Section 2,
  the factor $S$~(\ref{sq_br}) for functions
   that substantially  depend on energy
  is of the order of unity [see Eq.~(\ref{en_cons_law})],
  while  for functions $\Psi_m=\cos(m\varphi)$
  it is a value of the order of  $\zeta^2 $
 in the first non-vanishing order after integration
  over the energy variables $x_i$
 [see Eq.~(\ref{S3}) and (\ref{S3_})].

 In view of these estimates and as the length
 of this interval is approximately equal to $\pi$,
 the contribution $\mathrm{St}_1$
  to the collision operator $\mathrm{St}$
  for  $\Psi^F_m=F(x)\cos(m\varphi)$
with  $F(x) \neq \mathrm{const}$
 is estimated as:
 \begin{equation}
 \label{est_St_1_en}
    \mathrm{St}_1[\Psi^F_m]  \approx
     \mathrm{St}^{in}_1[\Psi^F_m]
     \, \sim  \, \frac{ T^2}{\varepsilon_F }
      \frac{1}{\hbar T}
    \left\{
    \begin{array}{l}
    1
    \,, \;
    \zeta \ll r_s
    \\\\
         r_s^2/\zeta^2  , \; \zeta \gg r_s
         \end{array}
     \right.
 \end{equation}
 at $\zeta \ll r_s$ and $\zeta \gg r_s$.
  For the pure angular harmonics  $\Psi_m=\cos(m\varphi)$
as well as   for the angular-dependent part
 $ \mathrm{St} ^ { dep , (m) } $  of the operator $\mathrm{St}$
 on the general functions   $\Psi^F_m=F(x)\cos(m\varphi)$
 [here we imply that $F(x)\sim 1 $]
 we have:
 \begin{equation}
   \label{est_St_1_ang}
      \mathrm{St}_1[\Psi_m]  \sim  \mathrm{St}_1
      ^{dep,(m)}[F]
      \, \sim  \, \frac{ T^2}{\varepsilon_F }
      \frac{1}{\hbar T}
    \left\{
    \begin{array}{l}
   \zeta^2
    \,, \;\;
    \zeta \ll r_s
    \\\\
       r_s^2 \,, \;\; \zeta \gg r_s
       \end{array}
     \right.
    \:.
 \end{equation}

  An analysis based on equations~(\ref{est_St_1_en}),
 (\ref{est_St_1_ang}) and the results of Sections 3.2, 3.3
  shows that, at $\zeta \ll r_s$, for the functions
  $\Psi=\Psi^F_m$~(\ref{X})  with $F(x) \neq \mathrm{const}$
  the collinear and the head-on collisions
   provide the two main similar contributions
 $\mathrm{St}_0^{in}$ and $\mathrm{St}_\pi^{in}$
  to the angular-independent   part of~$\mathrm{St}$ [the terms
  given by Eqs.~(\ref{St__at_0}), (\ref{int1}) and
 (\ref{St__at_pi_00_3}), (\ref{asy_angle_int_en_res})].
   Herewith  the angular-dependent  part of $\mathrm{St}$
 for           $\Psi_m^F$ at even   $m$ and  any   $F(x) $
 [in particular, $F(x) = \mathrm{cosnt }$]
   is mainly   determined    by the head-on collisions
  corresponding   to $\pi - |\varphi | \ll 1 $
 [the part $\mathrm{St}_\pi^{dep}$
 of the operator  $\mathrm{St}_\pi$   described
 by Eqs.~(\ref{St__at_pi_00_3}),  (\ref{res_angle_int0})].

In the case  $r_s \ll \zeta$,  the collinear scattering make
 the dominant contribution to the angular-independent part
 of $\mathrm{St} [\Psi]$  for the distribution functions
  $\Psi = \Psi_m^F$   with $F(x) \neq \mathrm{const}$
[the operator  $\mathrm{St}_0^{in}$ given by
 Eqs.~(\ref{St__at_0}), (\ref{int2})], while
the angular-dependent  part of $\mathrm{St}[\Psi_m^F]$
 for even  $m$ and  any $F(x)$ is again determined mainly
 by the head-on collisions  [the operator
   $\mathrm{St}_\pi^{dep}$ described by
 equations~(\ref{St__at_pi_00_3}), (\ref{res_angle_int0})],
 despite the fact that their contribution to
 the angular-independent  part  of $\mathrm{St}$   becomes
 insignificant [compare Eqs.~(\ref{int_est})
   and~(\ref{est_St_1_en})].

 The relative magnitudes of the angular-dependent
contributions  to $\mathrm{St}$ for
 the functions $\Psi=\Psi^F_m$ being proportional
  to odd harmonics $m$  will be studied in details
 further,  in Section~5.3 and 5.4.  Here we announce
that at low temperatures, $\zeta \ll r_s$,
 the angular-dependent   part  of $\mathrm{St } [\Psi^F_m]$
 also originate from the head-on collisions,
 corresponding to  $\varphi$ in the vicinity
 of the angle $\varphi=\pi$.

 \section{ 4. The kinetic equation  for 2D electron
  Fermi gas }
 \subsection{4.1. Transport regimes  and their description
 by the kinetic equation }
 For the problems of weakly nonequilibrium transport
 in a 2D electron gas, the electron
 distribution function has the form:
 \begin{equation}
 \label{delta_f}
   f _{\mathbf{p}}  =
   f_F(\varepsilon) + \delta f _{\mathbf{p}} \:,
  \end{equation}
 where the perturbed part is written as
  $ \delta f _{\mathbf{p}} =  -f_F'(\varepsilon)
  \Psi( \varepsilon, \alpha ) $.  The kinetic equation
  in the presence of an external electric field
  and a magnetic field perpendicular
 to the electron layer is:
 \begin{equation}
    \label{kin}
      \begin{array}{c}
 \displaystyle
-f'_F(\varepsilon)\,
 \Big[ \, \frac{ \partial  \Psi }{ \partial t }
 +
  \mathbf{v}_{\mathbf{p}}
   \cdot
  \frac{ \partial  \Psi }
   { \partial \mathbf{r} }
+
 \omega_c  \frac{ \partial \Psi }{ \partial \alpha }
 -
 \\
 \\
 -
  \displaystyle
 e\mathbf{E}(\mathbf{r},t) \cdot \mathbf{v}
  \, \Big]
=  \mathrm{St} [ \Psi ]\:,
    \end{array}
 \end{equation}
 where $\omega_c$ is the cyclotron frequency and
  $\mathbf{E}(\mathbf{r},t)$ is the electric field
   consisting  of the external field $\mathbf{E}_{ext}(t)$
 and the internal field $\mathbf{E}_{int}(\mathbf{r},t)$
   due to violation of the local charge neutrality
 in the electron gas.  The electric field term
 $ - e\mathbf{E}_{int}\cdot \mathbf{v}  _{\mathbf{p}} $ in Eq.~(\ref{kin})
 can be combined with the gradient term
   $\mathbf{v} _{\mathbf{p}}  \cdot  \partial  \Psi /
 \partial \mathbf{r} $ by the change
  $\Psi (\varepsilon , \alpha,  \mathbf{r}, t)
  \to \Psi(\varepsilon , \alpha ,  \mathbf{r}, t)
  + e \Phi_{el} (\mathbf{r},t)$, in which the zero harmonic
 by the angle $\alpha $ of the generalized
 distribution function  $\Psi$  takes into account
  the electric potential
  $\Phi_{el}$   ($\mathbf{E} _{int} =
 - \nabla \Phi_{el}$).

 One can distinguish the two types of the
  transport regimes, described
  by Eq.~(\ref{kin}) \cite{Alekseev_Alekseeva_}.

 First, when the cyclotron frequency $\omega_c$ and the
frequency $\omega$ of  variation of  $\Psi (\mathbf{r},t)$
 are slow as compared  with the characteristic value
of the operator $\mathrm{St} $ as well as
 the characteristic wavevector $k$ of the function
 $ \Psi (\mathbf{r},t)$ is small compared
  with $\mathrm{St}    /v_F$, the true hydrodynamic regime
  is realized. The function $\Psi$ is divided on the
 two parts:
 \begin{equation}
  \label{Psi_gen}
   \Psi = \Psi_0 + \Psi_1
       \:,
 \end{equation}
   where  $\Psi_0$ is  the locally equilibrium
 distribution, being    the correction
 to the Fermi function due to    slowly varying
 parameters $\mu$, $T$,  or the hydrodynamic velocity
 $\mathbf{V}$, while  $\Psi_1 \ll \Psi_0$
 is  the nonequlibrium part,   whose form depends
 on the considered transport problem.
 This part  $\Psi_1 $ describes dissipation and,
 for the given problem,  is proportional
 to  the fixed angular harmonic by $ \alpha $  and
has the dependence   on $\varepsilon $ of a fixed type.
 The kinetic equation (\ref{kin}) leads   to the two
  macroscopic equations. The first one is  the Fourier
 law-like equation  connecting the gradients of
$\mu$, $T$, or $\mathbf{V}$, defining
the function $\Psi_0$,  with the  flow values related
to the nonequilibrium part  $\Psi_1$
(the heat flow $\mathbf{q}$,
 the shear stress tensor $\sigma_{ik}$,   and so on)
 via the kinetic coefficients. The second one is
 the transport equation,   governing the  slow  evolution
 of $\mu$, $T$, or $\mathbf{V}$ and  containing
  the flow values  $\mathbf{q}$,  $\sigma_{ik}$.

 The macroscopic equation, connecting
  the gradient of equilibrium values with
  the inequilibrium  ones, in the case of
 low frequencies $ \omega,\,\omega_c \ll \mathrm{St} $,
 is derived from Eq.~(\ref{kin}) taking the form:
 \begin{equation}
   \label{kin_kin}
\begin{array}{c}
 \displaystyle
  -f'_F(\varepsilon)\: \mathbf{v} _ {\mathbf{p}} \cdot
  \frac{ \partial  \Psi }
   { \partial \mathbf{r} }
=  \mathrm{St} [ \Psi ]
\end{array}
\:.
  \end{equation}
 Here in the left part we should take into account
  only the locally equilibrium part $\Psi_0 \gg \Psi_1$,
 while the right part  contains only the nonequilibrium
 part $\mathrm{St}[\Psi_1]$,  as the collision integral
 conserve  any locally equilibrium distribution:
 \begin{equation}
    \mathrm{St}[\Psi_0]=0\,.
 \end{equation}
 So we obtain an inhomogeneous integral equation for $\Psi_1$,
 that should be solved to find the kinetic coefficient.
  Adding of the time-dependent and
  the magnetic field terms $\partial  \Psi / \partial t $
  and  $\omega_c \partial  \Psi /\partial \alpha $
 in Eq.~(\ref{kin_kin}) leads to the time dispersion
 and the magnetic field dependencies of kinetic coefficients,
 which both are  relatively weak
  at $\omega,\omega_c \ll \mathrm{St} $.

For the first time, the  problems of hydrodynamic-like
transport regimes in 2D electron systems
was considered within Eq.~(\ref{kin_kin})
 in Refs.~\cite{Gurzhi_et_al_}-\cite{Gurzhi_et_al_4_}.

 One of such regimes  is the low-frequency viscous transport
 in which  an inhomogeneous flow $\mathbf{V}(\mathbf{r},t)$
 leads to diffusion of momentum. The kinetic coefficient
 in this problem  is the viscosity $\eta $
 defined as the coefficient in the linear relation
 between the shear stress tensor,
  \begin{equation}
 \label{sigma_def}
 \sigma _{ik}  (\mathbf{r},t)
  = \int \frac{d ^2 \mathbf{p} }{2\pi^2 \hbar^2}\:
   f_F'(\varepsilon) \, \Psi (\varepsilon,\alpha) \, p_i \,
    \mathbf{v} _ { \mathbf{ p } \, k }
  \:,
\end{equation}
 and the gradients of the velocity, $\partial V_i / \partial x_k$:
  \begin{equation}
   \label{eta_def}
  \sigma _{ik} = \eta \,
   \Big( \,\frac{ \partial V_i }{ \partial x_k } +
  \frac{ \partial V_k }{ \partial x_i }\,\Big)
  \:.
 \end{equation}
 Equations (\ref{sigma_def}) and  (\ref{eta_def}) are written
 for the case of an incompressible flow
 for which $\mathrm{div }\mathbf{V} =0$.
  The perturbed  part $\Psi $ (\ref{Psi_gen}) of the distribution
 has the form:
 \begin{equation}
   \label{Psi_vis}
\Psi ( \varepsilon, \alpha ) = \mathbf{V} \cdot \mathbf{p} +
 \frac{mv_F^2 }{4}
 \,\Big(
 \frac{ \partial V_i }{ \partial x_k } +
  \frac{ \partial V_k }{ \partial x_i }
 \Big)
 \, F_{ik}(x,\alpha)
\:,
\end{equation}
 where, as usual,  $x = (\varepsilon - \mu  )/T$, the first term
 $\Psi_0 = \mathbf{V} \cdot \mathbf{p} $ describes the locally
 equilibrium flow,  and the functions  $F_{ik}(x,\alpha)$,
 determining the nonequilibrium part $\Psi_1$, are
 proportional to the second angular harmonics
by the velocity angle $\alpha$:
 \begin{equation}
   \label{Fxxxy}
\begin{array}{l}
F_{xx}(x,\alpha) = F(x)\cos(2\alpha)
\:,
\\
 F_{yy}(x,\alpha) = - F(x)\cos(2\alpha) \:,
\\
F_{xy}(x,\alpha)= F(x) \sin(2\alpha)
\:.
    \end{array}
 \end{equation}
The function  $F(x)$ should be found from
the inhomogeneous  integral  equation (M9):
   \begin{equation}
   \label{int_eq_vis}
  \mathrm{ St}[ \, F(x)\cos(2\phi) \, ](\varepsilon_1 , \alpha )
   = - f_F'(\varepsilon_1 ) \cos (2 \alpha)\:.
   \end{equation}
 As the right part of this equation   and   the kernel
 of the operator $\mathrm{St}$ are  even relative to
 the changes $ (\varepsilon_1 - \mu )  \to
 - (\varepsilon_1 - \mu ) $  and  $x_1,x' \to -
  x_1,-x' $ [see Eqs.~(\ref{St__at_0})
 and (\ref{St__at_pi_00_3})],  the resulting function $F(x)$
  will be also even  relative to $x \to - x$. Below   we will
 solve it and  calculate from the obtained $F(x)$ the shear stress
 relaxation time $\tau_{ee,2} $  which is determined
  via  the viscosity as:
 \begin{equation}
   \eta = mnv_F^2\tau_{ee,2}/4
    \:.
 \end{equation}
From Eqs.~(\ref{sigma_def})-(\ref{Psi_vis}) we have
 for this time parameter:
\begin{equation}
 \label{tau_2_ee}
  \tau_{ee,2}  =  \int d x \: f_F'(x) \,
     F(x)
  \:.
\end{equation}

In  the low-frequency heat transport, an inhomogeneous
temperature $T(\mathbf{r},t) = T+ \delta T (\mathbf{r},t)  $
induces a transfer  of heat due to diffusion of hot
and cool particles  in opposite directions.
  The thermal conductivity coefficient
 $\kappa$ is defined as the coefficient proportionality
  in the Fourier law,
 \begin{equation}
  \mathbf{q}= -\kappa \, \nabla  \delta T
   \:,
 \end{equation}
 connecting the gradient $\nabla  \delta T $
and the heat flow:
 \begin{equation}
  \label{q}
     \mathbf{q} (\mathbf{r},t)
  = \int \frac{d ^2 \mathbf{p} }{2\pi^2 \hbar^2}\:
  [-f_F'(\varepsilon)]
   \, \Psi (\varepsilon,\alpha) \,
    (\varepsilon - \mu)\, \mathbf{v}_{\mathbf{p}}
  \,.
 \end{equation}
The heat flow relaxation time $\tau_{ee,h}$ can be defined
 as the coefficient in the formula
 \begin{equation}
 \label{kappa}
 \kappa = T n v_F^2 \tau_{ee,h}
 \,.
 \end{equation}
The distribution function for this problem has the form:
 \begin{equation}
\label{psi_h_main}
\Psi ( \varepsilon, \alpha ) =\delta T
 \,  \frac{\varepsilon - \mu }{T} +
 v_F \, \frac{\partial  \, \delta T}{\partial x_i}
 \, G_i (x,\alpha)
\:,
\end{equation}
 where the first term $\Psi_0 =
 \delta T \,  (\varepsilon - \mu ) / T
 $ is the locally equilibrium part and
  the inequilibrium part is determined
  by $G_x (x,\alpha) = G(x) \cos \alpha $
  and $G_y (x,\alpha)  = G(x) \sin\alpha  $.
For such function $\Psi $, equation~(\ref{kin_kin})
 becomes the  inhomogeneous integral equation
  for $G (x) $:
 \begin{equation}
\label{int_eq_hea}
 \begin{array}{c}
 \displaystyle
   \mathrm{St} [\,G(x) \cos \phi \,]
   ( \varepsilon _1 , \alpha  )
= -  f_F'(\varepsilon _ 1 ) \,  x_1 \,
\cos \alpha
\:,
 \end{array}
\end{equation}
 where $x_1 = (\varepsilon _1 - \mu  )/T$.  After
 obtaining $G(x)$ from it,   we can calculate
 the time $\tau_{ee,h}$ by the formula:
  \begin{equation}
 \label{tau_h_ee}
  \tau_{ee,h}  =  \int d x \: f_F'(x) \,
    x \, G(x)
    \:.
\end{equation}

Note that in the case of a sufficiently strong inter-particle  interaction,
 2D electrons  should be treated as a Fermi liquid.
 Instead of the distribution function of electrons of type (\ref{delta_f}),
 one should consider a similar distribution function $f_{\mathbf{p}}$
 of quasiparticles
 and take into account the difference in the energy of a quasiparticle in the absence of other quasiparticles, $\varepsilon_p^0 =p^2/(2m)$,
 and its energy in the presence of non-equilibrium  quasiparticles with the distribution function $\delta f_{\mathbf{p}}$,
  having the form:
  \begin{equation}
  \label{e_Fermi_liq}
  \varepsilon_p = \varepsilon_p^0 +
  \sum _{\mathbf{p}'} F_{\mathbf{p}\mathbf{p}'} \delta f_{\mathbf{p}'}
  \:,
    \end{equation}
  where the second term is the Landau interaction term
  and we omit the spin variable \cite{LP_9_}.
   In this case, expressions (\ref{sigma_def})
    and (\ref{q}) for the viscous stress tensor $\sigma_{ik}$
   and the heat flux $\mathbf{q}$ change their forms and become proportional not to the integrals of the
   the full perturbation of the distribution function, $\delta f_{\mathbf{p}} = f_{\mathbf{p}} - f_F(\varepsilon_p^0)$,
    but to the nonequlibrium part $\delta \widetilde{f}_{\mathbf{p}} =  f_{\mathbf{p}} - f_F(\varepsilon_p) $ of
   the distribution function,
   where $\varepsilon_p$ includes the term with  the Landau function
   $ F_{\mathbf{p}\mathbf{p}'} $ according to Eq.~(\ref{e_Fermi_liq})
    \cite{tau_thermocond_1_}-\cite{LP_10_}.

 Second, if  the characteristic wavevector  $k$  of the function
  $\Psi (\mathbf{r},t)$  is small compared
   with the characteristic value of $\mathrm{St}  /v_F$
     or the values $\omega/v_F$, $\omega_c/v_F$,
  but the frequency $\omega$, $\omega_c$ are not small:
  \begin{equation}
 \omega ,\, \omega_c  \sim \mathrm{St}
  \;\; \mathrm{or} \;\;  \gg \mathrm{St}
  \:,
  \end{equation}
  the transport regime similar
  to the hydrodynamic one can be realized.
 In it, the distribution function $\Psi$
 is not divided on the equilibrium   and
 nonequilibrium parts.  Now its form is controlled by
  the inequalities $v_F k \ll \omega ,\, \omega_c ,
  \: \mathrm{or }\; \mathrm{St} $, dictating the relations
between the magnitudes of the angular harmonics
  of $\Psi$.  However, can be still valid  the decomposition
  $\Psi = \Psi_ 0 + \Psi_1$, similar to Eq.~(\ref{Psi_gen}),
 in which   the first term $\Psi_0$ describes
 the kinematic (not equilibrium) motion of
 the electron gas  in  fields and is responsible
 the values of particle flows,  while $\Psi_1$ describes
  dissipative processes  \cite{Alekseev_Alekseeva_}.
  The macroscopic  kinetic relations, following
  from Eq.~(\ref{kin_kin}),  between
 the flows proportional to $\Psi_1$ and
the gradients of the values related to $\Psi_0$
 are again valid, but become nonlocal in time.

The high-frequency viscous transport in
 a magnetic  field  is  an example of such
 hydrodynamic-like regime.  The relations between
 the shear stress tensor $\sigma_{ik}(\mathbf{r},t)$
  and the gradients
  of velocity $\mathbf{V}(\mathbf{r},t)$ are given in
 Refs.~\cite{Alekseev_Alekseeva_},\cite{je_visc_,vis_res_,Sem_}.
 In those works,  a simplified consideration neglecting
 the energy dependence of $\Psi_1$ was performed.
  The resulting viscosity coefficients becomes
  strongly dependent  on the frequencies
 $\omega$ and $\omega_c$.

Based on the exact form (\ref{St__at_pi_00_3}) of the part
 $\mathrm{St}_\pi$ of the collision operator $\mathrm{St}$,
 we will show in the next subsection that
  the inverse operator  $\mathrm{St}^{-1}$
  for the even functions $F(x)$,  $F(x) =F(-x)  $,
 turns out to be close  to the projector on the function
  $\Phi^0_0(x) = 1 $. Because of this,
 all the terms in the kinetic equation (\ref{kin})
  for the problems of the  viscous transport
 have, roughly speaking, approximately
  the same energy dependence.
This can be presented by the schematic equality
(``the relaxation time approximation''):
  \begin{equation}
  \label{simplif}
  \mathrm{St}^{-1}
  \, \to   \, - \tau_{ee,2}
  \:.
  \end{equation}
  Thus   the simplified consideration of
   the viscous transport  in Refs.~\cite{je_visc_}-\cite{Sem_}
   neglecting the energy dependence
    of $\Psi_1$ is, in fact, asymptotically exact.

  The problem of the heat transport (in zero
 as well as in nonzero  magnetic field)
   seems to cannot be simplified in a similar way.
 Its rigorous treatment   requires the solution
  of the kinetic equation (\ref{kin}), which becomes
  the integral equation for  $G_{x}(x)$ and $G_{y}(x)$,
 with the inhomogeneous term proportional to
 $ x_1\,f_F'(\varepsilon_1) $,
 the homogeneous terms with the factors $\omega$,
 $\omega_c$ and the integral term
 $\mathrm{St}[\Psi_1]$. Such inhomogeneous term
 is odd relative to the change $x _1 \to -x _1$.
  We will demonstrate in the next subsection that
  for the kinetic equation with   the inhomogeneous term
the simplification of $\mathrm{St}$ by the change:
\begin{equation}
  \label{simplif_bad}
  \mathrm{St}^{-1}
  \, \to \, - \tau_{ee,h}
  \:,
\end{equation}
is not possible. Therefore the functions $G_{i}(x)$,
  apparently,  are not universal for different relations
  between $\omega$, $\omega_c$, and  $\mathrm{St}$,
 as it takes place for the high-frequency viscous transport
 in magnetic field (see the above paragraph).
 Correspondingly, the thermal conductivity cannot be
 expressed in a closed form as a function of $\omega$
 and $\omega_c$, as it was done for the viscosity tensor
  $\hat{\eta}(\omega,\omega_c)$  in
 Refs.~\cite{je_visc_},\cite{vis_res_}.

One can be interested in the problem  of relaxation
 of different types of spatially homogenous perturbations.
 Such process should be studied
 on the base of Eq.~(\ref{kin})
with only the first and the last terms remained:
  \begin{equation}
    \label{kin_kin_kin}
  -f'_F(\varepsilon)\:  \frac{ \partial  \Psi }
   { \partial t }
    =  \mathrm{St} [ \Psi ]
   \:.
 \end{equation}
 This problem can be important, for example, for
 determination of the limits of applicability of
 the hydrodynamic description with respect to flow geometry
 and for numerical studies of the high-frequency
 heat transport,  when simplification (\ref{simplif_bad})
  of $\mathrm{St}$ is impossible.  In addition,
   a nontrivial transport regime related to
   different relaxation rates of the odd
   and even harmonics, the superdiffusion of electrons
 by the velocity angle,   was considered for 2D electrons
 in recent publications (see, for example,
   Ref.~\cite{Ledwith_3_})

 To solve this problem, one  should find the (weighted)
 eigenfunctions $ \Psi _i  $ and the eigenvalues
  $ \lambda_i $ of  $\mathrm{St}$ (for details of
the eigenvalue problem see the next section):
   \begin{equation}
\label{eig0}
  \lambda_i \Psi _i \, [-f'_F(\varepsilon)\:  ]
=  \mathrm{St} [ \Psi _i ]
\:.
\end{equation}
 Then the general solution of Eq.~(\ref{kin_kin_kin})
 is written as:
  \begin{equation}
\label{gen_rel}
  \Psi (t) = \sum \limits_{i}  c_i \Psi _i
  \,  e^{\lambda_i t}
 \:.
 \end{equation}

 Applicability of  the ``relaxation time approximation''
 (\ref{simplif}) for kinetic equation (\ref{kin_kin_kin})
  with the distribution functions  proportional
 to each harmonic $m$ means that $ \mathrm{St}$
has the eigenfunction:
 \begin{equation}
    \label{Psi_m_0}
      \Psi_m (x,\alpha)= \cos(m \alpha) \,,
 \end{equation}
 [and also $\sin(m \alpha)$] with no dependence
 on the energy variable $x$: $F_m (x) \equiv 1$.
 The corresponding eigenvalues
 $\lambda^{(m)}$, $\lambda^{(m)} \in \{\lambda_i\}$,
  are the rates of relaxation of
  the ``pure angular'' perturbations.

 We see from Eq.~(\ref{Fxxxy}) that such function (\ref{Psi_m_0})
 for  $m = 2$  has the form of  the nonequilibrium part
 of the distribution function  $\Psi$ (\ref{Psi_vis})
  for the viscous transport, if the rule (\ref{simplif})
   is applicable. Therefore, the problem  of calculating
 the viscosity $\eta$  and calculation of the relaxation rate
  of the functions (\ref{Psi_m_0}) at $m=2$ are directly
   related. Namely,   the $m=2$  eigenvalue   is
$ \lambda^{(2)} = -1/\tau_{ee,2} $, where $\tau_{ee,2} $
  is the shear stress relaxation time, defined above
   as the time parameter
 in the viscosity coefficient
 $ \eta = mnv_F^2 \tau_{ee,2}/4$.

\subsection{4.2. The mathematical structure
  and  the method of solution
    of the kinetic equation  }
 In Section 3, we examined the contributions to
the operator $\mathrm{St}$  from different types of
 collisions. In this section, we will use
  the obtained  properties  of these contributions
  to consistently construct  the  solution
   of the kinetic equation for the problems
   of the viscous transport, of  the heat transport,
 and of relaxation of higher angular  harmonics.

For the viscous transport,  the distribution function
 $\Psi_1$ is proportional to the second angular harmonics
  $m=2$: $F_{xx}(x,\alpha) = F(x) \cos(2\alpha) $
 [see Eqs.~(\ref{Psi_vis}) and (\ref{Fxxxy})].
 We must consider  that  $F(x)$ is even
  by $x$ due to the even inhomogeneous term
 in Eq.~(\ref{kin_kin}),   but, generally speaking,
 $F(x)$ could  substantially depend on $x$,
  as it takes place in the 3D case
   \cite{tau_thermocond_1_,tau_thermocond_2_}.

The main contribution to the angular-dependent part
of $\mathrm{St}$  at low temperatures, $\zeta \ll r_s$
 for such $\Psi_1 $  originates  from the head-on collisions
  [the term in Eq.~(\ref{St__at_pi_00_3}) proportional
 to $I_s$]. Herewith the collinear, $|\varphi | \ll 1$,
   and the head-on collisions, $\pi - | \varphi |\ll 1$,
   give the same contributions to  the angle-independent part
of $\mathrm{St}$ [Eqs.~(\ref{St__at_0}) and (\ref{St__at_pi_00_3})
   proportional to $I_0$ and $I_\pi$].  In the case of
the  weak interparticle interaction, when $\zeta \gg r_s$,
 the angular-dependent part of $\mathrm{St}$ again
  comes predominantly from the head-on collisions,
 but the angular-independent part  of  $\mathrm{St}$
 originates now originates
 only from the collinear scattering.

 The kinetic equation Eq.~(\ref{kin_kin}) with
 the collision operator   $\mathrm{St}=
 \mathrm{St}_{0}+\mathrm{St}_{\pi}$  for
 the odd functions $F(x)$, $F(-x) = F(x)$,
 can be presented as: $Qf=c$,  or, in the exact
  form:
\begin{equation}
    \label{vec_kin_eq}
    \int Q(x,x') \, f(x')\, dx'   = c(x)
 \end{equation}
where  the function $f(x)$ is proportional
 to the unknown  function $F(x)$:
\begin{equation}
\label{v}
f(x)   = \frac{F(x)}{2\, \cosh(x/2)}
\:;
\end{equation}
the function  $c(x)$ is proportional
 to the inhomogeneous term in Eq.~(\ref{kin_kin}):
\begin{equation}
\label{b}
c(x)= \frac{1}{2\, \cosh(x/2)}
\:;
\end{equation}
the kernel of the operator $Q$  in the case
 of low temperature, $\zeta \ll r_s$,
 is given by the formulas:
 \begin{equation}
   \label{A}
  Q(x,x')=2 Q_0(x,x') + Q_1(x,x')
\:,
\end{equation}
where the first term is
 the angular-independent part:
\begin{equation}
  \label{A0}
   \begin{array}{c}
\displaystyle
Q_0(x,x') =  K(x-x')-
\\
\\
\displaystyle
- \frac{\delta (x-x')}{c(x')} \int K(x'-x'')
 \, c(x'') \, dx''
\:,
    \end{array}
 \end{equation}
\begin{equation}
      \label{K}
   K(x)
   =
   \frac{ T^2 }{\varepsilon _F } \,
   \frac{I_\pi }{ 4 \pi \hbar } \,
   \frac{x/2}{ \sinh(x/2)}
\:,
\end{equation}
while the second term in Eq.~(\ref{A})
 is the angular-dependent part:
 $Q_1(x,x') = -  \xi \,  K(x-x')$, where
 \begin{equation}
   \label{xi}
    \xi = \xi(m)
     = m^2 I_s/I_\pi \ll1
 \end{equation}
  [$m=2$ for the viscous transport, but
 the further consideration is valid
 for any even $m \ll I_\pi/I_s$].
 These formulas imply that the kernel
 of the collision operator $\mathrm{St}_{\pi}$
 is related to the kernel  of the operator $Q = 2Q_0 +Q_1$
  as:
 \begin{equation}
   \label{relation}
 \mathrm{St}_{\pi} (x,x') =
 [Q_0(x,x')+Q_1(x,x')] \,c(x)\,c(x')/T \: .
 \end{equation}
 Doubling of the operator $Q_0$ in $Q$~(\ref{A})
 occurs due to the equal contributions from the collinear
 and the head-on collisions to the angle-independent  part
  of the collision  operator
$\mathrm{St}$: $ \mathrm{St} ^{in}= \mathrm{St}_0^{in} +\mathrm{St}_{\pi} ^{in} $
 [see Eqs.~(\ref{St__at_0}) and
  (\ref{St__at_pi_00_3})].

 For the case of the higher  temperatures, $\zeta \gg r_s$,
 we should make only one change, $2Q_0 \to Q_0$,
 in the above formula (\ref{A}) for  $Q $.
 All other formulas (\ref{vec_kin_eq})-(\ref{relation})
 remain valid.

 Note that the derivative of the Fermi function
 is connected with $c(x)$ as: $-f_F'(x) = c(x)^2$.

From  Eq.~(\ref{A0}) it is seen  that
\begin{equation}
    \int Q_0(x,x')\, c(x') \, dx' =0\,.
 \end{equation}
 This means that $c(x)$ is the eigenfunction $\Phi _0^0(x) $
  of the operator $Q_0$ corresponding to the zero eigenvalue
  $\lambda_0^0=0$.  The function $\Phi _0^0(x)  = c(x)$
  is normalized on unity as:
 \begin{equation}
    \label{norm}
      \int c(x)^2 \, dx =1.
 \end{equation}

 Owing to inequality $\xi \ll 1$, the operator $Q_1$,
 acting on the angular variable of $\Psi$,
  is a small perturbation in the operator $Q$ relative to
  the angular-independent part $2Q_0$.
  As it seen from Section 3, this fact,
 being crucial for our consideration,
  originates from the singular character
 of electron-electron collisions  in the 2D case reflected
 by the Fermi function factor (\ref{Fermi}), the energy
 delta-function factor (M7),   and the sharp
 dependence of  the interaction potential (M6)
 on the scattering angles.

  So the eigenvalues and the eigenfunctions
  of $Q$ can found by the perturbation  theory
  by $Q_1$.  It  can be proved by the analysis
of the operator $Q_0$ in the Fourier transformed form
 that the eigenvalues $\lambda _i^0$  and
 the eigenfunctions $\Phi_i^0 (x)$ of the operator $Q_0$
   are discrete and not degenerated  (see analogous
 considerations for the similar kernels in 2D and 3D cases
 in Refs.~\cite{tau_thermocond_2_} and \cite{Novikov_}).
For the eigenvalues $\lambda _i$ and the eigenfunctions
 $\Phi_i (x)$  of the operator $Q $
  we have in this case:
\begin{equation}
 \label{eig_main}
     \begin{array}{c}
  \lambda_i = \lambda_i ^0 + \Delta \lambda_i \:,
  \quad\quad
\Delta \lambda_i =   \langle \, \Phi _i^0 \, ,
 \, Q_1  \Phi _i^0  \,  \rangle
  \:,
  \\
  \\
  \Phi _i (x)\approx \Phi _i^0 (x)
  \:,
   \end{array}
 \end{equation}
where $ \langle  \Phi _1,\Phi_2 \rangle  = \int
 \Phi _1(x) \, \Phi_2(x)  \, dx $ is the scalar product
 in the Hilbert space $\{\Phi (x) \}$ of
   real functions $\Phi (x) $.

In particular, for the minimal eigenvalue $\lambda_0$
of the operator $Q$, corresponding
 to zero eigenvalue $\lambda_0 ^0 =0 $
  of the operator $Q_0$ we have:
  \begin{equation}
  \label{eig_zero}
  \begin{array}{c}
  \lambda_0 =  \langle \,   c\,,Q_1 c \, \rangle
   = \int dx \, d x'\, c(x) Q_1(x,x') c(x')=
    \\
  \\
  =\int dx\,  d x'\, c(x) Q(x,x') c(x')=
   \\
  \\
  =\int d\varepsilon_1 \, \mathrm{St}_{\pi}
  [\, F(x) \equiv 1\,](\varepsilon_1)
  \:.
   \end{array}
  \end{equation}

Solution of the kinetic equation (\ref{vec_kin_eq})
 can be now written  using  the basis $\Phi_i^0$
  of the eigenfunctions of $Q$.   The kernel $Q_{inv}(x,x')$
  of the inverse operator $Q^{-1}$
  is given by the usual formula:
\begin{equation}
\label{A_minus_1}
  \begin{array}{c}
  \displaystyle
Q_{inv}(x,x') = \sum \limits _{i} \frac{\Phi_i(x) \Phi_i(x') }
{\lambda_i} \approx    \\
  \\
  \displaystyle\approx
\sum \limits _{i} \frac{\Phi_i^0 (x) \Phi_i^0 (x') }
 {\lambda_i^0 + \Delta \lambda_i }
\:.
   \end{array}
\end{equation}
Thus the solution of Eq.~(\ref{vec_kin_eq})
 is written as:
\begin{equation}
\label{sol}
 f(x) = \sum \limits _{i} \frac{\Phi_i^0 (x)  }
 {\lambda_i^0 + \Delta \lambda_i }
 \int \Phi_i(x') \, c(x') \,dx' \:.
\end{equation}

Due to the discussed above properties of $\lambda_i^0$,
 in Eq.~(\ref{sol}) the term with $i=0$ dominates.
 Thus the function $f(x)$ is approximately expressed
  via the minimal eigenvalue $\lambda_0 =  \Delta \lambda_0 $
 and its eigenfunction $\Phi_0 \approx \Phi_0 ^0= c$.
 In this way, we have from Eqs.~(\ref{norm}) and (\ref{sol}):
\begin{equation}
  \label{sol_appr}
    f(x) \approx  \frac{c(x)}{ \Delta \lambda_0}  \:.
\end{equation}
 In terms of the function $F(x)$ result (\ref{sol_appr})
 and takes the form:
\begin{equation}
\label{sol_F}
 F(x) = \frac{1}{ \Delta \lambda_0}  \:.
\end{equation}
For the relaxation time $\tau_{ee,2}$ in the viscosity
 coefficient, according to  Eq.~(\ref{tau_2_ee}),
  we obtain:
\begin{equation}
   \label{tau2_secind__rev}
   \tau_{ee,2 } = \frac{- 1}{  \Delta \lambda_0} =
   - \Big \{  \int d \varepsilon _1  \: \mathrm{St} _{\pi}
    [ \, \cos(2 \phi ) \, ] (\varepsilon _1, 0)
   \Big \}^{-1}
  \:.
 \end{equation}

In this  way,  the special character of the kernel $Q$
[the existence of the eigenfunction $\Phi _0^0(x)= c(x)$
corresponding to the zero eigenvalue $\lambda_0^0=0$
 for the dominating part $Q_0 \gg Q_1$]
 greatly simplifies solution the kinetic equation.
 Its proper solution, being asymptotically
  exact by the parameter $\xi \ll 1$,
 is a constant [see Eq.~(\ref{sol_F})].

Now we briefly discuss the ways of solution of
the kinetic equations (\ref{kin}) and  (\ref{kin_kin})
 for the other problems of the hydrodynamic transport:
 the viscous transport at nonzero frequencies $\omega$ and $\omega_c$,
  the heat transport, and the auxiliary problem of relaxation
 of angular harmonics $\Psi_m$ (\ref{Psi_m_0}).

 For the problem of high-frequency viscous transport
 in magnetic field, the part $\Psi_1 (x,\alpha) $ of
the disturbed distribution function $\Psi(x,\alpha)$ is now
 proportional  to $ \cos(2\alpha)$ and $ \sin(2\alpha)$.
 The operator $Q$ remains the same for each term
  $F(x)\cos(2\alpha)$ and $F(x)\sin(2\alpha)$ as in
the discussed above case when $\omega,\omega_c=0$.
 The energy dependence  of the inhomogeneous term
 in equations  (\ref{kin}), is even, as
   in the case $\omega,\omega_c=0$:
 \begin{equation}
  \label{symm_b}
   c( x)  =  c(-x)
  \:.
\end{equation}
 One can again apply the operator $Q^{-1}$, expressed
via  formula (\ref{A_minus_1}), to all terms
 in the kinetic equation (\ref{kin}). As the result,
 all the terms of the transformed  kinetic equation,
 $ Q^{-1}c$, $\omega Q^{-1} f$, $ \omega _c Q^{-1}
  \partial f/ \partial \alpha $, and, thus,
 $Q^{-1} Qf =f$ will be proportional   to the eigenfunction
 of $ \Phi _0^0(x) =c(x)=1/[2\cosh(x/2)] $, corresponding to
 the zero eigenvalue $\lambda^0_0 =0$ of the operator $Q_0$.
 So the energy dependence of the solution $f (x)$ will be
 the same,  $ c(x)$,  as for the discussed above
  the problem of slow hydrodynamic
  at $\omega,\omega_c \to 0 $.

  This consideration is the proof that  the operator
 $\mathrm{St} $ in Eq.~(\ref{kin}) just should be replaced by
  its minimal eigenvalue $\lambda _0 = -1/\tau_{ee,2}$,
according to the ``relaxation time approximation rule''~(\ref{simplif}).
 So we  justify the approach of construction
of the high-frequency hydrodynamic equations
 in magnetic field, developed in
 Refs.~\cite{je_visc_}-\cite{Sem_}.

 For the problems of low and high-frequency thermal transport
 in zero and nonzero magnetic field,  the part $\Psi_1(x,\alpha)$
 of the disturbed function $\Psi(x,\alpha) $
  is proportional to the first angular harmonics
 $\cos \alpha$, $\sin \alpha$.   The kernel $Q$ is again consists
 of the two parts:  the main part $Q_0$, being the same
 as for the viscous transport,   and the angular-dependent
 correction $Q_1^h$:
 \begin{equation}
  \label{Heat_oper_ang_dep_est}
     Q= Q_1^h + Q_0
     \:  ,  \quad
     Q_1^h \ll Q_0
     \,,
  \end{equation}
 being different from the one for the viscous transport.
 The estimate (\ref{Heat_oper_ang_dep_est}), as the analogous estimate
 $\xi(m) \ll 1$ [see Eq.~(\ref{xi})] for relaxation of the
 even harmonics,   is due to special character
 of kinematics of 2D electron collisions.  The energy factor
 $d(x) = x\,c(x) $ in the inhomogeneous term  in Eqs.~(\ref{kin})
 and (\ref{kin_kin})  is now an odd function  relative
the change $x \to - x$ [see Eq.~(\ref{int_eq_hea})]:
 \begin{equation}
   \label{asymm_b}
   d(x)  = - d(-x)
  \:.
 \end{equation}
 Such  $d(x)$ turns out to be  orthogonal to the
   eigenfunction  $ \Phi _0^0(x) = c(x)  $ of $Q_0$:
 \begin{equation}
   \int dx \,  \Phi _0 ^0(x)\,d(x) =0
   \:.
\end{equation}
 Therefore, although the main term of the kernel $Q_{inv}(x,x')$
 of the the inverse operator $Q^{-1}$  is proportional to
  $ \Phi _0^0(x) \Phi _0^0(x') /\Delta \lambda_0$,
 where $\Delta \lambda_0 \sim Q_1 \sim \xi$,  the solution
  of the kinetic equation
 \begin{equation}
   \label{k_eq_heat}
   Q g= d
 \end{equation}
is not expressed the simple form similar to Eq.~(\ref{sol_appr}).
 In this case, one needs to perform  solution of
the kinetic equation (\ref{k_eq_heat}) analogous to
Eq.~(\ref{vec_kin_eq}),   but containing $d(x)$~(\ref{asymm_b})
 instead of $c(x)$,   by the method of the  decomposition
 of $ g(x) $ is series by some basis, for example, by
 the eigenfunctions $\Phi _i^0 (x)$ of $Q_0$.  Herewith,
 one should find, first of all,  the distribution function
 $ g(x) $  with the neglect of the small
 angular-dependent term $Q_1 ^h \ll Q_0$
  in the operator $Q$.

 A similar integral equation  for the thermal transport
  in  the 3D Fermi systems was solved
 in Ref.~\cite{tau_thermocond_2_}.

Finally, we present the exact formulas for the problem
 of relaxation of space-homogeneous functions proportional
  to the higher angular harmonics~(\ref{Psi_m_0}).

As we noted in previous subsection, applicability of Eq.~(\ref{simplif})
to the homogeneous kinetic equation (\ref{kin_kin_kin})
means that it has the solutions $\Psi _m e^{\lambda ^{(m)} t}$,
 where   $\Psi _m (x,\alpha)=  F_m(x)\cos (m \alpha) $ with
  $F_m(x) \approx 1 $ [Eq.~(\ref{Psi_m_0})] and $\lambda ^{(m)} $
 is the eigenvalue of the problem:
\begin{equation}
  \label{iterg_eq_rel_tau}
   \lambda^{(m)}  f_m(x) = \int Q^{(m)}(x,x') f_m(x') dx' \:.
\end{equation}
Here $ f_m(x)  = F_m(x) / [2\cosh(x/2)]$ and  $Q^{(m)}(x,x')$
 is the kernel~(\ref{A}) for even $m$ and  is the kernel
 of general collision integral $\mathrm{St}$~(M3)
for odd $m$.  The equality $F_m(x) \approx 1 $ is
 the consequence of  equality~(\ref{simplif}),
 so they both should be proved on base of properties
 of $Q^{(m)}(x,x')$ in the eigenproblem
 (\ref{iterg_eq_rel_tau}) formulated for arbitrary
   $F_m(x)$.

 Our analysis shows that for odd $m$ the operator $Q^{(m)}$
 also has the decomposition
 \begin{equation}
   \label{Q_odd_decomp}
     Q^{(m)}= Q _0+ Q^{(m)} _1\,,\quad Q^{(m)} _1 \ll  Q_0
\end{equation}
on the angular-independent term $Q_0$, identical with
 Eq.~(\ref{A0}) for the  discussed above case of
even $m$, and the smaller angular-dependent term
 $Q_1^{(m)}$.   Therefore inversion of $Q^{(m)}$
in Eq.~(\ref{iterg_eq_rel_tau})  by using of
 Eq.~(\ref{A_minus_1}) leads to  the eigenfunction
\begin{equation}
    f_m(x)   \approx c(x) =  1 / [2\cosh(x/2)]
\end{equation}
of the operator  $Q^{(m)}$, corresponding to the minimal
eigenvalue     $\lambda ^{(m)}  $ of $Q^{(m)}$,
 \begin{equation}
   \lambda ^{(m)} \sim Q^{(m)} _1
 \end{equation}
  [analogous to $ \Delta \lambda _0$ in the above consideration
 for even $m$].  Such eigenvalue provides
 the relaxation time $\tau_{ee,m } \equiv -1/\lambda^{(m)} $
 of pure angular harmonic  perturbations  $\Psi _m (x,\alpha) =
 \cos (m \alpha) $ of odd $m$.

 In this way, we conclude that the time
 of relaxation of even and odd harmonics
is given by the formula:
\begin{equation}
 \label{tau_m_even_secind__rev}
   \tau_{ee,m } =
   - \Big \{  \int d \varepsilon _1  \:
    \mathrm{St}_{\varsigma}  [  \Psi_m ] (\varepsilon _1, 0)
   \Big \}^{-1}
  \:.
\end{equation}
 Here      $   \mathrm{St}_{\varsigma}$ denotes the contribution
to $\mathrm{St}$, being most important  for
 the given $\Psi_m =  \cos(m\varphi) $. For even $m$
 $\mathrm{St }_ \varsigma  = \mathrm{St} _\pi $, as we have
 demonstrated in Section~3.3.  For odd $m$, the contribution
 $ \mathrm{St}_1 [\Psi_m] $  from the angles  $\varphi \sim 1 $
  dominates in the value of $ \mathrm{St}[\Psi_m] $
 (see the proof  of this point below
in Section 5.3).

The other eigenvalues $ \lambda_ i^{(m)} $ of $Q^{(m)}$
 for odd as well as even $m$  have significantly larger
  magnitudes, proportional to the value of
   the operator $Q_0$. They are responsible for relaxation
 of the nonequilibrium distribution both by angle
  and by energy, having the form
 $\cos(m\alpha) \Phi_i(x)$, where $\Phi_i  (x)$
  substantially depends on $x$.  Such eigenfunctions describe
  the effect of thermal conductivity (see the previous subsection)
  and a thermalization of inequilibrium excitations
   of a general type.

\section{ 5. Calculation of the
   relaxation times }
\subsection{ 5.1. Electron departure time
   (quantum lifetime) }
 First of all,
 we note that
 the distribution functions with  $ \Psi _j  = \sqrt{
2 m \varepsilon }  \, \cos \alpha $,   $ \Psi_{n}  = 1 $,  and
 $ \Psi_{\varepsilon }  = \varepsilon- \mu $, describing a
homogeneous particle  flow,   a perturbation of the particle density
and a perturbation of  the energy density,
 do not decay due to inter-particle collisions.
Indeed, for such functions   the expression $S$~(\ref{sq_br}),
 entering the exact collision integral~$\mathrm{St}$~(M3),  is zero. This fact
reflects  the conservation of  momentum, of  number
of particles,  and of energy   in inter-particle  collisions.

In this subsection we  calculate
 the reciprocal departure time
due to interparticle scattering, averaged by energy
 and  determining the width of the electron energy levels
in a Fermi gas:
 \begin{equation}
 \label{tau_q_def}
\begin{array}{c}
\displaystyle
\frac{1}{\tau_{ee,q}} =
\int \frac{d \varepsilon _1 }{T} \,
f_F(\varepsilon_1) \, \sum \limits _{2,3}
 W_{1,2 \to 3,4}  \,
 f_F(\varepsilon_2)
\times
\\
\\
 \displaystyle
\times
 [1-f_F( \varepsilon_ 3)] [1-f_F( \varepsilon_ 4) ]
 \:.
 \end{array}
   \end{equation}
 Here the scattering probability  $W_{1,2 \to 3,4} \equiv
  W (\varphi ,\theta ; \breve{\varepsilon })$ is the same
  as in Eq.~(M4) the main text.

  For the first time, the value $1/\tau_{ee,q}$  for 2D electrons
   was defined and calculated in Ref.~\cite{Chaplik_}.
For details of calculation of  $1/\tau_{ee,q}$   within the method of the density-density response function
see Ref.~\cite{Qian_}.

 From comparison of Eq.~(\ref{tau_q_def}) with the departure terms in
the collision integrals $\mathrm{St}^{in}_0$~(\ref{St__at_0})
and $\mathrm{St}_{\pi}$~(\ref{St__at_pi_00_3}),  one can see that,
 at low temperatures, $\zeta \ll r_s$, the  collinear
 and the head-on  collisions    provides
 the similar contributions   to the rate
  $1/ \tau  _{ee,q}  $:   $1/ \tau ^{0} _{ee,q} =
 1/ \tau ^{\pi} _{ee,q}  $   [namely, this is
 related to the equality $I_0=2I_\pi$]. Due to
  the presence of the singular factors
   \begin{equation}
\label{singul_0}
   \frac{1}{ \sqrt{ \varphi^2 + \zeta^2 \Delta}}
  \;\; \quad \mathrm{and }\;\; \quad
      \frac{1}{ \sqrt{s ^2 + \Delta} }
      \:
\end{equation}
 in $\mathrm{St}_0$  and $\mathrm{St}_{\pi}$, the contribution
  from the incident electrons  with the  angles
  $| \varphi | \sim 1 $ is negligibly small.
 Thus the departure  rate $1/\tau_{ee,q}$ is calculated
  via  the kernel $ K(x)$~(\ref{K}),
 describing the departure terms in both $ \mathrm{St}_{0} $
and $ \mathrm{St}_{\pi}$:
 \begin{equation}
   \label{tau_q_first_step}
 \frac{1}{\tau ^{0,\pi} _{ee,q}}
 =
 \int dx_1 \, dx' \:  c(x_1) \,  K(x_1-x') \, c(x')\:.
   \end{equation}

For calculation of the integral over the energy variables
 $x_1$ and $x'$ in Eq.~(\ref{tau_q_first_step}),
 it is convenient  to express this integral via
  the four primary variables $x_{i}$, $i=1,2,3,4$,
  expanding the value $K(x_1-x')$ backward
 as the integral of $c(x_i)$ [see, for instance,
  Eq.~(\ref{St__at_00})] and adding the energy delta
  function $ \delta(x_1+x_2 -x_3 -x_4) $. The departure
 rate~(\ref{tau_q_first_step}) takes the form:
  \begin{equation}
   \label{tau_q_first_step_}
  \frac{\hbar}{\tau  _{ee,q}}
 =   \frac{T^2}{\varepsilon _F }\,2\,
 \frac{ I_\pi I_E}{ 2  \pi  }
 \end{equation}
where
\begin{equation}
 \label{enegr_integr_res}
 \begin{array}{c}
 \displaystyle
I_E = \int \delta(x_1+x_2 -x_3 -x_4)
\prod \limits _ {i=1} ^4
\frac{ dx_i   }{
 \displaystyle 2\,
  \cosh( x_i/2 )
       }
 \:.
 \end{array}
\end{equation}
 Using the methods of the theory of functions
 of a complex variable, we obtain for this integral:
  \begin{equation}
 \label{int_E}
 I_E= 2 \pi^2/3 \:.
 \end{equation}
   Finally,  from Eqs.~(\ref{tau_q_first_step_})
     and (\ref{int_E}) we  obtain the result:
 \begin{equation}
 \label{tau_q}
 \frac{\hbar}{\tau_{ee,q}} = \frac{  2  \pi }{3}
  \frac{T^2}{\varepsilon _ F} \,   \ln (r_s/\zeta)
  \:.
   \end{equation}

At the weak interparticle interaction  (and not very
low temperatures), $ \zeta ^{3/2 }\ll r_s \ll  \zeta $,
 the factors  $I_0$~(\ref{int2})  and $I_\pi$~(\ref{int_est})
  in the angular-independent parts of
 $\mathrm{St}_0$ and $\mathrm{St}_\pi$ becomes
 strongly dependent on the interaction parameter $r_s$
  as $\sim r_s^2   $, and thus $1/ \tau_{ee,q}$ decrease.
From equations (\ref{int2})  and (\ref{int_est}) it is seen
 that the main contribution to $1/\tau_{ee,q}$
 now originates from the almost collinear collisions
 described by  $\mathrm{St}_0$.
  Thus we obtain:
\begin{equation}
 \label{tau_q_low_rs_init}
 \frac{\hbar}{\tau_{ee,q}} =  \frac{T^2}{\varepsilon _F }
 \frac{  I_0  I_E}{ 4  \pi  }
  \:,
\end{equation}
 that leads to
 \begin{equation}
 \label{tau_q_low_rs}
 \frac{\hbar}{\tau_{ee,q}}
 = \frac{    \pi }{3}
  \frac{T^2}{\varepsilon _ F} \,
    \frac{r_s^2 } { \zeta^2}
  \: \ln (\zeta/r_s)
 \:.
 \end{equation}
It is noteworthy that the obtained value depends
 on temperature only via the logarithm.

\subsection{ 5.2. Relaxation of
  the   shear stress  }
  Now  we complete the solution of the problem of relaxation
   of the  shear stress.  We calculate the time $\tau_{ee,2}$
    entering  in the viscosity $\eta = mnv_F^2\tau_{ee,2}/4$
  and being equal to the characteristic time of relaxation
   of the function $\Psi_2\equiv\Psi_{s0}$
  (as in was proved in Section~4.2):
 \begin{equation}
 \label{Psi_s_0}
  \Psi _{s0} (\varepsilon , \alpha ) = \cos (2\alpha)
    \: .
 \end{equation}

We have demonstrated in Sections 3 and 4,  that
 the main contribution to the relaxation rate
 $1/\tau_{ee,2}$ comes from head-on collisions
  with electrons whose angles lie
  in the interval $\zeta \ll \pi - |\varphi | \ll 1$.
 The rate $1/ \tau_{ee,2}$ is calculated just
 by integration of   the value
 $\mathrm{St}_{\pi}[\Psi_{s0}](\varepsilon _1,\alpha)$
    over the variable $\varepsilon _1$
 [Eq.~(\ref{tau2_secind__rev})].

For a more complete understanding, here we refine the estimates
of the quantity $S$~(\ref{sq_br})  in the whole interval
 of the angle $\varphi \in (0,\pi)$.

It can be seen from Eqs.~(\ref{varphi_4__at_0})
 and  (\ref{varphi'__at_0})   that   the expression  $S$~(\ref{sq_br})
 for the distribution function $ \Psi_{s0}$ (\ref{Psi_s_0})
 is a function of the order of   $\zeta^2 $
 at $|\varphi | \ll 1 $.  At the intermediate angles,
  $|\varphi | \sim 1$, we have derived the estimate (\ref{S3})
  saying that the factor  $S$  consists of the contribution
 of the order $\sim \zeta$ which linearly depends on  $x_{i}  $
  and the contribution of the higher powers  by $\zeta$
   and $x_i$, starting from $\sim \zeta^2 x_i^2$.
 So the nonvanishing terms in $S$ are estimated
 as $\sim\zeta^2 $.  At $ \pi - |\varphi| \ll 1 $
 the value  $S$ takes the form:
\begin{equation}
 \label{curl_Psi_start}
 \begin{array}{c}
 \displaystyle
 S _{\pm}
=
 4 \,  \Big( \,
 \frac{
 s \, b \pm a \sqrt {s ^2 + \Delta }
 }{
 s^2 + a^2
 }
 \, \Big)^2
 \:.
 \end{array}
\end{equation}
Such factors are about unity at $s \sim 1$
 and go to zero as $4(b \pm a)^2/s^2 $ at $s \gg 1 $.
 The last  values $4(b \pm a)^2/s^2 $
    are much greater at  $1 \ll s \ll 1/\zeta $
 [corresponding to    $ \pi - |\varphi  | \ll 1 $]
  than the magnitude of  $S$, $S \sim \zeta ^2 $,
 at the small and the intermediate angles,  $ \pi - |\varphi  | \sim 1 $.
 Note that the factors in the  operators
  $ \mathrm{St} _0$ and $ \mathrm{St} _{\pi}$,
   other than $S$,  have the similar order of magnitude
   as functions of  $|\varphi|$ and $(\pi - |\varphi | )$
   at $ |\varphi |  \lesssim r_s $ for $ \mathrm{St} _0$
    and  $ (\pi - |\varphi |  ) \lesssim  r_s $
   for~$ \mathrm{St} _{\pi}$.

 These estimates quantify the relative importance
 of head-on collisions for relaxation
  of the shear stress~(\ref{Psi_s_0}).

  In this way, according
   to Eqs.~(\ref{St__at_pi_00_3}), (\ref{tau2_secind__rev}),
    and  (\ref{enegr_integr_res}), we obtain:
\begin{equation}
 \label{St_final__at_pi}
 \frac{\hbar}{\tau_{ee,2}} =
  \frac{T^2}{ \varepsilon _F } \frac{I_s 2^2 I_E}{ 2 \pi  }
 \end{equation}
 Substitution of
 the angular integral $I_s$~(\ref{res_angle_int0})
 and the energy integral $I_E$~(\ref{int_E})
  yield the final result for the shear stress relaxation
 time:
\begin{equation}
 \label{tau_2_fin_res}
 \frac{ \hbar }{ \tau_{ee,2} }
 =
 \frac{ 8 \pi }{3}
  \frac{T^2}{ \varepsilon_F }
   \, r_s^2\, \ln
   \Big(\frac{ 1}{\zeta + r_s }\Big)
   \:.
 \end{equation}
  At low temperatures and moderately weak inter-particle
  interaction, when $\zeta \ll r_s $, the logarithm
   in Eq.~(\ref{tau_2_fin_res})
 takes the form $\ln(1/r_s)$  leading to the quadratic
  temperature dependence of the  relaxation rate:
$    1  / \tau_{ee,2} (T )  \sim T^2    $.
At the very weak interaction, corresponding to
 high 2D electron densities,  when $r_s  \ll \zeta $,
the temperature dependence also contains
 the logarithmic factor:
  $ 1  /  \tau_{ee,2} (T )  \sim T^2
   \ln ( \,  \varepsilon _F  /  T  \, ) $.
We remind the logarithm in Eq.~(\ref{tau_2_fin_res})
 arisen from integration of the function
 $1/(\pi - \varphi)$   over the interval
  \begin{equation}
      \zeta \ll \pi- \varphi \ll \zeta /r_s
 \end{equation}
in the case $ r_s \gg \zeta$
and the interval
 \begin{equation}
      \zeta \ll   \pi- \varphi \ll 1
  \end{equation}
in the case $r_s \ll \zeta$ [see Fig.~2(b)
in the main text].

 The   rate $ 1 / \tau_{ee,2  } $ (\ref{tau_2_fin_res})
  in the case of low temperatures, $ \zeta \ll r_s$,
 can be obtained, up to a numerical coefficient,
 from the results of Ref.~\cite{Ledwith_1_}, where the relaxation
 of distribution functions proportional
 to the even angular harmonics were studied for a general
 Fermi gas.  For this purpose, we substitutethe matrix element (M5)
   with the screened Coulomb potential (M6)  into equation~(11)
    from Ref.~\cite{Ledwith_1_}, which provides  the relaxation rate
 $1/\tau_{ee,2}$ with neglect of the change
 of the particle energy in the momentum conservation
  law.  However, such derivation of $1/\tau_{ee,2}$
 is incomplete and inconclusive.

First in Ref.~\cite{Ledwith_1_} (and in Ref.~\cite{Ledwith_2_})
  the  properties of the operator $\mathrm{St}$
  were not  fully studied,   in particular, it was not
   proved   that the energy factor  $F(\varepsilon)$
    in the  eigenfunction  $\Psi_{s0}$ of  $\mathrm{St}$  is
asymptotically equal to a constant. The solution
of the integral  equation~(\ref{int_eq_vis})
   is reduced  to calculation of the averaged
collision operator $\int d \varepsilon _1 \mathrm{St}[\cos(2\phi )] $
  exactly due to the singular character  of
  the eigenproblem for the operator $\mathrm{St}$,
namely, due to existing of the eigenvalue
  \begin{equation}
  \lambda ^{(2 )} = -1/\tau_{ee,2}
  \:,
  \end{equation}
  corresponding to $ F(x)\cos (2\alpha)$ with
   $ F(x) \approx 1 $   and   being much smaller
   by the absolute value than the other
 eigenvalues  $ \lambda _i^{(2 )} $ [corresponding
to the functions $ F_i(x)\cos (2\alpha)$ with $ F_i(x) \neq 1 $].
 It is the facts those allow us to use
 the ``relaxation time approximation''~(\ref{simplif})
  and the corresponding formula
 (\ref{tau2_secind__rev}).

 Second, the neglect of the transfer  of the particle energy
 in the momentum conservation law   at calculation
 of the relaxation rates of the even harmonics
   in Refs.~\cite{Ledwith_2_},\cite{Ledwith_1_} does not
   allow to  use the regularization  of
  the operators $\mathrm{St}_0$~(\ref{St__at_00})
  and $\mathrm{St}_{\pi}$~(\ref{St__at_pi})
       by the factors (\ref{singul_0}), which is crucial
  for obtaining $ 1 / \tau_{ee,2 } $~(\ref{tau_2_fin_res})
 at medium and moderately high temperatures,
  $r_s \sim \zeta$
   and $r_s \ll \zeta$.

In Ref.~\cite{Novikov_} the problem of relaxation
of the shear stress  was solved for the case of
  a strongly non-ideal 2D Fermi liquid   and formula (M1)
  for the shear stress relaxation time was derived.
   In this system the interaction parameter
$r_s$ is large:
 \begin{equation}
     r_s \sim 1
     \quad \mathrm{or \: even } \quad r_s \gg 1 \:,
 \end{equation}
up to the point when the Wigner crystallization
 of 2D electrons occurs.

In Fermi liquid the kernel of the integral equation
 for the energy part of the distribution function
 [$F(x)$ in our notation]  is rather similar
 to the kernel $Q = 2 Q_0 +Q_1 $ of  Eq.~(\ref{vec_kin_eq}).
  We traced the differences between Eq.~(\ref{vec_kin_eq})
  and the analogous equation
  in Ref.~\cite{Novikov_} for the Fermi liquid.
 Based on this analysis,
 we demonstrated that
our result for $\tau_{ee,2}$ (\ref{tau_2_fin_res})
 can be also derived from comparison of the
 formulas of our work and of Ref.~\cite{Novikov_}.
This method of obtaining $\tau_{ee,2}$ (\ref{tau_2_fin_res})
 is roundabout and much more complicated  than our method,
  but is important for additional
  checking of Eq.~(\ref{tau_2_fin_res}).

  The shear stress relaxation time in Ref.~\cite{Novikov_}
   is obtained  in a form of series [Eq.~(23)
  in that work].  The parameter $\alpha   $
   in the integral equation solved
   in Ref.~\cite{Novikov_}  for our case of a Fermi gas
  corresponds   to the parameter:
  \begin{equation}
\alpha _N = 2\,  \Big(\,  1
- \frac{ 8 \, r_s^2 \, \ln (1/r_s)}{ \ln (1/\zeta) }  \,\Big)
 \end{equation}
\{compare Eq.~(\ref{St__at_pi_00_3}) with Eq.~(22)
in Ref.~\cite{Novikov_}\}. For such $\alpha _N \approx 2 $,
equation~(23) from Ref.~\cite{Novikov_} together
 with Eq.~(\ref{tau_q}) for the scattering departure time
  $\tau_{ee,q}$ in our work lead to
  the shear stress relaxation
 time $\tau_{ee,2}$~(\ref{tau_2_fin_res}).

  At  large values of $r_s$,   Fermi-liquid effects
  become important and the random phase approximation,
in particular, Eq.~(M5) and (M6) for the scattering
 probability, turns out inapplicable.  Such scattering
 probability  $W(\varphi ,\theta , \breve{\varepsilon}  )$
  should be calculated via   the Fermi-liquid vertex
  function   $\Gamma ( \mathbf{p} _1 , \mathbf{p} _3 ,
    \mathbf{p} _1 + \mathbf{p}_2) $
 taking into account the Cooperon contribution \cite{Novikov_}.
The upper limit on the parameter $r_s$ when the
 consideration of the current work remains valid
 can be obtained from
  comparison of the decomposition
  of the Fermi-liquid vertex function  over the harmonics
 by the scattering angle $\theta$ $\{$Eq.~(15)
 in Ref.~\cite{Novikov_}$\}$ and the analogous decomposition
 of the RPA  screened potential (M6).

  We obtain from such procedure, using equations
  from Ref.~\cite{Novikov_},  that
   until  $r_s \ll   r_{ s , 0 } $,
 \begin{equation}
 \label{r_s_0}
 r_{ s , 0 }
  \sim
     \frac{1}{
   \ln(
   \, 1/\zeta \,
    )\:
     \ln[ \, \ln(
     \, 1/\zeta\,
    )  ]
    }
   \:,
 \end{equation}
 the  Cooperon renormalization
 of the scattering probability is negligible
 and we can use potential (M6).

At  $r_s \gtrsim 1 $ the  Cooperon renormalization
  of many angular harmonics of the scattering
  probability $W$  may be important. Equations~(M3),
    (M8), (M9) and the formulas   for
    $\theta_{\pm} (\varphi ,\breve{\varepsilon})$,
   $\psi_{\pm} (\varphi ,\breve{\varepsilon}) $
    at $\varphi \to \pi$ (from Section 1.1) lead to result
   the the Fermi liquid result~(M1), if we use
   the scattering probability with the  Cooperon renormalization,
  \cite{Novikov_}:
 \begin{equation}
 \label{W_Fem_liq}
 W  \sim \frac{1}{\ln^2(1/|\varphi - \pi|)}
 \:,
 \end{equation}
  instead of the RPA probability  given by
 Eqs.~(M4)-(M6).

We conclude that result~(\ref{tau_2_fin_res})  is valid
 in the interval of sufficiently low $r_s$,  $  r_s \ll r_{s,0} $,
 where the marginal value $  r_{s,0} \ll 1 $ is givenby Eq.~(\ref{r_s_0}).
  Apparently, relaxation processes due to
  the inter-quasiparticle scattering   in the intermediate interval
  of $r_s$,
 \begin{equation}
   r_{s0}\ll r_s \ll 1 \:,
   \end{equation}
has not been studied yet. In this range of $r_s$,
the Cooperon renormalization  of the scattering
probability $ W $ seems to be also substantial,
 but, apparently,
 the consideration is much more complex
 than ones of the current work and
 of Ref.~\cite{Novikov_} [in particular, for the scattering probability related to the vertex
  function   $\Gamma ( \mathbf{p} _1 , \mathbf{p} _3 ,
    \mathbf{p} _1 + \mathbf{p}_2) $, one should use some complex formula
     that interpolated the limiting forms of the probabilities  (M4)-(M6) and  (\ref{W_Fem_liq})
     at $r_s \ll r_{s0}$ and $r_s\sim 1 $, respectively].

In Ref.~\cite{Sem_}
 the viscous transport of
 strongly interacting  2D electrons (an electron Fermi liquid)
   in a magnetic field  was studied.
Taking into account the interaction of quasiparticles within the Landau Fermi-liquid theory
  leads to a renormalization of the relaxation times and the amplitude of viscosity \cite{tau_thermocond_1_},\cite{tau_thermocond_2_},\cite{Sem_}.
     The final results of Refs.~\cite{Novikov_} and \cite{Sem_}
  for the viscous stress relaxation rate  $1/\tau_{ee,2}$
   at $\alpha _N \approx 2$ ($r_s \sim 1$ and $\zeta \ll r_s$)
  can be presented in the form:
  \begin{equation}
  \label{tau_2_Ferm_liq}
   \frac{ 1}{\tau_{ee,2 } }  =-  (1+F_2)  \int d \varepsilon _1  \:
    \mathrm{St}_{\pi}^{liq}  [  \Psi_{s0} ] (\varepsilon _1, 0) \:.
  \end{equation}
  Here $\mathrm{St}_{\varsigma}^{liq}$ is the collision integral for the Fermi-liquid quasiparticles containing the scattering probability (\ref{W_Fem_liq})
  and
  $F_{m}$, $m=2,3,4...$, are the coefficients of the expansion of the Landau function $F_{\mathbf{p}\mathbf{p}'}$ in harmonics of the angle between  $\mathbf{p} $ and $\mathbf{p}'$.
   Herewith  the amplitude of the viscosity is renormalized by the factor
   $(1+F_2)(1+F_1)$~\cite{Sem_}:
   \begin{equation}
   v_F^2 \to (v_F^\eta)^2 =
    (1+F_1) (1+F_2) v_F^2 \:,
     \end{equation}
     thus the parameter $F_2$
   does not  actually enter the dc viscosity
    $ \eta = nm (v_F^\eta)^2 \tau _{ee,2}/4$.
It was also shown in Ref.~\cite{Sem_}
that interparticle interaction also induced a renormalization of the   cyclotron frequency entering the viscosity coefficients.
  It becomes different from the cyclotron frequency determining the cyclotron resonance,
  which is not renormalized by the interparticle interaction (Kohn's theorem \cite{Kohn}).

\subsection{ 5.3.  Relaxation of
     the higher angular harmonics }
 As it was discussed  in Section~4.1, it is
 instructive to study the relaxation   of
 the distribution functions   proportional
  to the higher harmonics by the velocity angle,
$   \Psi_m(\varepsilon , \alpha) = \cos (m\alpha) $
(\ref{Psi_m_0}) for even as well as for odd $m>2$.

 We present the results on this problem
 only for the case of low temperatures, $\zeta \ll r_s$.

  In Section 4.2 we demonstrated that relaxation
   of the even harmonics $\Psi_m$~(\ref{Psi_m_0}) with $m \geq 4 $
  is  mainly related to the head-on collisions
  and their relaxation rates $1/\tau_{ee,m}$
 are  calculated by Eq.~(\ref{tau_m_even_secind__rev})
  with $\mathrm{St}_\varsigma = \mathrm{St}_\pi$.
  From this formula and formula~(\ref{St__at_pi_00_3})
  for $\mathrm{St}_{\pi}$ we obtain for
 the rates $1/\tau_{ee,m}$:
 \begin{equation}
  \label{tau_ev_low}
 \frac{\hbar}{ \tau_{ee,m} }
  \sim
   \frac{T^2}{\varepsilon_F} \, r_s^2 \,
   \ln \Big( \, \frac{1}{r_s} \,  \Big)  \, m^2
  \end{equation}
 at not too large $m$, $m \ll 1/r_s$, and
 \begin{equation}
  \label{tau_ev_high}
     \frac{\hbar}{ \tau_{ee,m} }
     \sim
\frac{T^2}{\varepsilon_F} \, r_s^2 \, \ln (m)
 \end{equation}
 at rather large $m$, $1/r_s \ll m \ll 1/\zeta$.
   Equation~(\ref{tau_ev_high}) coincides with the result
 obtained in  Refs.~\cite{Ledwith_2_},\cite{Ledwith_1_}
 for  $ 1 / \tau_{ee,m}  $  at    $m \ll 1/\zeta $.
  Result~(\ref{tau_ev_low}) in the interval $m \ll 1/r_s$
 was not obtained in those works because of the neglect of
 the angle and the energy dependencies  of
 the  scattering matrix element   $\tilde{M} _ {\pm} ^2
  (\varphi, \breve{\varepsilon} ) $.

 A more sophisticated problem  is to evaluate the
 relaxation  rates  of  the distribution functions
 proportional  to the third and higher odd harmonics.
In Refs.~\cite{Gurzhi_et_al_}-\cite{Gurzhi_et_al_4_}
it was shown that   the relaxation of such perturbations
is a slow process as compared with the relaxation of
 the even harmonics.  In Ref.~\cite{Ledwith_2_}
  it was derived that  for odd $m$, $m \geqslant 3$
  the collision integral  $\mathrm{St}$ (M3)
  has eigenfunctions:
 \begin{equation}
 \label{eig}
 \Psi_m^{X} (\varepsilon, \varphi)
 = X_m(\varepsilon) \, \cos (m\varphi)
 \:,
 \end{equation}
with  a nontrivial dependencies on energy,
given by the factor $X_m(\varepsilon)$.
 The relaxation rates $ 1/\tau_{ee,m} $
 of the functions  $\Psi_m^{X}$  turn out to be
  much lower, in the factor $\zeta^2$,
than the relaxation rates of the even
  harmonics \cite{Ledwith_2_}.  This result is
due to   the peculiarities of the kinematics of 2D electron
collisions related  to the Fermi factors (\ref{Fermi})
 \cite{Gurzhi_et_al_}-\cite{Gurzhi_et_al_4_}.

In order to estimate  the rates $ 1 / \tau_{ee,m}$
 for odd $m$   within our approach,
 we below summarize the obtained above properties of $\mathrm{St}$
and study  the  behavior of $S$ for odd $m$ (for even $m$
it was described in details above in Section 5.2).

First, the consideration in Section 4 demonstrates
that the energy dependence  of the eigenfunction
of $\mathrm{St}$ proportional to an odd harmonic
and corresponding to the minimal eigenvalue
    $\lambda ^{(m)} $ is  close to $\Psi_m^F (x,\alpha)
    = F (x) \cos(m\alpha)$, where $F (x) \approx \mathrm{const}$,
  in the limit of smallness of the angular part
  of the corresponding operator $Q^{(m)}$
   [Eq.~(\ref{Q_odd_decomp})]. So   the following  question
 remains open:   to what extent the taking into account
  of the energy dependence  of the function
   $X_m(\varepsilon)$~(\ref{eig})   in Ref.~\cite{Ledwith_2_}
 corresponds or contradict
  to the fact that the eigenfunction $\Psi_m^F$
calculated in limit~(\ref{Q_odd_decomp})
  has    the energy factor $F (x) $ being  close to constant.

 %Second, from the decomposition of the factor
 % $S$~(\ref{sq_br})   in the Taylor series
 %  [see Section 2.2,  Eq.~(\ref{S3})]
% we obtain that  at    the intermediate angles,
%  $|\varphi  | \gg \zeta$  and $ \pi - | \varphi  | \gg \zeta$,
%this factor $S$ for  $\Psi_m^F$~(\ref{Psi_m_0})
% with odd $m$ and $F (x) \approx \mathrm{const}$
%should be estimated as $\sim \zeta^2$ in the integrals by $x_i$
% [as the term of the order of $\sim \zeta$ in Eq.~(\ref{S3})
%  is asymmetric relative to the change
%   $x_j \leftrightarrow  - x_j $].

Second, an analysis shows that  the  factor $S$~(\ref{sq_br}), (\ref{S3})
 for  the functions   $\Psi_m^F$~(\ref{Psi_m_0}) with odd $m$
and $F (x) \approx \mathrm{const}$
should be estimated as    $ \zeta^2 a_{ij} $
 in the integrals by $x_i$
 in the interval $\zeta \ll \pi - |\varphi  | \lesssim 1 $,
 where $a_{ij}  \sim 1 $
  [as the term of the order of $\sim \zeta$ in Eq.~(\ref{S3})
  is asymmetric relative to the change
   $x_j \leftrightarrow  - x_j $]..
 So the contribution $\mathrm{St}_\pi^{dep}[\Psi_m^F]$
  to the collision operator from  the angles $\varphi \to\pi $
 is proportional to the integral of the product
 of the part $[S]_2 \sim \zeta^2 a_{ij}$ of the factor $S$,
  the singular divergent factor $1/\sqrt{s^2 +\Delta}$,
and the matrix element $\tilde{M}_{\pm}^2(\varphi,\breve{x}) $
with the static screened Coulomb potential~(M6). This
leads  to the logarithmic divergence by the variable $\varphi$,
 being   similar to one in Eq.~(\ref{asy_angle_int_en})
up to the factor    $\zeta^2 $:
 \begin{equation}
   \label{int_pi_odd}
    I_{odd,\pi} (\breve{x}) =
    \int \limits _1 ^{1/\zeta} \frac{ds }{s}
    \frac{[\,S(\varphi ,\breve{x})]_2}
    {[1+ |a-b|/(\sqrt{2} \,r_s s) \, ] ^2 }
    \:.
 \end{equation}
 In view of the relation   $[S]_{2} \sim \zeta ^2$,
this integral at typical $ x_i \sim 1 $
 is estimated as:
\begin{equation}
   \label{I_pi_odd}
     I_{odd,\pi}  \sim \zeta^{2}   \ln(r_s/\zeta)
   \:.
  \end{equation}
 We remind that in this subsection
 we consider only the case $\zeta \ll r_s$.

 Third, an analysis based on Eqs.~(\ref{varphi'__at_0})
 and (\ref{varphi_4__at_0_short}) yields
  that in the vicinity of the angle $\varphi =0$,
  $|\varphi| \ll1$, the factor $S$ for the functions
 $\Psi_m = \cos(m\varphi)$ becomes independent on $\varphi$
 in the main order by $|\varphi| $
  and takes the form:
  \begin{equation}
       \label{S_0_odd}
   S (\varphi , \breve{x}) = m^2 \,
    \frac{\zeta^2 \Delta(\breve{x})}{4}
        \:.
 \end{equation}
 In follows from ~(M3) that the resulting angle-dependent
 contribution $\mathrm{St}^{dep}_0$ to the collision integral
 from the vicinity of $\varphi =0$  becomes
 proportional to the expression:
 \begin{equation}
  \label{int_0_odd}
     m^2 \, \zeta^2  \int  \frac{dx_2dx_3}
     {f_1f_2f_3f_4}\int _\zeta^{r_s}
      \frac{d\varphi}{\sqrt{\varphi^2
   + \zeta^2 \Delta(\breve{x})}} \, \Delta(\breve{x})
   \:.
   \end{equation}
 In view of result~(\ref{tau_m_even_secind__rev}),
the relaxation rate  of the odd harmonics,
  $1/\tau_{ee,m}$, is expressed via the integral
$\int dx_1 \mathrm{St}^{dep} [\Psi_m]$,
 where $\mathrm{St}^{dep} $ contains the sum of the
  contributions  $\mathrm{St}^{dep} _0$  and
$\mathrm{St}^{dep}_\pi $  being proportional
to factors~(\ref{int_pi_odd})  and (\ref{int_0_odd}).
Owing to  the antisymmetry   of the value
 $\Delta(\breve{x}) = x_3x_4 - x_1x_2 $ relative to
 the change  $x_3,x_4 \leftrightarrow x_1,x_2$,
the integral of expression (\ref{int_0_odd}) by $x_1$
turns out  to zero.  Thus the main part of the integral
 $\int dx_1 \mathrm{St}^{dep}[\Psi_m]$   originates
 from  $\mathrm{St}_\pi^{dep}$    and is proportional
 to the factor (\ref{I_pi_odd}).

We conclude that  the relaxation of the odd harmonics
 is mainly due to the head-on collision,
corresponding to $\zeta \ll \pi - |\varphi|\ll1 $
  and being described by the operator
 $\mathrm{St}^{dep}_\pi$. The relaxation rate $1/\tau_{ee,m} $
 is calculated by equations~(\ref{St__at_pi}),
(\ref{tau_m_even_secind__rev}), and (\ref{int_pi_odd}).
The result of such calculation for $\zeta \ll r_s$ and
 the odd harmonic number $m$ in the range
 $3 \leq m \ll 1/r_s$   takes the form:
\begin{equation}
 \label{res_tau_3}
\frac{\hbar}{ \tau_{ee,m} } \sim
    \frac{T^2}{ \varepsilon_F }
    \, \zeta ^2 \, \ln(r_s/\zeta) \, m^2
 \:.
\end{equation}
This result differs from the one obtained
in Ref.~\cite{Ledwith_2_}  by another way on
 the logarithmic factor $ \ln(r_s/\zeta) $.
 We think that this difference is due to
 taking into account of the exact dependence of
 the scattering matrix element $\tilde{M}^2_\pm$
on the variables  $\varphi$ and $\breve{x}$.

Additionally, let us note that
calculation of relaxation times
 of the distribution function
 proportional to the higher angular harmonics, $m\geq3$,
  for strongly interaction electrons, forming a Fermi liquid,
  should be performed
  with taking into account the results obtained in Ref.~\cite{Sem_} and presented in Subsection 5.2 for the case $m=2$.
    A preliminary analysis shows that equation~(\ref{tau_2_Ferm_liq})
  is generalized to all higher harmonics for the typical Fermi-liquid regime
   when $r_s\sim 1 $, $\zeta \ll 1$ and, thus, $\alpha _N \approx 2$.
    Namely, the relaxation rates of   the functions proportional to higher harmonics $\Psi_m$~(\ref{Psi_m_0}) are calculated by Eq.~(\ref{tau_m_even_secind__rev}) with the added factor $(1+F_m)$ and the substitution $\mathrm{St} \to \mathrm{St}^{int} $:
      \begin{equation}
        \label{tau_m_Ferm_liq}
    \frac{ 1}{\tau_{ee,m } }=
   -(1+F_m)     \int d \varepsilon _1  \:
    \mathrm{St}^{liq}  [  \Psi_m ] (\varepsilon _1, 0)
    \:.
  \end{equation}
An explicit calculation of these rates is
 a difficult problem for future studies. For its solution one needs, generally speaking,
  a more detailed information
 about the kernel of the collision operator $\mathrm{St}^{liq}$ that it was presented in Ref.~\cite{Novikov_} for calculation of
 $\tau_{ee,2}$.

\subsection{ 5.4.  Relaxation of the heat flow}
  Using the results of Section 4,
 one can study the processes in a 2D electron gas
 that involves the relaxation of
the heat flow: the thermal conductivity  and
 the thermoelectric effects an zero
and nonzero frequencies $\omega$ and $\omega_c$.
  Here we consider only the stationary problem
 of   the thermal conductivity
at zero magnetic field.

In section 4.1 it was shown that, in the problems
  of heat transport, the kinetic equation
  (\ref{kin_kin})   for the function
 $\Psi$~(\ref{psi_h_main}) with the nonequilibrium term
 $\Psi_h = v_F \, dT/dx \, \cos \alpha \, G(x)$
is transformed into the closed
form~(\ref{int_eq_hea}). The heat flow relaxation time
 and  the  thermal conductivity coefficient,
should be calculated via $\Psi_h$
 by Eqs.~(\ref{tau_h_ee}) and (\ref{kappa}).

 The energy factor $G(x)$ in the nonequilibrium term
$\Psi_h $ in the distribution function
 $\Psi$~(\ref{psi_h_main}) can be presented
in  the form of series:
 \begin{equation}
  \label{series_Psi_def}
    \Psi _h (x, \alpha) =
    G(x)  \cos \alpha
    \:, \quad\quad
    G(x ) = \sum \limits _{l=1} ^{\infty}
    a_l   x^l
   \:.
\end{equation}
The first term in this formula, being
 proportional to
\begin{equation}
  \label{Psi_1}
  \Psi_{h1}(x, \alpha)
 = x   \cos \alpha
    \:,
\end{equation}
and other term with odd numbers $l$ provides
the contributions to the value of the heat flow (\ref{q}).
 The solution of  the integral equation~(\ref{int_eq_hea})
implies finding all the coefficients $a_l$
 in the decomposition of $G(x)$~(\ref{series_Psi_def}).
  Unlike the case of relaxation  of the ``pure
 angular harmonics'' (Sections~5.2 and 5.3),
in the problem of heat transport all the terms
 in Eq.~(\ref{series_Psi_def})  are substantial
 for constructing the proper solution $G(x)$.

 The main role in the collision operator $\mathrm{St}$
  in  the equation for  $G(x)$ is played by the
 angular-independent part $\mathrm{St}^{in}$ of $\mathrm{St}$,
which contains the equal contributions $\mathrm{St}^{in}_0$
 and $\mathrm{St}^{in}_\pi$ at $\zeta \ll r_s$
 and only the contribution  $\mathrm{St}^{in}_0$
 at $\zeta \gg r_s$ [see Eq.~(\ref{Heat_oper_ang_dep_est})
 and Section 3]. The angular-dependent part
  of the collision integral for $\Psi_h$ and $\Psi_{h1}$
is much smaller and  is not substantial
 for their relaxation.

 The problem of heat transport in a 3D Fermi liquid was solved
  in Refs.~\cite{tau_thermocond_1_},\cite{tau_thermocond_2_}.
 The kernels of the main integral equations for $G(x)$
 in the 2D and 3D cases  differ only by the density
 of states.  According to Ref.~\cite{tau_thermocond_2_},
  the solution of Eq.~(\ref{int_eq_hea}) for $ G (x ) $
 should be performed by using the Fourier transform
   of Eq.~(\ref{int_eq_hea}) by  the variables $x$, $x_1$
 and solving the resulting differential equation
 for the transformed  function
 \begin{equation}
 G(\xi)=\frac{1}{\sqrt2\pi}
  \int dx \,e^{-ix\xi}\,G(x)
 \:.
 \end{equation}

 In this work,  we only estimate the time $\tau_{ee,h}$
 by an order of magnitude. For this purpose, we  truncate
 the  distribution function $\Psi$ (\ref{series_Psi_def})
 down to the first term, being proportional
 to the lowest term  $\Psi_{h1}$~(\ref{Psi_1}):
 \begin{equation}
   \label{first}
    \Psi (x,\alpha ) \to a_1 \Psi_{h1}(x,\alpha )
   \:.
 \end{equation}
 Such function $a_1 \Psi_{h1}  $  is the simplest
 distribution function  carrying a heat flow.
 From Eqs.~(\ref{int_eq_hea}) and (\ref{tau_h_ee})
   we can deduce the estimates for  $a_1\Psi _ {h1}$ and
  the corresponding relaxation time
  $\tau_{ee,h}$.

 The resulting expression   for  $\tau_{ee,h} $ turns out
to be  similar the expression   for the angular harmonics
 relaxation rates (\ref{tau2_secind__rev})  and
(\ref{tau_m_even_secind__rev}):
 \begin{equation}
\label{aver_St__heat_fl}
  \tau_{ee,h}
         \sim
   - \Big\{ \int d \varepsilon _1  \, x_1 \,
   \mathrm{St}[ \,  \Psi_{h1} \, ] (\varepsilon _1 ,0)
   \Big\}^{-1} \:,
\end{equation}
 where the odd factor $x_1 =  ( \varepsilon_1 - \mu) / T \sim 1  $
is  important  as the function  $\Psi_{h1}(x,\alpha)$~(\ref{Psi_1})
 is odd by  $x \to -x $  and therefore the  value
 $\mathrm{St}[ \,  \Psi_{h1} \, ] (\varepsilon _1,\alpha)$
is also odd relative the change $(\varepsilon _1-\mu)
 \to -(\varepsilon _1-\mu)$.

 As it was shown in Section~2.2,
the factor $S(\varphi,\breve{x})$~(\ref{sq_br})
 for $\Psi_{h1}$ being  substantially dependent on $x$
 is estimated as  unity at the intermediate  angles
and for the head-on collisions,  corresponding to
  $ 1 \lesssim  | \varphi | <\pi$.    For the collinear collisions,
 when    $ |\varphi |\ll 1$, it follows from Eqs.~(\ref{Psi_1})
  and (\ref{en_cons_law}) that:
\begin{equation}
   \label{est_S_heat}
     S(\varphi,\breve{\varepsilon}) \sim \varphi ^2 \ll 1
     \:.
\end{equation}
 Based on these estimates of $S$ and  Eqs. (\ref{St__at_00}),
(\ref{St__at_pi}), and (\ref{est_St_1_en})
 for the contributions $ \mathrm{St} _ \varsigma $
 we obtain that at very low temperatures, $\zeta \ll r_s$,
the main effect on  the rate $1/\tau_{ee,h}$~(\ref{aver_St__heat_fl})
 originates from the angles:
 \begin{equation}
 \label{int_tau_h}
  \zeta/r_s\ll \pi - |\varphi |  \ll 1
  \:.
  \end{equation}
 In this interval,   the inequalities $|\theta_{\pm} (\varphi)|
 \gg \zeta $  and  $| \psi_{\pm} (\varphi)|  \gg \zeta $
 are fulfilled,   thus we can use the static inter-particle
  potential~(M6)   and all the formulas from Section 3.3
   for the operator~$\mathrm{St}_\pi$.
 As the result,   the value $1/\tau_{ee,h}$ is calculated
 by angular-independent part   $\mathrm{St}_\pi^{in}$
of $\mathrm{St}_\pi$~(\ref{St__at_pi_00_3})  with the factor
  $I_\pi$~(\ref{asy_angle_int_en_res}).  Estimating
 the energy variables $x_1$ and $x'$ in Eq.~(\ref{St__at_pi_00_3})
  as unity, we obtain from Eq.~(\ref{aver_St__heat_fl}):
\begin{equation}
 \label{res_tau_heat_fl}
  \frac{ \hbar } {\tau_{ee,h} } \sim
  \frac{T^2}{ \varepsilon_F }  \,  \ln(r_s /\zeta)
 \end{equation}
for the current case  $\zeta \ll r_s$.

At moderately low temperatures, $r_s \ll \zeta \ll1 $,
according to the analysis in the Section 3.3,
the head-on collisions, corresponding to
$\pi-|\varphi |\ll 1$, do not provide the most significant
contribution $\mathrm{St}_\pi^{in}$
 in the angular-independent part of $\mathrm{St}^{in}$
as compared with the scattering with electrons
  with the angles $|\varphi | \sim 1$.
In view of Eq.~(\ref{est_S_heat}),
  the factor $S$~(\ref{sq_br}) for $\Psi_{h1} $ is small
for the collinear collisions, when  $ |\varphi| \ll 1 $,
therefore they also does not provide
 the substantial contribution     to $\mathrm{St}^{in}$.
Thus the main contribution to  $1/\tau_{ee,h}$
 comes from the scattering by the incident electrons
with the angles $|\varphi |\sim 1$, that is described
by the contribution~$\mathrm{St}_1^{in}$,
estimated by Eq.~(\ref{est_St_1_en}).
In view of Eqs.~(\ref{est_St_1_en}) and (\ref{aver_St__heat_fl}),
 in the case  $\zeta \gg r_s$ we have the estimate:
\begin{equation}
 \label{res_tau_heat_fl2}
 \frac{ \hbar } {\tau_{ee,h} } \sim
  \frac{T^2}{ \varepsilon_F }  \,  \,\frac{r_s^2}{\zeta^2} =  r_s^2 \varepsilon_F \:.
\end{equation}
It is noteworthy that such rate is independent
  on temperature.

At very low temperatures, $\zeta \ll r_s$,
the derived heat flow relaxation rate
   $1/ \tau_{ee,h}$~(\ref{res_tau_heat_fl})
 has the magnitude of  one half of the departure rate
  $1/\tau_{ee,q}$ (\ref{tau_q}).
This is related to the following facts: (i) the heat flow
 being transferred by an electron is completely lost in each head-on
 collision; (ii) these collision  provide the main contributions
 to the rate  $ 1/ \tau_{ee,h} $  and  the one-half contribution
   to the rate $1/ \tau_{ee,q} $.

At moderately low temperatures, $r_s \ll \zeta \ll1 $, the rate
 $1/ \tau_{ee,h}$~(\ref{res_tau_heat_fl2}) turns out
  to be much smaller than the the departure rate
  $1/\tau_{ee,q}$~(\ref{tau_q_low_rs}), as for the rate $1/ \tau_{ee,q}$
   the main effect  originates from the collinear collisions,
    $|\varphi| \ll  1$, while their role in the relaxation
     of the heat flow function $\Psi_{h1}$~(\ref{Psi_1}) is weakened
due to the particular form of the factor $S$
 for $\Psi=\Psi_{h1}$ [see Eq.~(\ref{est_S_heat})].

For the case of a strongly interaction electrons
    with the parameters $r_s\sim 1 $, $\zeta \ll 1$ and  $\alpha _N \approx 2$,
   result (\ref{tau_2_Ferm_liq})
  can be generalized to the relaxation of the heat flow.
  As this flow within our approximation
   is proportional to the integral of the function $\Psi_{h1} \propto \cos\alpha$,
 it is calculated in Fermi liquid by the analog of Eq.~(\ref{aver_St__heat_fl})
  with the added factor $(1+F_1)$ and the substitution $\mathrm{St} \to \mathrm{St}^{liq} $,
  similarly as it was done for $\tau_{ee,m}$ in Eqs.~(\ref{tau_2_Ferm_liq}) and (\ref{tau_m_Ferm_liq}):
      \begin{equation}
        \label{tau_h_Ferm_liq}
    \frac{ 1}{\tau_{ee,h } }=
   -(1+F_1)     \int d \varepsilon _1  \: x_1 \,
    \mathrm{St}^{liq}  [  \Psi_{h1} ] (\varepsilon _1, 0)
    \:.
  \end{equation}
An explicit calculation of these rate is also
 a difficult problem for the future studies that requires
 a detailed information about the kernel of $    \mathrm{St}^{liq} $.

\section{6. Difficulties in comparison
   of experiment and theory }
 A hydrodynamic  mechanism for
 negative magnetoresistance, similar to the one used
 in the current paper to explain the experiments,
 was proposed  for the first time  in Ref.~\cite{Gurzhi_Kopeliovich_}
 for a Poiseuille flow of electron-phonon fluid in pure bulk 3D metals
 with the  strong electron-phonon interaction.
The viscosity of the electron fluid in magnetic field,
 similar to one in formula (M13), was obtained for the first time
 in Ref.~\cite{Steinberg_} for bulk 3D metals,
 in which electrons scatter on phonons.

The possibility of the hydrodynamic regime
in real samples, apparently,  is associated
with the substantial relaxation  of  the odd
 and the even harmonics of $\Psi$ in scattering of electrons
on disorder   with the rates $1/\tau_{imp,m}$,
 being much greater than $1/\tau_{ee,m}$ for odd $m \geq 3$
and comparable  with $1/\tau_{ee,m}$ for even  $m$
 (see Section 5.3 and Fig.~3 in the main text).
 If there were only relaxation due to electron-electron scattering, a strong difference in the relaxation times of even and odd harmonics
   could lead to realization of anomalous non-hydrodynamic transport regimes \cite{Gurzhi_Kopeliovich_,Ledwith_3_}.

  In Fig.~3 in the main text
   the experimental data on $1/\tau_{ee,2 } $ are presented
    for the samples for which      the Loretzian  profile (M13)
  of the hydrodynamic  magnetoresistance  is observed
   very well   in a wide range of magnetic fields,
  including    the interval
     \begin{equation}
     \omega_c \lesssim 1/\tau_{2}
      \:.
 \end{equation}
  This situation corresponds that
  the hydrodynamic
   regime of the electron flow is formed
  in the whole range of magnetic field.
  For realization of such regime, at least, the inequality
 $l_{2 }   < W $    for the width $W$ of the conducting channel  must be fulfilled
  Here $l_{2 } = v_F \tau_{2}$ is the shear stress relaxation length
     and  the time
     $\tau_{2}$ corresponds
     to the total rate  of the relaxation of the shear stress:
     \begin{equation}
     \label{Matthiessen}
     \frac{1}{ \tau_{2} } = \frac{1}{ \tau_{imp,2} } +  \frac{1}{  \tau_{ee,2}  }
     \:,
     \end{equation}
as it was implied in Eq.~(M14) in the main text.

However, if the mean free path $l_{2}$ is greater  than the
  width of the conducting channel $W$, the ballistic regime is realized
   in small magnetic fields $B< B_0$, corresponding to
 \begin{equation}
 2R_c >W \
  \:,
 \end{equation}
  while  the hydrodynamic one becomes possible
   only at the higher fields $B>B_0 $ corresponding to
    \begin{equation}
   2R_c <W
   \:,
 \end{equation}
   when a substantial part of electrons in the cental part of the sample does not
   scatter on the channel edges.

    In this case, only
    the segment of the curve $\varrho_{xx} (B) \propto 1/B^2 $~(M13)
  at large magnetic fields,  $|B|>B_0$,
  can be  observed, and therefore it becomes impossible to extract
 relaxation time $\tau_2$ from the widths of magnetoresistance curve.
At the fields, $|B|<B_0$, the magnetoresistance is
 determined by the ballistics
 effects controlled by the relations between the magnitudes of
 the scattering length, the sample width $W$ and lengths $L$,
  and the cyclotron radius $R_c$ \cite{we_5_},\cite{we_52_}.

 Matthiessen's rule for the shear stress
 relaxation rate, Eq.~(\ref{Matthiessen}) [Eq.~(M14) in the main text], is applicable both in the cases the 2D electron gas ($r_s \ll 1$)
 and the electron Fermi liquid ($r_s \sim 1$).
 Indeed, to describe both weakly nonequilibrium electrons and weakly nonequilibrium quasiparticles
in a disordered sample,
 kinetic equation  (\ref{kin}) is applicable, in which, we should include, in addition to the interparticle collision integral $\mathrm{St} = \mathrm{St}^{ee}$ leading to relaxation of the $m\geq 2$ harmonics, the collision integral of scattering on disorder, inducing the relaxation of the $m\geq 1$ harmonics:
     \begin{equation}
\mathrm{St} \to \mathrm{St} ^{ee} + \mathrm{St}^{imp}
   \:.
 \end{equation}
 In accordance with Ref.~\cite{Sem_} and of Subsections~5.2 and 5.3,
  both contributions $1/\tau_{ee,2}$ and $1/\tau_{imp,2}$
  to the relaxation rate $1/\tau_2$ are
   proportional to the products of the Landau parameter factor $(1+F_2)$
    and
    the quasiparticle collision integrals $\mathrm{St} ^{ee} $ and  $ \mathrm{St}^{imp}$
    averaged by the energy variable $\varepsilon_1$.
   For the inter-quasiparticle relaxation of the second
   harmonic $
   \Psi_2 (\phi) = \Psi_{s0} (\phi) = \cos(2\phi) $~(\ref{Psi_s_0})
    such rate is given by Eq.~(\ref{tau_2_Ferm_liq}).
   For the relaxation of  $\Psi_{s0}$ due scattering of quasiparticles on disorder,
   analogously  Eqs.~(\ref{tau_2_Ferm_liq}) and (\ref{tau_m_Ferm_liq}), we have:
   % [provided... ] NOVIKOV sdetja ssilku vishe , chto Q1 << Q0 i dlja Fermi-liquid
   \begin{equation}
   \frac{1}{\tau_{imp,2}}  = - (1+F_2)\int d \varepsilon_1 \, \mathrm{St}^{imp}[\Psi_{s0}](\varepsilon_1, 0)
   \:.
 \end{equation}

Finally, we note that in the limit of lowest temperatures and high magnetic field
the momentum  relaxation time on bulk disorder, $\tau_{imp,1} $, can
be estimated  from the resistance $\varrho (\infty)$ \cite{je_visc_}.
Although  it is usually presumed
that   $\tau_{imp,1} $ and $\tau_{imp,2} $
are of the same order
 of magnitude, the simple fitting procedure performed in Ref.~\cite{je_visc_}
 (and in Refs.~\cite{Gusev_1_},\cite{Gusev_3_} by the method of \cite{je_visc_})
 led to a surprisingly large difference  between
  these two times in high-mobility GaAs quantum wells,
   $\tau_{imp,1} $ turns out to be longer than $\tau_{imp,2} $  in 10-100 times.
 We believe that  the  resolution of this issue
  will provide a very important information
  about the nature of the 2D electron system
  in the high-mobility structures.

\end{document}